\documentclass[acmsmall,screen]{acmart}

\usepackage{listings}

\usepackage{algorithm} %format of the algorithm
\usepackage{algorithmic} %format of the algorithm

\usepackage{multirow} %multirow for format of table
\usepackage{enumitem} %缩进
\usepackage{amsmath}
\usepackage{xcolor}
\usepackage{subfigure}
\usepackage{mwe}
\usepackage{booktabs}
\usepackage{threeparttable}
\usepackage{algorithm}
\usepackage{algorithmic}
\usepackage{amsmath}
\usepackage{amsfonts}
\usepackage[switch]{lineno}
\usepackage{tabularray}

\usepackage{xpatch}
\usepackage{array}
\usepackage{verbatimbox}

\AtBeginDocument{%
  }

\setcopyright{acmlicensed}
\copyrightyear{2018}
\acmYear{2018}
\acmDOI{XXXXXXX.XXXXXXX}

\acmJournal{JACM}
\acmVolume{37}
\acmNumber{4}
\acmArticle{111}
\acmMonth{8}

\begin{document}

\title{Collaborative Knowledge Fusion: A Novel Approach for Multi-task Recommender Systems via LLMs}
% \author{Chuang Zhao}
% % \authornote{Both authors contributed equally to this research.}
% \email{czhaobo@connect.ust.com}
% \affiliation{%

\author{Chuang Zhao}
\affiliation{%
  \institution{Department of Electronic and Computer Engineering, The Hong Kong University of Science and Technology}
  \city{Hong Kong}
  \country{Hong Kong}}
\email{czhaobo@connect.ust.hk}

\author{Xing Su}
\affiliation{%
  \institution{AI Lab at Lenovo Research	}
  \city{Beijing}
  \country{China}}
\email{suxing2@lenovo.com}

\author{Ming He}
\authornote{Both authors are corresponding authors.}
\affiliation{%
  \institution{AI Lab at Lenovo Research	}
  \city{Beijing}
  \country{China}}
\email{heming01@foxmail.com}

\author{Hongke Zhao}
\affiliation{%
  \institution{College of Management and Economics, Tianjin University}
  \city{Tianjin}
  \country{China}}
\email{hongke@tju.edu.cn}

\author{Jianping Fan}
\affiliation{%
  \institution{AI Lab at Lenovo Research	}
  \city{Beijing}
  \country{China}}
\email{jfan1@lenovo.com}

\author{Xiaomeng Li}
\authornotemark[1]
\affiliation{%
  \institution{Department of Electronic and Computer Engineering, The Hong Kong University of Science and Technology}
  \city{Hong Kong}
  \country{Hong Kong}}
\email{eexmli@ust.hk}

\renewcommand{\shortauthors}{Chuang Zhao et al.}

\begin{abstract}
  %  Rec
  Owing to the {impressive general intelligence} of large language models (LLMs), there has been a growing trend to integrate them into recommender systems to gain a more profound insight into human interests and intentions. 
  % existing work
  Existing LLMs-based recommender systems primarily leverage item attributes and user interaction histories in textual format, improving {the single task} like rating prediction or explainable recommendation.
  % to enhance performance in specific recommendation tasks, such as rating prediction or explainable recommendation.
  % weakness
    Nevertheless, these approaches overlook the crucial contribution of traditional collaborative signals in discerning users' profound intentions and disregard the interrelatedness among tasks.
  % overview+
  To address these {limitations}, we introduce a novel framework known as \emph{CKF}, specifically developed to boost multi-task recommendations via personalized collaborative knowledge fusion into LLMs. 
  % specific
    Specifically, our method synergizes traditional collaborative filtering models to produce collaborative embeddings, subsequently employing the meta-network to construct personalized mapping bridges tailored for each user. {Upon mapped, 
    the embeddings are incorporated into} meticulously designed prompt templates and {then fed} into an advanced LLM to represent user interests.
    To investigate the intrinsic relationship among diverse recommendation tasks, we develop Multi-Lora, a new parameter-efficient approach for multi-task optimization, adept at distinctly segregating task-shared and task-specific information. This method forges a connection between LLMs and recommendation scenarios, while simultaneously enriching the supervisory signal through {mutual knowledge transfer among various tasks}.
      % exp
      Extensive experiments and in-depth robustness analyses across four {common recommendation} tasks on {four} large public data sets substantiate the effectiveness and superiority of our framework.
\end{abstract}





%%
%% This command processes the author affiliation and title
\maketitle

\begin{figure}[!ht]
  \centering
   \includegraphics[width=0.7\linewidth]{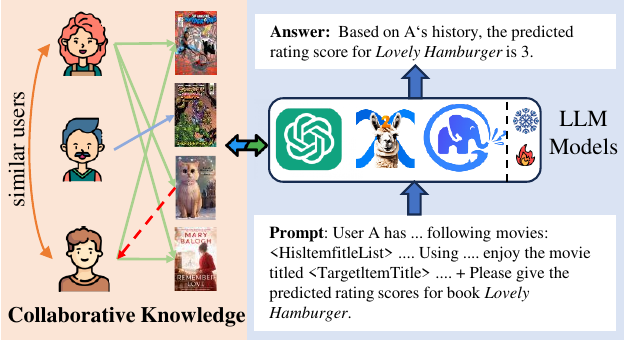}
   \caption{The combination of collaborative knowledge and LLMs. Lines in distinct colors depict various user-item interactions, while icons of fire and snow symbolize trainable and frozen LLM parameters, respectively.}
   \label{fig:motiv}
\end{figure}

\section{Introduction}\label{sec:intro}
% rec + LLM
Recommender systems, tailored to identify user-specific preferences, are essential in easing the challenge of information overload~\cite{zhu2022personalized, liu2023graph,zhao2024dense}.
Conventional recommendation models strive to capture better user and item collaboration signals using advanced deep learning architectures, such as graph neural network{s} or attention mechanisms~\cite{qi1,zhaowww,dang2023ticoserec}.
{However, these models encounter challenges in effectively understanding users’ interests, capturing textual information, and reasoning on their predictions~\cite{zhao2024recommender}}.
With the advent of {pre-trained language models} ~\cite{wu2023survey,min2023recent,zhao2024lane}, the focus on harnessing their potential to improve recommender systems has gained considerable interest.

{
Early efforts focus on integrating small-scale pre-trained language models (PLMs) into recommender systems~\cite{qiu2021u,zhao2022resetbert4rec}, which feature parameter counts ranging from several hundred thousand to several hundred million~\cite{zhao2023survey,zhao2024recommender}. 
Recently, efforts to integrate large language models (LLMs) into recommendation systems have flourished as researchers discover that scaling PLMs (e.g., increasing model size or data size) often enhances the model capacity for downstream tasks~\cite{zhang2023collm,wu2023survey}.
These LLM-based methods~\cite{bao2023bi,talrec} employ substantially larger language models like Llama~\cite{llama} or GPT-3~\cite{brown2020language}, with parameter sizes spanning from billions to several hundred billion.
% The main distinction between PLMs and LLMs is that the latter boast a greater number of parameters and emergent capabilities~\cite{ding2023parameter}. 
Following~\cite{wu2023survey,lin2023can}, we discuss PLM-based methods as part of the LLMs review.
Technically}, recommender systems employing LLMs can be primarily classified into two distinct categories~\cite{wu2023survey,lin2023can,hao2023multi}: discriminative and generative.
Discriminative approaches treat LLMs, especially BERT-based variants, as embedded backbones~\cite{qiu2021u, zhang2022gbert}, concentrating efforts on aligning learned {text} representations with {those of} the recommendation domain.
In contrast, generative methodologies predominantly reinterpret recommendation tasks as natural language tasks, typically employing in-context learning~\cite{brown2020language} or prompt learning~\cite{li2023personalized}. This practice enables LLMs to grasp recommendation task requirements, thus facilitating {the} direct generation of recommendations.
Given GPT's remarkable capabilities in recent studies~\cite{li2023hamur,di2023retrieval}, the latter is gaining increasing attention and spurring further exploration into its potential applications.

% overview + specific
Despite notable advancements in prior work, there remain two critical limitations that hinder further performance enhancement. 
First, existing research primarily emphasizes the generative capabilities of LLMs, often neglecting the proven value of traditional collaborative knowledge~\cite{wu2022survey} in uncovering deeper user intentions~\cite{geng2022recommendation}.
Envisage a scenario where two items, despite possessing comparable profiles, are favored by divergent consumer cohorts. This subtle distinction in consumption patterns presents a significant challenge for LLMs, given their predominant focus on textual semantics rather than on the intricate co-purchase patterns of consumer behavior.
While several efforts, including CoLLM~\cite{zhang2023collm}, have been made, they often fail to distinguish between user and item semantic spaces and overlook personalized transfer, potentially leading to noise.
Second, the majority of current LLM-based systems are tailored for specific recommendation tasks~\cite{acharya2023llm}, like rating prediction~\cite{zhao2022resetbert4rec} or Click-Through Rate (CTR) estimation~\cite{talrec}, resulting in an inability to perceive {intricate}
correlations between various tasks. However, emerging evidence suggests that systems trained in multiple related tasks~\cite{yuan2022tenrec,wang2023single,wang2023multi} can more effectively interpret user behavior.
For instance, CTR estimation discerns user's immediate interests, and rating prediction gauges content quality and satisfaction, together providing a holistic view of user preferences.
This underscores the necessity for empowering LLMs to handle multi-task recommendations.

% Our naive approach
A straightforward solution is leveraging traditional collaborative filtering (CF) models for user-item embedding generation, followed by their incorporation into multi-task prompt templates for joint parameter efficient fine-tuning~\cite{lora}, as shown in Figure~\ref{fig:motiv}.
Nevertheless, the semantic disparity between these embeddings and those employed by LLMs, arising from variances in training methodologies and data sets, precludes their direct incorporation into prompts. 
Moreover, employing a uniform set of fine-tuning parameters for diverse tasks may introduce noise or omitting crucial information, given the intricate interactions and correlations between  tasks~\cite{wang2023multi}. 
% {This is demonstrated by the limited performance of CKF-NML, as shown in Table~\ref{tab:aba}.}

% further approach
In light of these considerations, we propose an innovative framework known as CKF (Collaborative Knowledge Fusion) to address the aforementioned challenges. 
To tackle the semantic misalignment in knowledge fusion, we propose training the meta-networks on historical interaction records. These networks are designed to generate personalized mapping functions for user and item embeddings derived from collaborative filtering models. Subsequently, these embeddings are seamlessly integrated into carefully designed prompt templates via mapping functions, thereby enriching collaborative knowledge.
{Unlike general linear mapping~\cite{zhang2023collm}, our approach  distinguishes the semantic space of each user/item and enhances recommendation ability for new users/items, as evidenced in Section~\ref{sec:5.1}. For instance, if the distribution of a user in the test set significantly deviates from that in the training set, a generic collaborative signal mapping might prove ineffective. In contrast, our approach considers the user's specific historical interactions and performs an effective mapping.}
To exploit the heterogeneous relationships across diverse recommendation tasks, we develop `Multi-Lora', a novel parameter-efficient fine-tuning strategy. 
Our approach incorporates a shared low-rank matrix to effectively capture the associations between tasks while employing multiple unique low-rank matrices to pinpoint specific attributes of each task.
% This explicit decoupling method helps the model gain in-depth insight into task correlations and alleviate the phenomenon of negative transfer.
{Compared to adopting a set of Lora for each task, this explicit decoupling method not only gains in-depth insight into task correlations but also significantly reduces the required fine-tuning parameters. 
}
Furthermore, our experiments in {S}ection~\ref{sec:5.3} indicate that uniformly training on mapping functions alongside fine-tuning LLM may lead to performance decline. 
This is ascribed to the initial marked progress in collaborative knowledge, prompting models to concentrate on honing mapping functions, consequently impacting their capacity for semantic comprehension.
% This is attributed to the initial advancements from collaborative knowledge, which concentrate on the former, potentially resulting in a diminished capacity for semantic comprehension, thereby inducing the model to take shortcuts.
To counteract this, we develop a curriculum learning strategy, designing a smooth weighting function to mitigate such dependency.
% Additionally, XXX

% our contributions
Our key contributions can be summarized as follows:
\begin{itemize}[leftmargin=12pt]
    \item To the best of our knowledge, \textit{CKF} stands as the pioneering LLMs-based recommendation framework that incorporates collaborative knowledge. 
    This novel approach forms a bridge between traditional collaborative filtering models and generative recommendation frameworks leveraging LLMs.
    % This comprehensive exploration enhances the overall performance of drug recommendation, marking a novel advancement in the field.
    \item We devise an innovative parameter{-}efficient fine-tuning strategy for multi-task recommendations, focusing on the explicit decoupling of task-shared and task-specific information. This approach yields deeper insights into the intricate interrelationships among tasks and enhances the effectiveness of {mutual} knowledge transfer.
    \item Extensive experiments and detailed discussions on {four} large public data sets, showcasing substantial advancements achieved by CKF. We have released the code and data sets for follow-up and reproducibility.
\end{itemize}

The remainder of this paper is structured as follows, {S}ection~\ref{section2} reviews the closet-related work done. Section~\ref{section3} formalizes the definition of diverse recommendation tasks and elaborates each sub-module of \textit{CKF}. In {S}ection~\ref{section4}, ~\ref{sec:5}{,} and~\ref{sec:6}, we present
the experimental results, various robust tests{,} and hyper-parameter tests. Finally, we draw a conclusion in {S}ection~\ref{section7}.

\section{Related Work}\label{section2}
% overview
In this section, we review the three most relevant research areas, collaborative filtering-based recommender system, LLMs-based recommender system{,} and parameter efficient learning.
We first provide an overview of their classic genres and representative works, and then clarify the connections and distinctions between the proposed framework and them.

\subsection{Collaborative Filtering-based Recommender System}
Collaborative filtering-based recommender systems are a foundational approach within the broader sphere of recommender systems, designed to predict the preferences of users by collecting and analyzing data from many individuals~\cite{luo2024collaborative,liu2023megcf}. This genre hinges on the assumption that those who agreed in the past will agree in the future about other preferences. 

Three prominent technical solutions that have shaped the development of these systems include matrix factorization, graph neural networks, and sequential recommendation models~\cite{wu2022survey,peng2024less}. Each of these approaches brings unique strengths to handling the complexities of recommendation scenarios, such as capturing user-user/item-item similarities~\cite{chen2020efficient}, leveraging high-order collaborative signals~\cite{limcn4rec}, and understanding dynamic interest evolution~\cite{gao2024smlp4rec}. 
Matrix Factorization (MF)~\cite{wu2022survey} is a cornerstone technique in collaborative filtering, primarily used to discover latent features underlying the interactions between users and items.
LightGCN~\cite{he2020lightgcn} further introduce{s} a powerful way to exploit the rich relational information inherent in recommender systems, where interactions can be naturally represented as graphs. 
DIN~\cite{zhou2018deep} and SASRec~\cite{kang2018self} differ from the static modeling methods employed in the first two approaches. Instead, they view the user's sequential interaction process as an evolving chain of interests, integrating dynamic information to recommend the subsequent item.
{Beyond technique advancement, several studies demonstrate that multi-task learning enhances the recommendation performance of collaborative models~\cite{song2024multi,zhang2024m3oe}. For example, M3oE~\cite{zhang2024m3oe} draws on the idea of the mixture of experts, leveraging three modules to learn common, domain-aspect, and task-aspect user preferences respectively to address the complex dependencies among multiple domains and tasks.}

% {Multi-task recommendation}

Our approach aims to harness the power of meta-learning's mapping function to project collaborative signals into a large language model for multi-task recommendations. We have rigorously tested this idea across three commonly encountered types of collaborative signals. {Additionally, leveraging the powerful language understanding capabilities of LLMs, different tasks can be interconnected using open-world semantics to uncover deeper user preferences. This approach contrasts with CF-based multi-task recommendations, which rely solely on co-occurrence signals. Our Multi-Lora design offers new insights and practical solutions for LLM-based multi-task recommendations.
}

\begin{figure*}[!ht]
  \centering
   \includegraphics[width=0.98\linewidth]{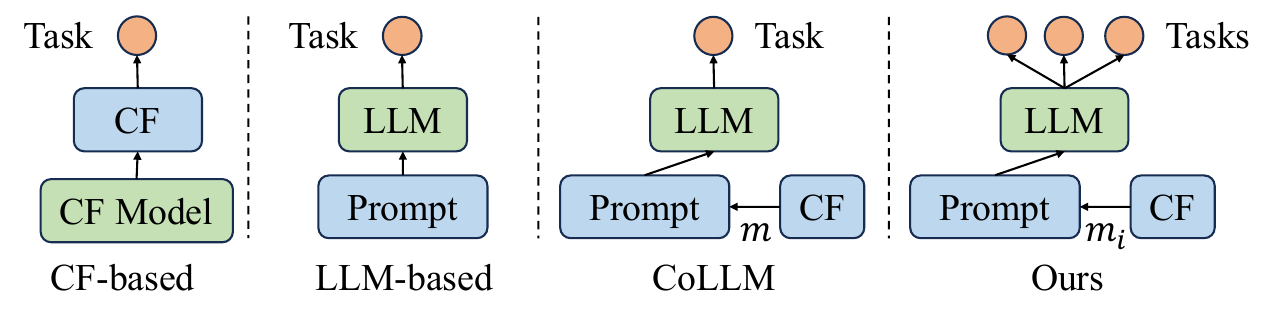}
   \caption{Difference in related work. $m$ denotes the generalized mapping function and $m_{i}$ refers to the personalized mapping function. {The} comparison reveals that the majority of LLM-based research focuses on the individual task without incorporating joint optimization across multiple tasks. CoLLM's collaborative knowledge mapping ignores the semantic gap and personalized {nature} of users and items.}
   \label{fig:diff}
\end{figure*}

\subsection{LLMs-based Recommender System}
% why llm, one sentence for the overall approach
The exceptional proficiency of LLMs in knowledge processing and reasoning has garnered significant interest in developing recommender systems that leverage their nuanced understanding of user context and behavior~\cite{knowfu}.

{Early language model-based recommender systems focus on leveraging small-scale pre-trained language models (PLMs, $\sim$ million parameters), such as BERT variants~\cite{qiu2021u,zhao2022resetbert4rec}, to discern subtle differences in contextual text. Despite their success, limitations in model size and pre-training corpora still render them less effective in handling unseen recommendation scenarios and complex semantic interpretations~\cite{zhao2024recommender,zhao2023survey}. 
Recently, with the demonstrated outstanding language understanding and emergent capabilities of LLMs—a type of PLM that scales significantly in data and parameter volume (over a billion parameters)—research has shifted toward leveraging LLMs for recommendations~\cite{zhao2024recommender}. Due to the similarity in architecture and close connection in training, following~\cite{wu2023survey,lin2023can}, we discuss PLM-based methods as part of the LLM review. Below, we refer to both approaches as LLM-based methods.}
% traditional pipeline
LLMs-based recommender systems can be {technically} classified into discriminative and generative genres, with the latter further branching into non-tuning and tuning approaches~\cite{wu2023survey}.
Discriminative approaches focus on aligning representations from large pre-trained models with domain-specific data, employing fine-tuning or prompt tuning techniques.
As an example, UniSRec~\cite{kddunirec} utilizes BERT for fine-tuning on recommendation data spanning various domains, with the goal of generating universal sequence representations. Prompt4NR~\cite{kprompt} transforms click prediction into a cloze mask prediction task, facilitating the direct application of a fine-tuned language model in recommendation scenarios.
In contrast, generative methods leverage LLMs as task schedulers or recommender systems to generate recommendation results.
For example, ChatRec~\cite{chatrec} operates as a dialog-based scheduler, interpreting user intentions through iterative conversations and collaborating with conventional recommender systems to provide accurate recommendations.
TALLRec~\cite{talrec} undergoes fine-tuning on Llama~\cite{llama} using few-shot recommendation data, thereby augmenting its ability to discern user preferences, including both likes and dislikes. 
{PEPTER~\cite{li2023personalized} designs two prompt templates, discrete and continuous, and simultaneously learns explainable and recommendation tasks, thereby improving performance on explainable tasks.}
{BIGRec~\cite{bao2023bi} additionally incorporates a grounding stage from the recommendation space to the item space, enhancing LLM performance on the Top-K task.}

% our approach difference multi; cf k
Our approach falls under the generative {genre}.
Unlike previous research, our emphasis lies in the integration of traditional collaborative knowledge into the LLMs-based paradigm and the pursuit of joint optimization via a multi-task approach. 
Even though CoLLM~\cite{zhang2023collm} adopts a similar integration idea, its rudimentary mapping design overlooks the semantic gap between users and items and lacks a perspective of multi-task joint optimization. 
A key difference between our work and previous work can be seen in Figure~\ref{fig:diff}.

\subsection{Parameter Efficient Learning}
% why pel, one sentence for the overall approach
With the escalating parameter count in pre-trained models, the costs associated with transfer learning have surged significantly.
This trend has underscored the growing importance of parameter efficient fine-tuning (PEFT)~\cite{pef1,peft0}.

% traditional pipeline
Existing PEFT methods can be classified into three categories~\cite{pef2}: addition-based, specification-based{,} and reparameterization-based. The addition-based approach integrates supplementary parameter networks into existing neural architectures, among which the Adapter~\cite{poth-etal-2023-adapters} and prompt designs~\cite{promptliu} stand out as particularly innovative and effective.
In contrast, specification-based methods fine-tune some intrinsic parameters while keeping most parameters unchanged during model adaptation. These methods typically employ heuristic designs or training-adaptive techniques to selectively prune parameters or streamline the model structure~\cite{spec}. Recently, reparameterization-based methods have emerged, aiming to transform adaptive parameters in the optimization process into a more parameter-efficient format. These methods often operate under the assumption of low-rank characteristics during model tuning, leading to the decomposition of the original large matrices into several low-rank matrices for efficient weight updates. Lora and its variants stand as the most representative methods in this domain~\cite{lora,wang2023multilora}.

% our approach difference multi; 
Our strategy for tuning falls into the category of reparameterization.
In contrast to prior research, our primary emphasis is on designing a novel PEFT to explicitly disentangle task-shared from task-specific semantics, thereby facilitating knowledge transfer.

\subsection{{Distinction and Novelty}}
{Our proposed CKF significantly differs from other solutions in terms of motivation, technical approach, and model performance.
 At the motivation level, CKF focuses on personalized mapping of collaboration signals, treating both users and items individually, unlike methods such as CoLLM that apply a uniform mapping universally. This differentiation is crucial as it acknowledges the distinct semantics derived from users and items. Uniform mapping methods can obscure these semantics, as shown in Figure~\ref{fig:tsne}. Additionally, our approach allows the model to learn {how to construct an effective mapping function}, significantly {enhancing its adaptability} to new users/items. For instance, if the distribution of a user in the {test set} significantly {deviates from} training distribution, a {general} collaborative signal mapping may fail. Conversely, our approach can not only learn from the meta-network {how other users} map the collaborative space to the LLM semantic space but also {make corrections} based on their own interaction history. 
 Section~\ref{sec:5.1} serves as strong evidence for this.
 % Our design creates a unique semantic space for each user, transforming CF signals into LLM space in alignment with the personalized nature of recommendation systems. During inference, our approach allows the mapping function to be tailored to new users or items, rather than uniformly applying a generic function.
Regarding Multi-Lora, our goal is to explicitly separate shared and independent knowledge across multiple recommendation tasks. While similar motivations exist in natural language processing, our adaptation of this concept to the recommendation domain is substantial. Our framework addresses multiple recommendation tasks simultaneously, unlike CoLLM, which is primarily designed for CTR tasks. This capability enables our framework to capture comprehensive user preferences and perform robustly across various recommendation scenarios.
}

{
As for the technical approach, our solution shows significant differences. For the mapping function, we use a meta-network to generate a unique parameter matrix for each user and item, in contrast to CoLLM’s linear mapping approach. 
% This matrix is customized based on their historical sequence data, processed by the meta-network.
More specifically, our meta-network utilizes an attention mechanism that integrates information from a user's historical interactions. During training, each user’s (or item’s) interaction history influences the generation of unique mapping parameters. These parameters are not fixed but are continuously refined as the model encounters different data, allowing the mapping function to evolve in tandem with changes in user behavior or item attributes.
With Multi-Lora, we diverge from typical PEFT practices by adopting a more granular perspective on parameter sharing, as illustrated in Eq.~\ref{eq12}. This strategy requires fewer sets of parameters compared to traditional methods, leveraging the semantics of the attention mechanism to mine related knowledge from shared sources during inference. To ensure the semantics of different Lora, we implement \(\mathrm{q}\)-orthogonalization in Eq.~\ref{eq13} for different tasks.
Additionally, our training methodology employs a curriculum learning strategy, simplifying CoLLM's two-stage training into a streamlined end-to-end process. This strategic revision simplifies the training approach and ensures alignment with our unified model objectives, promoting coherent learning outcomes.
}

{
Concerning the experiments, our method outperforms all CF-based and LLM-based baselines across four data sets, as shown in Table \ref{tab:exp1}-\ref{tab:exp-sup}. We also conduct comprehensive   robustness experiments to assess our model's efficacy in complex scenarios, including warm-cold scenarios, few-shot applications, contextual examinations, etc. These experiments examine our model's performance across different levels of data sparsity and diverse recommendation scenarios. The results, depicted in Section~\ref{sec:5}, consistently show our method surpassing other baselines, validating our innovative approach and practical value.}

{
These distinctions make our framework stand out and underscore our contributions.
}

\section{Methodology}\label{section3}
We first detail the problem definition, then provide an overview of CKF, followed by elaborating {on} each sub-module.

\subsection{Problem Formulation}
% basic typo + warm/cold start
We represent a user as $u\in \mathcal{U}$ and an item as $v \in \mathcal{V}$, where $\mathcal{U}$ and $\mathcal{V}$ are the complete sets of users and items, respectively. Each user $u$ is associated with a historical behavior sequence $\mathcal{I}_{u} = \{v_{1}, v_{2}, \cdots, v_{\mathcal{N}_{u}}\}$, arranged in chronological order and of length $\mathcal{N}_{u}$. Correspondingly, we have the user's ratings and comments sequence, i.e., $\mathcal{R}_{u} = \{r_{u,v_{1}}, r_{u,v_{2}}, \cdots, r_{u,v_{\mathcal{N}_{u}}}\}$ and $\mathcal{C}_{u} = \{c_{u,v_{1}}, c_{u,v_{2}}, \cdots, c_{u,v_{\mathcal{N}_{u}}}\}$.
For multiple commonly recommendation tasks, we give the following definition{s},

% multi task definition
\noindent{\textbf{Rating Prediction~\cite{zhang2023collm} (RP):}}
Given a sequence of user historical behavior $\mathcal{I}_{u}$, we intercept the last item $v_{\mathcal{N}_{u}}$ as the candidate item, and the rest $\mathcal{I}_{u}-v_{\mathcal{N}_{u}}$ are used as input to predict the user's rating $r_{u,v_{\mathcal{N}_{u}}}$, where $r \in [1,5]$.

\noindent{\textbf{Click-Through Rate Estimation~\cite{zhou2018deep} (CTR):}}
Given a sequence of user historical behavior $\mathcal{I}_{u}$, we select the last item $v_{\mathcal{N}_{u}}$ as the ground-truth, and randomly select one item that the user has not interacted with before as the negative item, and combine them into a candidate set $S_{u}$. For each item in $S_{u}$, we ask the model to answer whether to click or not. %Here, we set $\mathcal{N}_{neg}=1$.

\noindent{\textbf{Top-K Ranking~\cite{he2020lightgcn} (Top-K):}}
Given a sequence of user historical behavior $\mathcal{I}_{u}$, we select the last item $v_{\mathcal{N}_{u}}$ as the ground-truth, and randomly select $\mathcal{N}_{neg}$ items that the user has not interacted with before as negative items, and combine them into a candidate set $S_{u}$. We ask the model to pick the interacted items from $S_{u}$. Here $\mathcal{N}_{neg}=10$. {We also examine the impact of Top-K scalability, as discussed in Section~\ref{sec:5.5.2}.}

\noindent{\textbf{Explainable Recommendation~\cite{geng2022recommendation} (Explain):}}
Given sequences of user historical behavior $\mathcal{I}_{u}$, comments $\mathcal{C}_{u}$, and ratings  $\mathcal{R}_{u}$, we select $c_{u,v_{\mathcal{N}_{u}}}$ for the candidate item $v_{\mathcal{N}_{u}}$ and predict the user's rating $r_{u,v_{\mathcal{N}_{u}}}$ for it, where $r \in [1,5]$. Please be aware that this task leverages the sequence of comments instead of the user's rating history, which is utilized in the RP task, with an emphasis on understanding the content. {We also explore the impact of various explainable recommendation settings in Section~\ref{sec:5.5.3}.}
% For clarity, we mathematically form them as follows,

% Our objective is to conduct a joint optimization for prevalent tasks within recommendation systems. 
Our goal is to incorporate collaborative knowledge into LLM and jointly optimize prevalent tasks within recommender systems.
This entails bridging the semantic disparity between CF models and LLMs, while exploiting the intrinsic correlations and shared characteristics among recommendation tasks, thus yielding a more thorough understanding of user preferences and behaviors.
% This will involve harnessing the inherent correlations and overlapping characteristics among these recommendation tasks, thereby providing a more comprehensive insight into user preferences and behaviors compared to models focused on single tasks.
For clarity, henceforth, $\mathcal{I}_{u}$ will denote the historical sequence, excluding {the} candidate item. We also further test the model performance in warm and cold start recommendation scenarios, see {S}ection~\ref{sec:5.1} for details. Important mathematical symbols are shown in Table~\ref{tab:math}.

\begin{table}[!h]
\centering
\caption{Mathematical Notations.}
\label{tab:math}
\begin{tabular}{c|l} 
\toprule
\textbf{Notations}                                                                                                     & \textbf{Descriptions}                                              \\ 
\hline
$\mathcal{U}, \mathcal{V}  $                                                                                            & user set , item set                                         \\
$\mathcal{I}, \mathcal{R}, \mathcal{C}  $   & item sequence, rating sequence, comment sequence                               \\
$S $                                  & candidate item set                    \\
$\mathcal{N}_{u}$            & length of behavior sequence of $u$                       \\ 
$u, v, r, c$            & user, item, rating, comment                       \\ 
\hline
$\mathbf{E}_{\mathcal{U}}, \mathbf{E}_{\mathcal{V}} $                                                                                          & collaborative embeddings of users, items                                   \\
$\mathbf{E}_\mathcal{T}$                                                                                          & embeddings of token                                  \\
$\mathcal{L} $                                                                                          & loss function    \\

$\mathcal{T} $                                                                                          & prompt template for the task    \\
$B, A $                                                                                          & low rank parameters    \\
$\mathrm{e}_{u}$, $\mathrm{e}_{v}$, $\mathrm{e}_{I_{u}}$, $\mathrm{e}_{t}$                                                                                          & embedding of user, item, historical behavior, token                                                        \\
                 
$f(\cdot;\theta)   $                                                                                          & collaborative model                     \\
$\text{Meta}(\cdot;\theta_{2})   $                                                                                          & meta network for personalized mapping function                     \\
$\alpha$                                                                                      & attention score                            \\
$\mathrm{w}$                                                                                  & personalized mapping function                                   \\              
$\mathrm{p}_{u}$                                                                                          & user preference                                              \\
$\varphi, \Delta \varphi$                                                                                          & frozen LLM parameters, trainable LLM parameters                         \\
$\mathrm{q}, \mathrm{k}, \mathrm{v}, \mathrm{o}$                                                                                          & query, key, value, output                         \\
$y, \hat{y}$                                                                                          & ground truth, estimations                                        \\
\bottomrule
\end{tabular}
\end{table}
% 插入表格

% \noindent{\textbf{Warm/Cold Start:}} We examine model performance in both warm and cold scenarios. Specifically, we divide the test set into warm subsets and cold subsets: for the warm subset, we select users and items with more than K interactions, while the cold subset is other interactions with less content and collaborative filtering information.

% \noindent{\textbf{Multi-task Recommendation:}}

% To sum up, multitask

\begin{figure*}[!ht]
  \centering
   \includegraphics[width=0.98\linewidth]{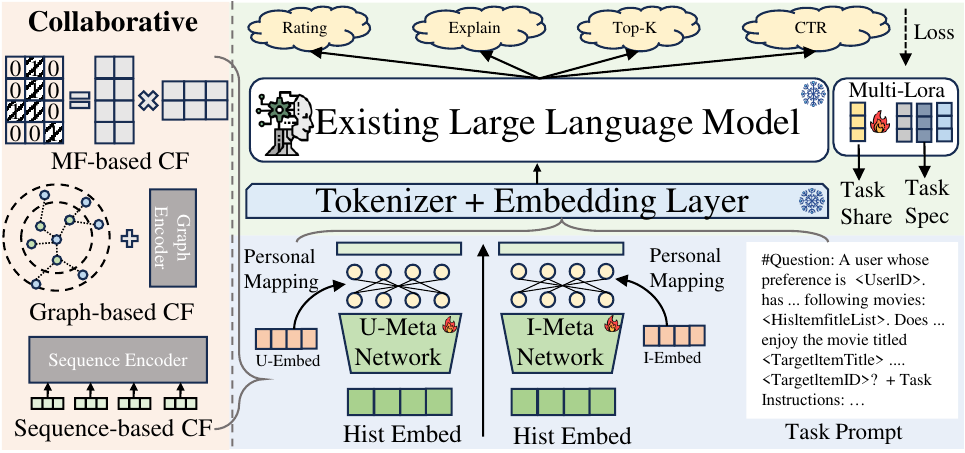}
   \caption{Overview of {the} framework, where orange, blue, and green backgrounds represent CKEM, KFM, and MTM parts respectively. Initially, user and item embeddings are generated via the collaborative filtering model, constituting what is referred to as collaborative knowledge. Subsequently, historical embeddings are utilized as input to two meta-networks, yielding personalized mapping functions for both user and candidate item. These mapping functions facilitate the transmission of collaborative knowledge to the LLM, in conjunction with prompt templates for multi-task recommendation. The whole optimization process is carried out using {the} designed Multi-Lora strategy.}
   \label{fig:framework}
\end{figure*}

\subsection{Overview of CKF Framework}
CKF is comprised of three distinct modules: the Collaborative Knowledge Extraction Module (CKEM), the Knowledge Fusion Module (KFM), and the Multi-Task Tuning Module (MTM). In CKEM, a well-established collaborative filtering model is employed to derive user and item embeddings. These embeddings are then merged into carefully crafted prompt templates via personalized mapping functions generated by meta-networks in KFM, ensuring tailored inputs for the MTM. Within the MTM, an chosen LLM is utilized to generate varied recommendation outcomes, aligned with distinct tasks. This module is tuned by the Multi-Lora strategy to accommodate both task-shared and task-specific requirements, ensuring that the recommendations are not only precise but also contextually relevant.
The overall framework of the model is shown in Figure~\ref{fig:framework}.

\subsection{Collaborative Knowledge Extraction {Module}}
In the initial phase, we employ state-of-the-art CF filtering techniques, such as MF~\cite{koren2009matrix,long2023decentralized} or DIN~\cite{zhou2018deep}, to obtain collaborative knowledge.
Given a user $u$ and his historical behavior sequence 
$\mathcal{I}_{u}$, traditional collaborative filtering model initially employs embedding layers to transform these inputs into latent vectors. Subsequently, these embeddings undergo further processing via additional feature extraction networks, aimed at elucidating the collaborative dynamics between users and items. The final stage involves the refinement of model parameters through the application of a backward gradient propagation algorithm. This pipeline can be mathematically represented as follows,
\begin{equation}\label{eq1}
\mathrm{e}_{u} = \mathbf{E}_{\mathcal{U}}(u), \quad
\mathrm{e}_{v_{i}} = \mathbf{E}_{\mathcal{V}}(v_{i}), \quad
\mathrm{e}_{\mathcal{I}_{u}} = \mathbf{E}_{\mathcal{V}}(v_{j}), \forall v_{j} \in \mathcal{I}_{u},
\end{equation}
\begin{equation}\label{eq2}
\text{min} \ \ \mathcal{L}(y_{u,i}, f(\mathrm{e}_{u}, \mathrm{e}_{v_i}, \mathrm{e}_{I_{u}}; \theta)),
\end{equation}
where $\mathrm{e}_{u}$, $\mathrm{e}_{v_i}$, $\mathrm{e}_{I_{u}}$ denotes embeddings of user, candidate item{,} and historical behavior sequence respectively. The matrix $\mathbf{E}$ denotes the embedding layer, while $f(\cdot;\theta)$ represents the chosen CF model, $\mathcal{L}$ refers to the loss function, and $y$ signifies the ground truth. Finally, the well-trained $\mathbf{E}_{\mathcal{U}}$ and $\mathbf{E}_{\mathcal{V}}$ contain rich collaborative knowledge, which could be retrieved by user or item id.

% \subsubsection{Conventional CF-based Models}

\subsection{Knowledge Fusion Module}
In this stage, we strive to incorporate obtained collaborative knowledge into the LLMs-based recommender system.
\subsubsection{LLMs-based Recommendation}\label{llm}
Prompt tuning is essential in LLM-based recommendation frameworks, translating user behaviors into text for LLM fine-tuning~\cite{talrec}.
Drawing inspiration from this,
we carefully design a prompt template, incorporating specific ID blanks within the textual description to facilitate the fusion of collaborative knowledge, as depicted in Figure~\ref{fig:framework}.
The training data set, structured based on this template, will be initially fed into the tokenizer and embedding layers of the LLM. Formally, for user $u$, the embedding of his prompt template $\mathcal{T}$ is,
\begin{equation}\label{eq3}
     \mathrm{e}_{u}^{\mathcal{T}} = [\mathrm{e}_{t_1},\cdots,\mathrm{e}_{t_k},\mathrm{e}^{p}_{u},\mathrm{e}_{t_k+1},\cdots,\mathrm{e}^{p}_{v_i},\cdots,\mathrm{e}_{\mathcal{N}_{u}^{\mathcal{T}}}],
\end{equation}
where ${\mathrm{e}_{t_k}}$ denotes the token embedding of ${t_k}$th token, i.e., ${\mathrm{e}_{t_k}} = \mathbf{E}_{\mathcal{T}}(t_{k})$.
${\mathrm{e}^p_{u}}$ and ${\mathrm{e}^p_{v_i}}$ signify embedding for the user-id placeholder and candidate item-id placeholder, respectively. $\mathcal{N}_{u}^{\mathcal{T}}$ refers to the total number of tokens. A straightforward approach involves directly substituting the placeholders with the corresponding entities $\mathrm{e}_{u}$ and $\mathrm{e}_{v_i}$, as indexed by $\mathbf{E}_{\mathcal{U}}$ and
$\mathbf{E}_{\mathcal{V}}$. However, a semantic discrepancy arises between $\mathbf{E}_{\mathcal{T}}$ and ($\mathbf{E}_{\mathcal{U}}$, $\mathbf{E}_{\mathcal{V}}$), arising from distinct training data sets and processes~\cite{li2023ctrl}. {Meanwhile, the embedding dimension commonly used by CF, usually 64 or 128, is much smaller than the matching size of 4096 for LLM, making direct insertion difficult~\cite{zhang2023collm,llama}.}
A more feasible strategy involves employing a mapping layer to project the embeddings acquired via CF into the textual semantic space used by CoLLM~\cite{zhang2023collm}. Formally,
\begin{equation}\label{eq:gm}
     \mathrm{e}^p_u = m_{\phi}(\mathrm{e}_u), \quad \mathrm{e}^p_{v_i} = m_{\phi}(\mathrm{e}_{v_i}),
\end{equation}
where $m_{\phi}$ is a trainable mapping function.
Nonetheless, this mapping approach presents two notable constraints. Firstly, as the user and item information reside in disparate semantic spaces, applying a uniform mapping function risks distorting their respective semantic meanings. Secondly, this method employs a singular mapping function for all users, thereby overlooking the nuances of personalized considerations. 
{For example, if the collaborative signals corresponding to users in a test set differ significantly from the training distribution, the general mapping function may fail.}
This necessitates a more meticulous design of the mapping function to address these concerns effectively.

\subsubsection{Personalized Mapping Function}
Given that a user's personalization is mirrored in their past behavior, we have developed an attention mechanism to accurately identify and represent their unique preferences. Formally,
\begin{equation}
\mathrm{p}_{u} = \sum_{{v_j}\in {\mathcal{I}_{u}}}\alpha_{j}{\mathrm{e}_{v_j}},
\end{equation}
where $\alpha_{v_j} = \frac{\mathrm{exp}(s(\mathrm{e}_{u},\mathrm{e}_{v_j}))}{\sum_{{v_i} \in {\mathcal{I}_{u}}} \mathrm{exp}(s(\mathrm{e}_{u},\mathrm{e}_{v_i}))}$. $s(\cdot)$ represents the scoring function, and without loss of generality, we use the dot product. Subsequently, a two-layer fully connected meta-network $\text{Meta}(\cdot;\theta_2)$ is employed to transform 
$\mathrm{p}_{u}$ into a $d^{2}$ dimensional vector, which is then reshaped into a $d\times d$ parameter matrix $\mathrm{w}$, serving as a personalized mapping function. Formally,
\begin{equation}\label{eq6}
     \mathrm{w}_{u} = \text{Meta}(\mathrm{p}_{u};\theta_2),
\end{equation}
where $\mathrm{w}_{u}$ denotes the personalized mapping function. A meta-network with the same architecture could also be prepared for items, and the generated mapping function is recorded as $\mathrm{w}_{v_i}$.
Ultimately, the derived mapping functions serve to bridge the semantic gap between the LLM and collaborative knowledge.
Formally,
\begin{equation}\label{eq7}
     \mathrm{e}^p_u = \mathrm{e}_u \mathrm{w}_{u},\quad \mathrm{e}^p_{v_i} = \mathrm{e}_{v_i} \mathrm{w}_{v_i}.
\end{equation}
In this way, collaborative knowledge is conveyed alongside the encoded text information to the LLM for various downstream tasks.
{Compared to the general mapping described in Eq.~\ref{eq:gm}, this meta-network parameter generation allows the model to learn how to construct an effective mapping function, significantly enhancing its mapping ability for new users.
More specifically, users can not only learn from the meta-network how other users map the collaborative space to the LLM semantic space but also make corrections based on their own interaction history to achieve a more personalized mapping.
}

\subsection{Multi-task Tuning Module}
Differing from CF-based methods, which typically acquire preference scores directly via dot or sigmoid layers, LLM-based approaches predominantly employ the Transformer~\cite{vaswani2017attention} architecture to directly generate textual content, i.e., $\hat{y}_{k}=\text{LLM}(\mathrm{e}^p_u)$, where $\hat{y}_{k}$ denotes the generated recommendation results for task $k$. For instance, Llama~\cite{llama}, as an illustration, comprises $L$ stacked decoder layers, where each block encompasses two submodules: a multi-head attention (MHA) and a fully connected feed-forward network (FFN). For clarity, we only focus on the MHA module. Given the input prompt $\mathcal{T}$, MHA executes the attention function across parallel heads. Formally,
\begin{equation}
     \mathrm{h}_{i} = \sigma(\frac{\mathrm{e}^{\mathcal{T}}_{u}{W^\mathrm{q}_{i}} (\mathrm{e}^{\mathcal{T}}_{u}{W^\mathrm{k}_{i}})^{\mathbf{T}}}{\sqrt{d}})(\mathrm{e}^{\mathcal{T}}_{u}{W^\mathrm{v}_{i}}),
\end{equation}
\begin{equation}
     \text{MHA}(\mathrm{e}^{\mathcal{T}}_{u};\varphi) = (\mathrm{h}_{1} ||\cdots || \mathrm{h}_{n})(\mathrm{e}^{\mathcal{T}}_{u}{W^\mathrm{o}_{i}}),
\end{equation}
where $\mathrm{h}_n$ represents the representation of a specific head {$n$}, $\sigma$ denotes the softmax activation function, $W$ signifies the pre-training parameter, $||$ indicates the concatenation operation, ${\mathbf{T}}$ is the transpose operation, and $\varphi$ encompasses all pre-trained parameters within the MHA.
Early LLMs-based methods usually handle recommendation tasks in isolation, and their typical approach is to optimize and store multiple sets of Lora~\cite{lora} parameters based on efficient parameter tuning. Formally,
\begin{equation}
     \text{MHA}(\mathrm{e}^{\mathcal{T}}_{u};\varphi, \Delta\varphi_{k})
     =||_{1}^{n} 
     [\sigma(\frac{(\mathrm{e}^{\mathcal{T}}_{u}{\tilde{W}^{\mathrm{q},k}_{i}}) (\mathrm{e}^{\mathcal{T}}_{u}{\tilde{W}^{\mathrm{k},k}_{i}})^{\mathbf{T}}}{\sqrt{d}})(\mathrm{e}^{\mathcal{T}}_{u}{\tilde{W}^{\mathrm{v},k}_{i}})] (\mathrm{e}^{\mathcal{T}}_{u}{\tilde{W}^{\mathrm{o},k}_{i} }),
\end{equation}
     \begin{equation}
     \tilde{W}_{i}^{*} = W_{i}^{*} + {B}_{i}^{*}{A}_{i}^{*},
\end{equation}
where $\varphi$ denotes the frozen parameters of LLM, $\Delta\varphi_{k}$ refers to the low-rank trainable parameters $\{B^{*}, A^{*}\}$ for task $k$, i.e., $A^{*} \in \mathbb{R}^{{d}\times \tilde{r}}, B^{*} \in \mathbb{R}^{\tilde{r}\times d}$. Please note that $\Delta\varphi_k$ are equipped with $\{\mathrm{q}$, $\mathrm{k}$, $\mathrm{v}$, $\mathrm{o}\}$ in each LLM attention layer.
However, such an approach fails to recognize the inherent interrelationships among tasks, an aspect that has been demonstrated to be crucial in enhancing overall recommendation performance and the user experience~\cite{wang2023multi}.
{This is further illustrated by the performance decline of the CKF-S variant in Table \ref{tab:exp1}-\ref{tab:exp-sup}.}
A naive approach is to employ the same Lora parameters for fine-tuning across multiple recommendation tasks. 
However, the semantic disparity among these tasks results in blurred decision boundaries, as depicted in Figure~\ref{fig:tsne:losp}.
Therefore, we shift our focus towards a fine-grained differentiation of Lora parameters, emphasizing task-shared and task-specific semantics. Formally,
\begin{equation}\label{eq12}
     {\text{MHA}}(\mathrm{e}^{\mathcal{T}}_{u};\varphi_{k}, \Delta\varphi_{share}, \Delta\varphi_{k})
     =||_{1}^{n} 
     [\sigma(\frac{(\mathrm{e}^{\mathcal{T}}_{u}({W^\mathrm{q}_{i} + {B}_{i}^{\mathrm{q},k}{A}_{i}^{\mathrm{q},k})}) (\mathrm{e}^{\mathcal{T}}_{u}{\tilde{W}^\mathrm{k}_{i}})^{\mathbf{T}}}{\sqrt{d}})(\mathrm{e}^{\mathcal{T}}_{u}{\tilde{W}^\mathrm{v}_{i}})] (\mathrm{e}^{\mathcal{T}}_{u}{\tilde{W}^\mathrm{o}_{i}}),
\end{equation}
% \begin{equation}
%      W_{i}^{*} = W_{i}^{*} + \dot{B}_{i}^{*}\dot{A}_{i}^{*} + \tilde{B}_{i}^{*}\tilde{A}_{i}^{*}
% \end{equation}
% \begin{equation}
%      \hat{y} = f_{(\varphi,\ \Delta\varphi_{share}, \ \Delta\varphi_k)}(\mathrm{e}^{\mathcal{T}}_u),
% \end{equation}
% \begin{equation}
% F(\phi, \phi_{share}, \phi_{spec}; \mathcal{A}, \mathcal{T}) = g( \sum_{i=1}^{N} \mathrm{h}_i(\phi, \phi_{share}) + \sum_{j=1}^{M} k_j(\phi, \phi_{spec_j}; \mathcal{T}_j))
% \end{equation}
where $\Delta\varphi_{share}$ is shared among tasks and $\Delta\varphi_k$ is for task $k$. To elaborate, during the training, $\Delta\varphi_k$ is trained and updated exclusively on the data corresponding to that task, whereas $\Delta\varphi_{share}$ is trained and updated encompassing all tasks. 
To mitigate excessive parameter introduction and ensure semantic decoupling, we opt to equip $\{\mathrm{k,v,o}\}$ with task-shared Lora, while $\{\mathrm{q}\}$ is configured with task-specific Lora. Compared with the previous method of Lora tuning that required the use of four sets of $\{\mathrm{q, k, v, o}\}$ for multi-task recommendations, we only need to use four $\{\mathrm{q}\}$ Lora and one $\{\mathrm{k, v, o}\}$ Lora. 
This method draws on the semantics of {the} attention mechanism, where using a specific query to mine related knowledge from sharing sources during the inference stage. To ensure task-specific semantics, we force $\mathrm{q}$-orthogonalization of different tasks.
Formally, for each MHA layer,
\begin{equation}
\mathcal{L}_{orth} = \sum_{k_1, k_2, \ k_1 \neq k_2} \sum_{\substack{i,j \ i \neq j}} ||({A^{\mathrm{q},k_1}})^{\mathrm{T}} A^{\mathrm{q},k_2}||^{2},
\end{equation}
where $i, j$ denotes the value of $({A^{q,k_1}})^{\mathrm{T}} A^{q,k_2}$ in row $i$ and column $j$, and $||\cdot||^{2}$ refer to the 2-norm.
Note that here we only perform orthogonal loss calculations for the $A$ matrix, because $B$ can be regarded as a linear combination of $A$~\cite{lora,promptliu,wang2023orthogonal}.
% Formally,
% \begin{equation}
%      Att(query,key,value) = FFN(\frac{query*key}{\sqrt{d}}*value),
% \end{equation}
% where FFN represents the full connected network.
{In general, our semantic decoupling approach facilitates the model's quick understanding of the invariance and uniqueness of each task, promoting mutual knowledge transfer and reducing training costs.}

\subsection{Training Strategies}
Our training process involves updating Multi-Lora $\{\Delta\varphi_{share}, \Delta\varphi_{k}\}$ and meta-networks parameters $\theta_{2}$, which are utilized for fine-tuning the LLM and integrating collaborative knowledge, respectively. The straightforward strategy is to optimize both components concurrently; however, our experiments in {S}ection~\ref{sec:5.3} reveal that this approach may lead to an overreliance on collaborative knowledge, at the expense of the model's comprehension of textual semantics. Therefore, we adopt a curriculum learning strategy, gradually fusing the {collaborative} knowledge.
More specifically, for each user, we generate two distinct prompts for the task: $\mathcal{T}_{1}$ and $\mathcal{T}_{2}$. $\mathcal{T}_{1}$, devoid of $\mathrm{e}^{p}_{u}$ and $\mathrm{e}^{p}_{v_i}$, depends exclusively on text semantics for generating recommendations. In contrast, $\mathcal{T}_{2}$ incorporates user and item collaborative embeddings, as described in {S}ection~\ref{llm}. 
During the training phase, the LLM will yield recommendation $\hat{y}_{k}$ and $\tilde{y}_{k}$ based on different inputs $\mathcal{T}_{1}$ and $\mathcal{T}_{2}$, respectively.
Then we compute a distinct loss in Eq.~\ref{eq14} and then employ a smooth weighting function in Eq.~\ref{eq13} to dynamically balance these two losses within the model. Formally,
\begin{equation}\label{eq13}
     \mathcal{L}_{all} = \beta\mathcal{L}_{\mathcal{T}_{1}} + (1-\beta)\mathcal{L}_{\mathcal{T}_{2}} + \mathcal{L}_{orth},
\end{equation}
\begin{equation}\label{eq14}
     \mathcal{L}_{\mathcal{T}_{1}} = \sum_{n=1}^{\mathcal{N}}\mathcal{L}_{n}(y_{n}, \hat{y}_{n}),\quad
     \mathcal{L}_{\mathcal{T}_{2}} = \sum_{n=1}^{\mathcal{N}}\mathcal{L}_{n}(y_{n}, \tilde{y}_{n}),
\end{equation}
\begin{equation}
      \beta = \frac{1}{1+\exp(\frac{i}{\mathrm{z}}-1)/\tau},
\end{equation}
where $\mathcal{L}_{n}$ denotes the loss function for task $n$, for instance, cross-entropy in the context of rating prediction. $\mathcal{N}$ denotes the total number of tasks. 
$i$ denotes the number of step{s} currently being trained, $\mathrm{z}$ refers to the number of steps required for the entire training process, {and} $\tau$ is temperature coefficient.
To elaborate, controlled by $\beta$, $\mathcal{L}_{\mathcal{T}_{1}}$ will predominantly influence the initial stage of training, thereby assisting the LLM in acquiring knowledge about user preferences from textual semantics and developing a basic capability for making recommendations. As training progresses, $\mathcal{L}_{\mathcal{T}_{2}}$ will increasingly become the dominant factor in the training process. This shift enables the progressive incorporation of collaborative signals into the LLM, thereby augmenting its recommendation capabilities.
Algorithm~\ref{alg1} presents a detailed overview of the flow and optimization process employed by the \textit{CKF} algorithm.

\begin{algorithm}[!h] %算法开始
\caption{The Algorithm of \textit{CKF}} %算法的题目
\label{alg1} %算法的标签
\begin{algorithmic}[1] %此处的[1]控制一下算法中的每句前面都有标号
\REQUIRE Prompt Templates $\mathcal{T}$, Behavior Sequence $\mathcal{I}$, Comment Sequence $\mathcal{C}$;
\ENSURE Parameters $\theta$, $\theta_{2}$, $\Delta \varphi_{share}, \Delta \varphi_{spec}$;
%输出结果(此处的ENSURE默认关键字为Ensure在上面已自定义为Output)
% if-then-else
\STATE Initialize CF model parameters $\theta$, trainable LLM parameters $\Delta \varphi_{share}, \Delta \varphi_{spec}$;
\STATE Extract collaborative knowledge $\mathbf{E}_{\mathcal{U}}$, $\mathbf{E}_{\mathcal{V}}$ in Eq.~\ref{eq1};
% while-loop
\WHILE{not converged}
\STATE Sample a batch of data from $\mathcal{U}$;
\STATE Construct prompts $\mathrm{e}_{u}^{\mathcal{T}_{1}}$ and $\mathrm{e}_{u}^{\mathcal{T}_{2}}$  in Eq.~\ref{eq3};
\STATE Calculate personalized mapping function  $\mathrm{w}_{u}$ in Eq.~\ref{eq6};
\STATE Calculate embedding for placeholder $\mathrm{e}^p_u, \mathrm{e}^p_{v_{i}}$ in Eq.~\ref{eq7};
\FOR{prompt in \{$\mathcal{T}_{1}$, $\mathcal{T}_{2}$\}}
\STATE Multi-Task prediction in Eq.~\ref{eq12};
\ENDFOR
\STATE Weighted Optimization in Eq.~\ref{eq13};
\STATE Update the parameters;
\ENDWHILE
\RETURN Parameters $\theta$, $\theta_{2}$, $\Delta \varphi_{share}, \Delta \varphi_{spec}$;
\end{algorithmic}
\end{algorithm}
% \IF {$i\geq maxval$}
%         \STATE $i\gets 0$
% \ELSE
%         \IF {$i+k\leq maxval$}
%                 \STATE $i\gets i+k$
%         \ENDIF
% \ENDIF

\section{Experiments}\label{section4}
In this section, we initially outline the requisite experimental settings, followed by conducting comparative experiments. We strive to solve these questions,
\begin{itemize}[leftmargin=12pt]
 	\item \textbf{RQ1:} What is the comparative efficacy of CKF's performance relative to prior studies?
	\item \textbf{RQ2:} Is the sub-module effective, and does it have a broad application value?
	\item \textbf{RQ3:} How do variations in critical parameters influence the performance of the model?
\end{itemize}

\subsection{Experimental Settings}
\subsubsection{Data Sets}
We utilize {four widely-used} data sets: Movie-Lens~\footnote{https://grouplens.org/datasets/movielens/}~\cite{harper2015movielens} and {three} subsets of the Amazon data set~\cite{talrec}, specifically Movies-TV~\footnote{https://datarepo.eng.ucsd.edu/mcauley\_group/data/amazon\_v2/categoryFilesSmall/Movies\_and\_TV\_5.json.gz}, Books~\footnote{https://datarepo.eng.ucsd.edu/mcauley\_group/data/amazon\_v2/categoryFilesSmall/Books\_5.json.gz}, {and Kindles}~\footnote{{https://datarepo.eng.ucsd.edu/mcauley\_group/data/amazon\_v2/categoryFiles/Kindle\_Store.json.gz}}. 
 Movie-Lens is renowned for its comprehensive collection of user movie ratings and Amazon encompasses a wide array of user interactions, including reviews and ratings across multiple product categories.
These data sets are extensively used in recommender systems and each contains a wealth of user historical behaviors. 
Following~\cite{zhou2018deep,zhu2022personalized,zhang2023collm}, we apply a dual filtering approach to ensure data quality, which only retains those users and items with interactions greater or equal to 20. {We use the common leave-one-out method to construct the training, validation, and test datasets~\cite{zhou2018deep}. During the test phase, each user generates data for a single task.}
The statistics of these {four} data sets are detailed in Table~\ref{tab:sta}. 
{We also examine the impact of various filtering rules in Section~\ref{sec:5.4}.}
Please note that Movie-Lens has no comment data and cannot make explainable recommendations. 
Consequently, $\mathcal{N}$ is set to 3 for it, while for the others, $\mathcal{N}$ is established at 4.
% At the same time, for the CTR task, we regard scores of 1-4 as 0, and scores of 5 as 1, as in the literature~\cite{zhou2018deep,zhang2023collm,talrec}.
\begin{table}
\centering
\caption{Data sets statistics. {\# Interactions refers to the number of user-item pairwise interactions}. \#Train, \#Valid, \#Test {represent the number of sampled sequences used for training, validation, and testing, respectively}. Avg-U and Avg-I represent the average number of user/item interactions respectively.} % Add records Interaction;
\label{tab:sta}
\resizebox{0.9\textwidth}{!}{
\begin{tabular}{lcccccccc} 
\hline
Data sets & {\#Interactions} & \#Train & \#Valid & \#Test & \#User & \#Item & \#Avg-U & \#Avg-I \\ 
\hline
Movie-Lens & {12,649} & 36,900    &   2,789   &   5,125   &   729   &   2,643   & 17.35 & 6.16\\
Amazon-M{ovies-TV}    &  {126,482} & 38,178    &    3,025   &  9,476    &   6,416   &   13,718    & 19.71& 9.22\\
Amazon-B{ooks}   &{284,980}   &   76,248    &    6,367   &   20,887   &    14,057  &  25,082     & 20.27 & 11.36\\
Amazon-K{indles}    &{1,023,030} &  {252,645}     &  {10,327}  &  {76,664}   &   {51,616}  &  {79,870}    & {19.82}  & {10.52} \\
\hline
\end{tabular}}
\end{table}

\subsubsection{Baselines}
To demonstrate the superiority of the proposed CKF, we employ the subsequent baselines for comparative analysis.
\begin{itemize}[leftmargin=12pt]
     \item {\textbf{GAR:} Global average ratings. We use the average rating of all users as the prediction for the RP and Explain Tasks. This is a method that requires no training.}
	\item \textbf{MF~\cite{koren2009matrix}:} Decompose the user-item co-occurrence matrix to extract low-dimensional embeddings of users and items.
	\item \textbf{LightGCN~\cite{he2020lightgcn}:} Acquire high-order collaboration through the linear propagation across a user-item interaction graph.
	\item \textbf{SASRec~\cite{kang2018self}:} Employ self-attention to effectively capture the interactions among various items within a sequence.
	\item \textbf{DIN~\cite{zhou2018deep}:} Enhance the conventional multi-valued feature pooling approach by introducing a weighting factor, derived from the correlation between the item and the candidate item.
     \item {\textbf{M3oE~\cite{zhang2024m3oe}:} Use three mixture expert modules to adaptively learn user preferences in general, domain, and task aspects. It is the state-of-the-art multi-task recommendation system.}
        \item \textbf{ICL~\cite{zhangicl}:} Utilize in-context learning to augment the recommendation efficiency of LLMs.
        \item \textbf{SoftPrompt~\cite{softp}:} Leverage an optimizable prompt, complemented by the integration of Lora for fine-tuning LLMs, thereby bolstering the efficacy of the recommendation.
        \item \textbf{TALLRec~\cite{talrec}:} Fine-tune the LLM within the recommendation domain, following a structured instruction learning approach.
	\item \textbf{P5~\cite{geng2022recommendation}:} Utilize personalized prompts to reformulate various recommendation task{s} into NLP tasks and then perform the multi-task learning method. 
    \item {\textbf{BIGRec~\cite{bao2023bi}:} Design the bi-step grounding paradigm to migrate LLMs first to the recommendation space and then to the item space for Top-K recommendations.}
    \item {\textbf{PEPLER~\cite{li2023personalized}:} Employ user and item IDs as discrete or continuous prompts and perform sequential tuning to bridge the gap between LLMs and explainable recommendations.}
        \item {\textbf{LLARA~\cite{liao2023llara}:} Develop a hybrid prompting strategy to drive LLM to learn Top-K tasks from easy to difficult.}
      \item \textbf{CoLLM~\cite{zhang2023collm}:} Employ the concept akin to mini-GPT 4~\cite{zhu2023minigpt}, the approach integrates the collaborative filtering signal into the LLMs using a unified mapping function.
    \item \textbf{CKF-S:} A variant of CKF that targets one task at a time.

\end{itemize}
Please note that these algorithms are categorized as follows, 1) CF-based: \textit{MF, LightGCN, SASRec, DIN} 
2) LLMs-based: \textit{ICL, SoftPrompt, TALLRec, P5, {PEPLER, BIGRec}} and 3) (LLMs+CF embedding)-based: {\textit{LLARA}}, \textit{CoLLM}, \textit{CKF-S}. For 3) and CKF, we adopt Llama~\footnote{https://llama.meta.com/}~\cite{llama} as LLM backbone. {For all Lora-based fine-tuning works, such as CoLLM and LLARA, we choose the same rank size as CKF to ensure fairness.}
{M3oE,} P5, and CKF are multi-task recommendation frameworks and only require a single training, while the rest of the algorithms are trained separately on multiple tasks. 
To be fair, we tune the hyper-parameters of each model to achieve the best results.
% {Since some work, such as BIGRec, has not been tested on other recommendation tasks, we use the same generation head as CKF while keeping other architectures unchanged.}
We employ MF~\cite{koren2009matrix} and Llama~\cite{llama} as CF model and LLM model for {LLARA}, CoLLM, CKF-S, and CKF to avoid unfair comparisons caused by collaborative knowledge quality. {We also examine the impact of different CF and LLM backbones on these algorithms in Section~\ref{sec:5.7}.} 
% For all LLM-based models, unless otherwise stated, we use Llama~\cite{llama} as the backbone.
\begin{figure*}[!ht]
  \centering
   \includegraphics[width=\linewidth]{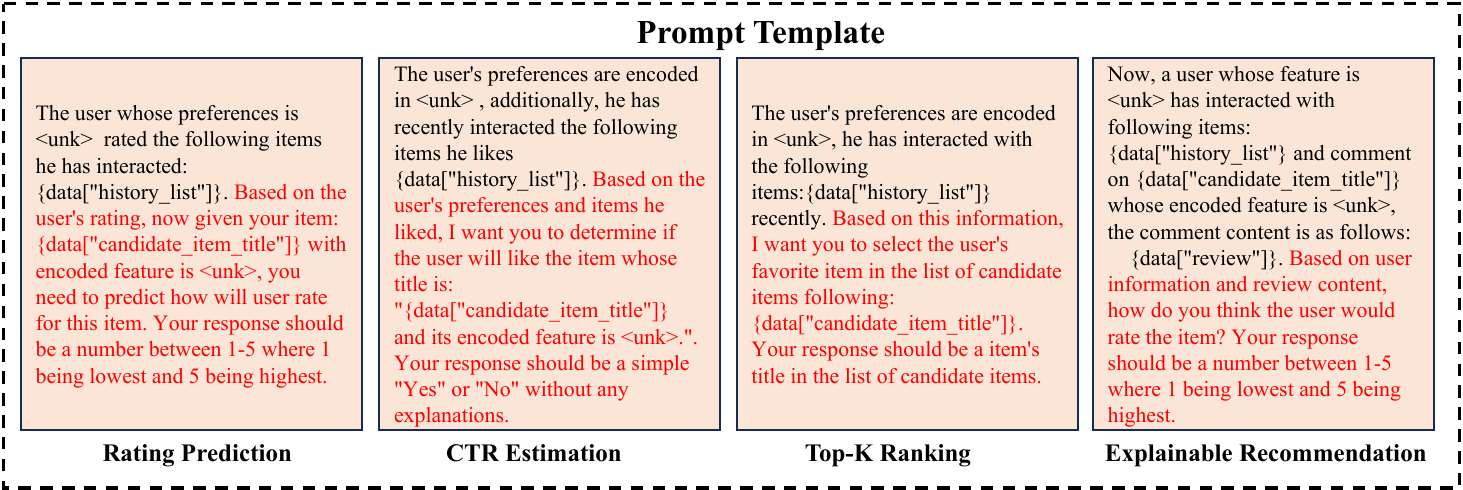}
   \caption{{Task Prompt. Black fonts are input content, red fonts are task instructions. The two unk are placeholders for the user and item collaboration vectors respectively.}}
   % Please note that for Explainable recommendations, data['history\_list'] contains a sequence of comments, while other tasks contain a sequence of titles.
   \label{fig:taskp}
\end{figure*}
\subsubsection{Parameter Settings \& Evaluation Metrics}
Our experiments are conducted on a system equipped with an Intel(R) Xeon(R) Platinum 8358 CPU @ 2.60GHz, eight NVIDIA A800 GPUs, and the environment is running Torch 2.0.1, all within an Ubuntu 20.04 LTS operating system. The prompt templates employed for various tasks are depicted in Figure~\ref{fig:taskp}.
% For all baseline models, such as LightGCN, we use consistent training settings and ensure their convergence.
For all data sets, we employ the AdamW optimizer with a weight decay of 1e-3 {and} a learning rate of 1e-4.
The training epochs and batch size are set to 3 and 8, respectively.
For critical parameters, like meta-network size $d$, we conduct hyperparameter tuning to guarantee optimal performance, as shown in {S}ection~\ref{sec:6}.

For various recommendation tasks, distinct evaluation metrics are utilized. Specifically, for the task of RP and Explain, 
we employ MAE and MSE. In the context of CTR, we adopt AUC and U-AUC~\cite{zhang2023collm}. Furthermore, for Top-K, we apply Hit@1-E and Hit@1-H, where the former adopts the uniform negative {sampling} and the latter employs the nearest neighbors-based negative sampling. 
Except for MAE and MSE, higher values indicate better performance. 
% \begin{equation}
%     % AUC (Area Under the ROC Curve)
% \text{AUC} = \frac{1}{N^+ \times N^-} \sum_{i=1}^{N^+} \sum_{j=1}^{N^-} \mathbb{I}(r_i > r_j)
% \end{equation}
% \begin{equation}
%     \text{UAUC} = \frac{1}{N} \sum_{u=1}^{N} \text{AUC}(u)
% \end{equation}
% \begin{equation}
%     \text{Hit@K-E} = \frac{1}{N} \sum_{u=1}^{N} \frac{|R(u) \cap T(u)|}{K}
% \end{equation}
% \begin{equation}
%     \text{Hit@K-H} = \frac{1}{N} \sum_{u=1}^{N} \frac{|R(u) \cap T(u)|}{K}
% \end{equation}
% \begin{equation}
% % MAE (Mean Absolute Error)
% \text{MAE} = \frac{1}{n} \sum_{i=1}^{n} |y_i - \hat{y}_i|
% \end{equation}
% \begin{equation}
% \text{MSE} = \frac{1}{n} \sum_{i=1}^{n} (y_i - \hat{y}_i)^2
% \end{equation}

\begin{table*}
\centering
\caption{Performance comparisons across multiple recommendation tasks (Movie-Lens data set). $\downarrow$ indicates that lower values are preferable, while $\uparrow$ signifies that higher values represent better outcomes. Bold font represents the best performance on the task. Movie-Lens, lacking comment data, is unsuitable for the explainable recommendation task. {GAR has no learning process and is unsuitable for CTR and Top-K tasks.}}
\label{tab:exp1}
\resizebox{0.9\textwidth}{!}{
\begin{tabular}{c|cc||cc||cc||cc} 
\toprule
\multirow{2}{*}{TASK}  & \multicolumn{2}{c||}{RP}          & \multicolumn{2}{c||}{CTR}         & \multicolumn{2}{c||}{Top-K}       & \multicolumn{2}{c}{Explain}                \\ 
\cline{2-9}
                       & MAE$\downarrow$             & MSE$\downarrow$               & AUC$\uparrow$             & U-AUC$\uparrow$           & Hit@1E$\uparrow$          & Hit@1H$\uparrow$          & MAE$\downarrow$                    & MSE$\downarrow$                     \\ 
\hline
{GAR}  &         {0.8987}        &     {1.1745}            &    {-}             &  {-}                &  {-}                & {-}                 & {-} & {-}  \\
MF                     & 0.8892          & 1.4504          & 0.6361          & 0.6286          & 0.2913          & 0.1202          & -                   & -                    \\
LightGCN               & 1.0346          & 1.5246          & 0.5783          & 0.5982          & 0.2420          & 0.0980          & -                   & -                    \\
SASRec                 & 0.9759          & 1.4802          & 0.6605          & \textbf{0.6572} & 0.3160          & 0.1579          & -                   & -                    \\
DIN                    & 0.9432          & 1.2666          & 0.6405          & 0.5753          & 0.3024          & 0.1696          & -                   & -                    \\
{M3oE} &       {0.8413}          &      {1.2132}           &     {0.6559}            &     {0.6023}            &      {0.3084}           &       {0.1736}          & {-} & {-}  \\ 
\hline
ICL                    & 1.2464          & 1.7783          & 0.5242          & 0.5204          & 0.1223          & 0.0427          & -                   & -                    \\
SoftPrompt             & 0.9884          & 1.6732          & 0.6337          & 0.5856          & 0.1786          & 0.0641          & -                   & -                    \\
TALLRec                & 0.9300          & 1.5049          & 0.6720          & 0.6017          & 0.2631          & 0.1391          & -                   & -                    \\
P5                     & 0.9448          & 1.5448          &  0.4792       & 0.5049          & 0.2009          & 0.0934          & -                   & -                    \\
{PEPLER}    &    {0.9612}             &        {1.5632}         &        {0.4736}         &        {0.4715}         &        {0.1826}         &       {0.1017}         & {-} & {-}  \\
{BIGRec}    &     {0.9287}           &      {1.4983}        &    {0.6786}             &       {0.6084}         &       {0.2887}         &        {0.1587}         & {-} & {-}  \\ 
\hline
{LLARA}        &       {0.9003}          &    {1.3812}             &   {0.6472}            &     {0.5931}              &    {0.3086}            &      {0.1594}            & {-}                   & {-}                    \\
CoLLM                  & 0.9052          & 1.3823          & 0.6463          & 0.5858          & 0.3018          & 0.1259          & -                   & -                    \\
CKF-S                  & 0.9251          & 1.3267          & 0.6487          & 0.5973          & 0.2798          & 0.1087          & -                   & -                    \\
\textbf{CKF}           & \textbf{0.8012} & \textbf{1.1562} & \textbf{0.6914} & 0.6332          & \textbf{0.3302} & \textbf{0.1920} & \textbf{-}          & \textbf{-}           \\
\bottomrule
\end{tabular}}
\end{table*}

\begin{table*}
\centering
\caption{Performance comparisons across multiple tasks (Amazon Movies-TV data set). The CF-based baselines are incapable of processing text data, resulting in the absence of results for explainable recommendations.}
\label{tab:exp3}
\resizebox{0.9\textwidth}{!}{
\begin{tabular}{c|cc||cc||cc||cc} 
\toprule
\multirow{2}{*}{Task} & \multicolumn{2}{c||}{RP} & \multicolumn{2}{c||}{CTR} & \multicolumn{2}{c||}{Top-K} & \multicolumn{2}{c}{Explain}                \\ 
\cline{2-9}
                      & MAE$\downarrow$     & MSE$\downarrow$              & AUC$\uparrow$     & U-AUC$\uparrow$            & Hit@1-E$\uparrow$ & Hit@1-H$\uparrow$           & MAE$\downarrow$                  & MSE$\downarrow$                  \\ 
\hline
{GAR} &    {0.8499}    &  {\textbf{1.1675}}               & {-}       &{-}                   & {-}         &  {-}                  &   {0.9015}                   &  {1.2990}                     \\
MF                    & 0.7480 & 1.5672          & 0.7545 & 0.7353           & 0.3146  & 0.1222            & -                   & -                    \\
LightGCN              & 1.1230 & 2.1550          & 0.7595 & 0.7524           & 0.4249  & 0.0967            & -                   & -                    \\
SASRec                & 1.0042 & 1.9308          & 0.8425 & 0.8002           & 0.4463  & 0.2788            & -                   & -                    \\
DIN                   & 0.8490 & 1.3672          & 0.8411 & 0.7988           & 0.4261  & 0.2447            & -                   & -                    \\
{M3oE}   &   {0.9172}     &     {1.5706}            &    {0.8446}    &       {0.8025}           &     {0.4418}    &    {0.2763}               & {-} & {-}  \\ 
\hline
ICL                   & 1.5235 & 2.4123          & 0.5012 & 0.4998           & 0.0970  & 0.0591            & 0.8012              & 1.0231               \\
SoftPrompt            & 1.2902 & 1.9802          & 0.6589 & 0.6474           & 0.2053  & 0.1251            & 0.7219              & 0.8310               \\
TALLRec               & 0.9034 & 1.5234          & 0.8012 & 0.8035           & 0.3905  & 0.2762            & 0.5425              & 0.6208               \\
P5                    & 0.8686 & 1.4865          & 0.4992 & 0.4972           & 0.2257  & 0.1041            & 0.5722              & 0.6705               \\
{PEPLER}   &    {0.9016}    &       {1.5011}          &   {0.4927}     &       {0.4877}           &     {0.2032}    &     {0.0946}              &           {0.5846}          &        {0.6983}              \\
{BIGRec}   &     {0.9078}   &       {1.5587}          &     {0.8242}   &          {0.8197}        &    {0.4287}     &        {0.2931}           &        {0.5218}             &       {0.6076}               \\ 
\hline
{LLARA}         &  {0.8724}    &       {1.4519}          &               {0.8106}        &   {0.8156}    &       {0.4397}           &          {0.3248}           &   {0.5517}    &   {0.6511}            \\
CoLLM                 & 0.8782 & 1.4509          & 0.8123 & 0.8072           & 0.3998  & 0.2886            & 0.5912              & 0.7012               \\
CKF-S                 & 0.9176 & 1.6012          & 0.8098 & 0.8021           & 0.4210  & 0.2479            & 0.4426              & 0.6104               \\
CKF                   & \textbf{0.6063} & 1.2071          & \textbf{0.8654} & \textbf{0.8642}           & \textbf{0.4883}  & \textbf{0.3671}            & \textbf{0.3624}              & \textbf{0.4987}               \\
\bottomrule
\end{tabular}}
\end{table*}

\begin{table*}
\centering
\caption{Performance comparisons across multiple tasks (Amazon Books data set).}
\label{tab:exp4}
\resizebox{0.9\textwidth}{!}{
\begin{tabular}{c|cc||cc||cc||cc} 
\toprule
\multirow{2}{*}{Task}   & \multicolumn{2}{c||}{RP} & \multicolumn{2}{c||}{CTR} & \multicolumn{2}{c||}{Top-K} & \multicolumn{2}{c}{Explain}                \\ 
\cline{2-9}
                        & MAE$\downarrow$    & MSE$\downarrow$             & AUC$\uparrow$    & U-AUC$\uparrow$            & Hit@1-E$\uparrow$ & Hit@1-H$\uparrow$           & MAE$\downarrow$                 & MSE$\downarrow$                  \\ 
\hline
{GAR}   &    {0.7514}    &        {\textbf{0.8792}}         &  {-}       & {-}                  & {-}         &  {-}                  &  {0.7406}                    &  {0.8522}                     \\
MF                      & 0.8340 & 1.2946          & 0.7469 & 0.7466           & 0.3238  & 0.1172            & -                   & -                    \\
LightGCN                & 0.9698 & 1.8924          & 0.7182 & 0.7264           & 0.3777  & 0.1037            & -                   & -                    \\
SASRec                  & 0.8419 & 1.4209          & 0.7336 & 0.7361           & 0.3993  & 0.1904            & -                   & -                    \\
DIN                     & 0.7589 & 1.1045          & 0.7407 & 0.7389           & 0.3817  & 0.2125            & -                   & -                    \\
{M3oE}     & {0.8050}      &        {1.2160}              &     {0.7483}             &    {0.7407}     &        {0.3993}       & {0.2046}   & {-} & {-}  \\ 
\hline
ICL                     & 1.1034 & 1.6357          & 0.4982 & 0.4882           & 0.1107  & 0.0516            & 0.8909              & 1.1438               \\
SoftPrompt              & 1.1427 & 1.2923          & 0.7143 & 0.7078           & 0.2690  & 0.0824            & 0.7065              & 0.8923               \\
TALLRec                 & 0.7023 & 1.2150          & 0.7980 & 0.7605           & 0.3634  & 0.2373            & 0.4436              & 0.5509               \\
P5                      & 0.9023 & 1.3231          & 0.4998 & 0.4893           & 0.2920  & 0.1012            & 0.4764              & 0.7823               \\
{PEPLER}     &    {0.9417}    &         {1.4285}            &      {0.4817}            &    {0.4715}     &        {0.2317}          &           {0.0961}          &     {0.5016}    &      {0.8252}       \\
{BIGRec}     &    {0.6915}    &       {1.1932}          &   {0.8032}     &           {0.8012}       &    {0.3915}     &       {0.2603}            &       {0.4219}              &     {0.5360}                 \\ 
\hline
{LLARA} &   {0.7415}     &        {1.1824}         &     {0.8185}   &         {0.7998}         &    {0.3969}     &     {0.2357}              &         {0.4316}            &       {0.5057}               \\
CoLLM                   & 0.7473 & 1.1709          & 0.8149 & 0.7978           & 0.3868  & 0.2559            & 0.4411              & 0.5680               \\
CKF-S                   & 0.7895 & 1.2601          & 0.8243 & 0.8097           & 0.4005  & 0.2402            & 0.4026              & 0.4823               \\
CKF                     & \textbf{0.6289} & 0.9841          & \textbf{0.8579} & \textbf{0.8452}           & \textbf{0.4673}  & \textbf{0.2954}            & \textbf{0.3534}              & \textbf{0.4269}               \\
\bottomrule
\end{tabular}}
\end{table*}

\begin{table}
\centering
\caption{{Performance comparisons across multiple recommendation tasks (Amazon Kindles data set).}}
\label{tab:exp-sup}
\resizebox{0.9\textwidth}{!}{
\begin{tabular}{c|cc||cc||cc||cc} 
\toprule
\multirow{2}{*}{{Task}} & \multicolumn{2}{c||}{{RP}}          & \multicolumn{2}{c||}{{CTR}}         & \multicolumn{2}{c||}{{Top-K}}         & \multicolumn{2}{c}{{Explain}}        \\ 
\cline{2-9}
                                        & {MAE$\downarrow$}    & {MSE$\downarrow$}    & {AUC$\uparrow$}    & {U-AUC$\uparrow$}  & {Hit@1-E$\uparrow$} & {Hit@1-H$\uparrow$} & {MAE$\downarrow$}    & {MSE$\downarrow$}     \\ 
\hline
{GAR}                   & {0.6359} & {\textbf{0.7046}} &   {-}                        &  {-}                         &  {-}                          &  {-}                          &  {0.6519}                         &  {0.7377}                          \\
{MF}                    & {0.7156} & {1.1997} & {0.7407} & {0.7387} & {0.2578}  & {0.1342}  & {-}      & {-}       \\
{LightGCN}              & {0.8145} & {1.4912} & {0.8476} & {0.8412} & {0.3123}  & {0.0979}  & {-}      & {-}       \\
{SASRec}                & {0.5825} & {0.9019} & {0.8710} & {0.8658} & {0.3416}  & {0.2319}  & {-}      & {-}       \\
{DIN}                   & {0.6098} & {0.9168} & {0.8602} & {0.8211} & {0.3219}  & {0.2034}  & {-}      & {-}       \\
{M3oE}                  & {0.7094} & {0.9056} & {0.8790} & {0.8415} & {0.3346}  & {0.2087}  & {-}      & {-}       \\ 
\hline
{ICL}                   & {1.1389} & {1.8614} & {0.4846} & {0.4802} & {0.1504}  & {0.1035}  & {0.9045} & {1.1032}  \\
{SoftPrompt}            & {1.0425} & {1.6832} & {0.5312} & {0.5248} & {0.1723}  & {0.1132}  & {0.7425} & {0.9424}  \\
{TALLRec}               & {0.5295} & {0.9570} & {0.8537} & {0.8526} & {0.3382}  & {0.2223}  & {0.3877} & {0.4530}  \\
{P5}                    & {0.8034} & {1.2412} & {0.5023} & {0.4982} & {0.1832}  & {0.1206}  & {0.4935} & {0.6421}  \\
{PEPLER}                & {0.8125} & {1.3141} & {0.4952} & {0.4936} & {0.1664}  & {0.0956}  & {0.5023} & {0.6720}  \\
{BIGRec}                & {0.5319} & {0.9704} & {0.8632} & {0.8589} & {0.3427}  & {0.2383}  & {0.3825} & {0.4532}  \\ 
\hline
{LLARA}                 & {0.5548} & {0.9751} & {0.8618} & {0.8556} & {0.3473}  & {0.2410}  & {0.3642} & {0.4307}  \\
{CoLLM}                 & {0.5531} & {0.9836} & {0.8610} & {0.8576} & {0.3405}  & {0.2369}  & {0.3619} & {0.4286}  \\
{CKF-S}                 & {0.5448} & {0.9752} & {0.8646} & {0.8587} & {0.3489}  & {0.2426}  & {0.3569} & {0.4205}  \\
{CKF}                   & {\textbf{0.4909}} & {0.8690} & {\textbf{0.8897}} & {\textbf{0.8836}} & {\textbf{0.3811}}  & {\textbf{0.2842}}  & {\textbf{0.2929}} & {\textbf{0.3585}}  \\
\bottomrule
\end{tabular}}
\end{table}

\subsection{Experimental Results (RQ1)}
Table~\ref{tab:exp1}-{\ref{tab:exp-sup}} displays the performance of each algorithm across all tasks.
{We conduct an in-depth analysis from three perspectives: algorithms, genres, and tasks.}

% Evidence shows that CKF achieves the best performance on all tasks, marking an enhancement of up to 15 \% over the most robust baseline in Top-K (Amazon Books).

{
In terms of algorithms comparison, M3oE is a highly effective collaborative recommendation algorithm, achieving notable improvements of 2\%, 13\%, 5\%, and 4\% compared to DIN on the four data sets (Top-K). This underscores the efficacy of multi-task joint training, which effectively captures comprehensive user interests by leveraging the distinct co-occurrence signals of different tasks. 
P5 also employs multi-task learning to achieve good performance, but it is constrained by T5's~\cite{raffel2020exploring} limited pre-training knowledge and the semantic ambiguity introduced by ID encoding.
CoLLM and LLARA can capture collaborative associations, but they suffer from an inability to distinctively represent user and item semantics, trailing behind CKF by a minimum 3\% gap. Additionally, the objectives of CoLLM's two-stage mapping training process are not aligned, further impacting its effectiveness.
CKF-S stands as a robust baseline, sharing a similar design with our framework except for its lack of a multi-task structure. Despite its strong performance, we observe that it is slightly weaker than CoLLM on some tasks, e.g., RP / CTR on Amazon Movies-TV. This may be because CKF-S does not exert its personalized mapping capabilities in some simpler tasks. More specifically, in scenarios where data features or sequence patterns are straightforward, the majority of user and item collaborative signal transformations adhere to similar patterns that can be readily transferred to the semantic space of the LLM. In such cases, the complex functionalities of the meta-network might not yield substantial performance improvements. For instance, Amazon Movies-TV, which has less data and fewer sequential patterns, may show smaller improvements when utilizing complex meta-network functions. In contrast, in the Amazon Books and Kindles data set, which contains richer data and complicated behavior patterns, the meta-network approach can offer greater value.
}

{
In genre aspects, we find that several LLM-based baselines, such as 
ICL and SoftPrompt, exhibit limited performance compared to CF-based models like SASRec and DIN, as they rely exclusively on textual semantics for discerning user preferences. This reliance, however, falls short in revealing co-purchase patterns among items. 
This aligns with the research findings of MoRec~\cite{yuan2023go} that in certain scenarios, ID-based recommendation systems still hold advantages, achieving results that LLM-based systems, which primarily rely on semantic signals, cannot attain. 
% This aligns with the research findings of MoRec~\cite{yuan2023go}, suggesting that when user interaction data is plentiful, ID-based baselines may achieve better results. 
CKF shows significant improvement over M3oE, with up to 14\% enhancement on the Amazon-Book data set (CTR). This progress stems from M3oE's continued reliance on co-occurrence signals, which limits its ability to discern semantic associations between items. 
In contrast, CKF not only incorporates collaborative signals but also leverages item semantic similarities from a pre-trained LLM. 
Furthermore, our Multi-Lora design enhances mutual knowledge transfer through task sharing and task-specific disentanglement, outperforming single-task algorithms like LLARA and CoLLM in signal richness.
% The LLM with collaborative signals is much stronger than that without them. This shows that the co-occurrence relationship of collaborative signals can effectively convey user preference information.
}

In task aspects, for the Top-K, performance is assessed in both uniform and nearest neighbor negative sampling settings, with the latter posing greater difficulty. CKF outperforms the baseline in both cases, notably increasing its lead in the more challenging scenario, aided by robust multi-task supervision signals. 
{BIGRec is a strong baseline for Top-K tasks. Compared with TALLRec, it adds a grounding strategy for recommendations to the item semantic space. However, its non-optimization in the second stage limits its performance improvement.}
Furthermore, CKF excels in CTR across various data sets, with the exception of a marginal underperformance compared to SASRec in Movie-Lens (UAUC), which can be attributed to the data set's strong collaborative signal meeting the task's needs.
In RP and Explain, {GAR is a strong baseline, performing well in terms of MSE. This is because the rating distribution in the data set exhibits a strong central tendency (e.g., most ratings cluster around a certain average), as evidenced in Section~\ref{sec:5.5.1}.  In contrast, CKF, which is optimized for personalized predictions, identifies subtle variations in the rating distribution. This allows it to provide predictions that are closer to the actual ratings overall, as reflected by the MAE metric. However, due to MSE's greater sensitivity to outliers, CKF may not perform as well on this metric compared to the GAR model}.  {PEPLER shows impressive results on the RP task, demonstrating that incorporating signals from other recommendation tasks helps explainable tasks. However, its smaller foundation model limits its capabilities, and indiscriminate use of multi-task knowledge may lead to potential signal confusion.  Conversely, CKF shows a {remarkable} improvement over other LLM baselines, attributable to enhanced collaborative knowledge.
In addition, CKF is also more consistent with the actual distribution in terms of the overall uniformity of the scores, rather than being easily biased towards the popular score distribution, as described in Section~\ref{sec:5.5.1} for details.
}
Remarkably, the Explain exhibits a reduced regression loss compared to the RP. This is because reviews contain nuanced user sentiment, extending the context for content understanding.

To sum up, CKF integrates the complementary strengths of CF-based and LLMs-based recommender systems, offering the potential for broader applicability across various scenarios.

\section{Robust Testings (RQ2)}\label{sec:5}
This section undertakes comprehensive robustness testing and discussion, aiming to {gain} a deeper insight into {the} model's mechanism {and its performance in various challenging scenarios.}

\subsection{Warm-Cold Scenarios}\label{sec:5.1}
\begin{figure*}[!ht]
\centering
\subfigure[RP (MAE $\downarrow$)]{
\begin{minipage}[t]{0.24\linewidth}
\centering
\includegraphics[width=\linewidth]{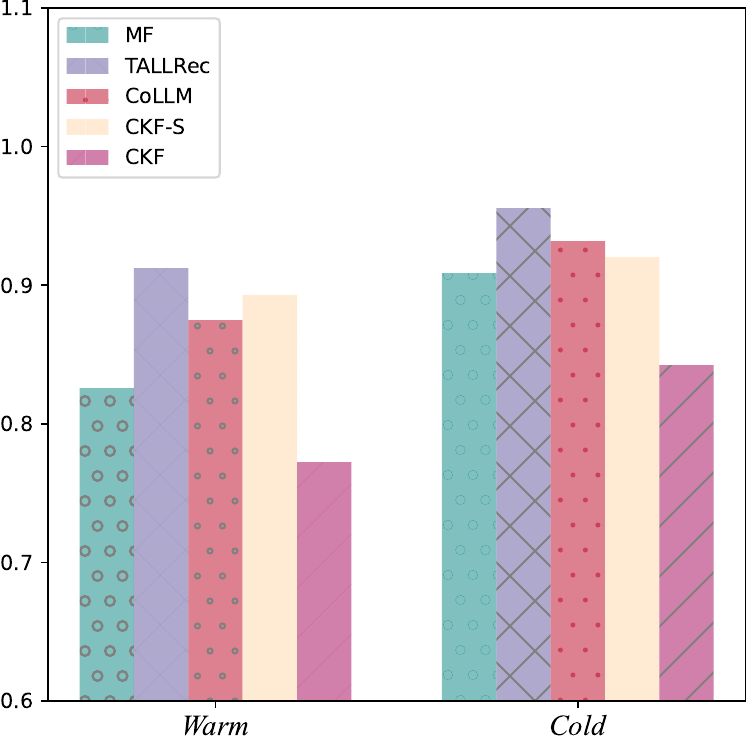}
%\caption{fig1}
\label{fig:wc-rp-m}
\end{minipage}%
}%
\subfigure[CTR (AUC $\uparrow$)]{
\begin{minipage}[t]{0.24\linewidth}
\centering
\includegraphics[width=\linewidth]{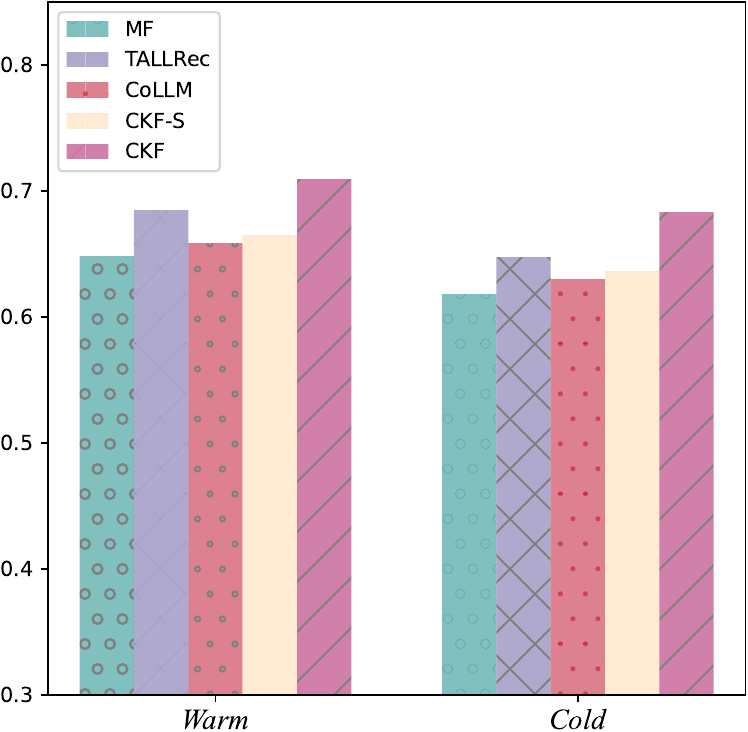}
%\caption{fig1}
\label{fig:wc-ctr-m}
\end{minipage}%
}%
\subfigure[Top-K (Hit@1-E $\uparrow$)]{
\begin{minipage}[t]{0.24\linewidth}
\centering
\includegraphics[width=\linewidth]{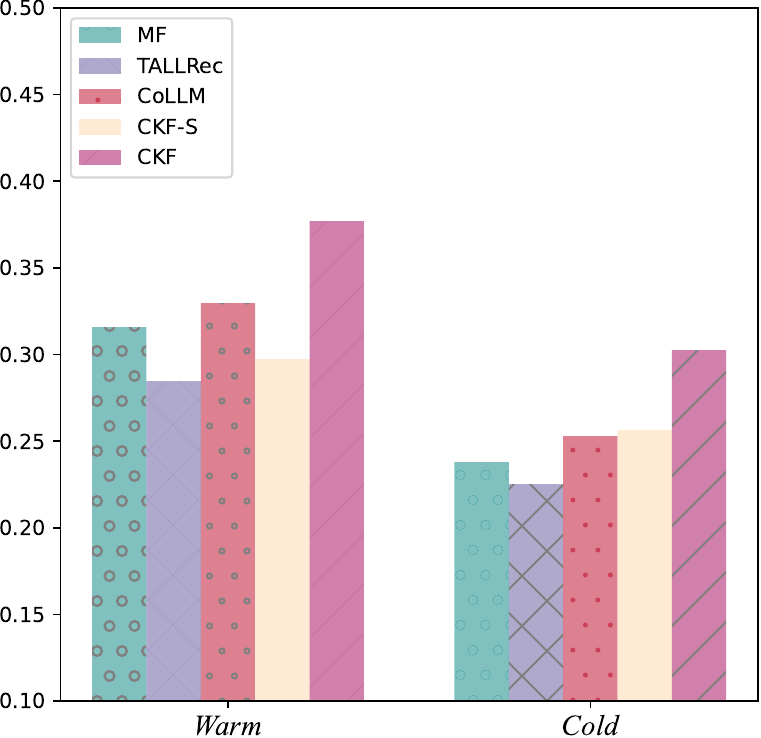}
%\caption{fig1}
\label{fig:wc-topk-m}
\end{minipage}%
}%
% \subfigure[Explain (MAE $\downarrow$)]{
% \begin{minipage}[t]{0.24\linewidth}
% \centering
% \includegraphics[width=\linewidth]{fig/wc-exps-m.pdf}
% %\caption{fig1}
% \label{fig:wc-exp-m}
% \end{minipage}%
% }%
\centering
\caption{Warm-cold scenarios (Movie-Lens data set). We select the strong baselines MF, TALLRec, {CKF-S},  CoLLM{,} and CKF for comparison. The Movie-Lens data set lacks comment data, precluding comparisons of explainable recommendations.}
\label{fig:wc-m}
\end{figure*}

\begin{figure*}[!ht]
\centering
\subfigure[RP (MAE $\downarrow$)]{
\begin{minipage}[t]{0.24\linewidth}
\centering
\includegraphics[width=\linewidth]{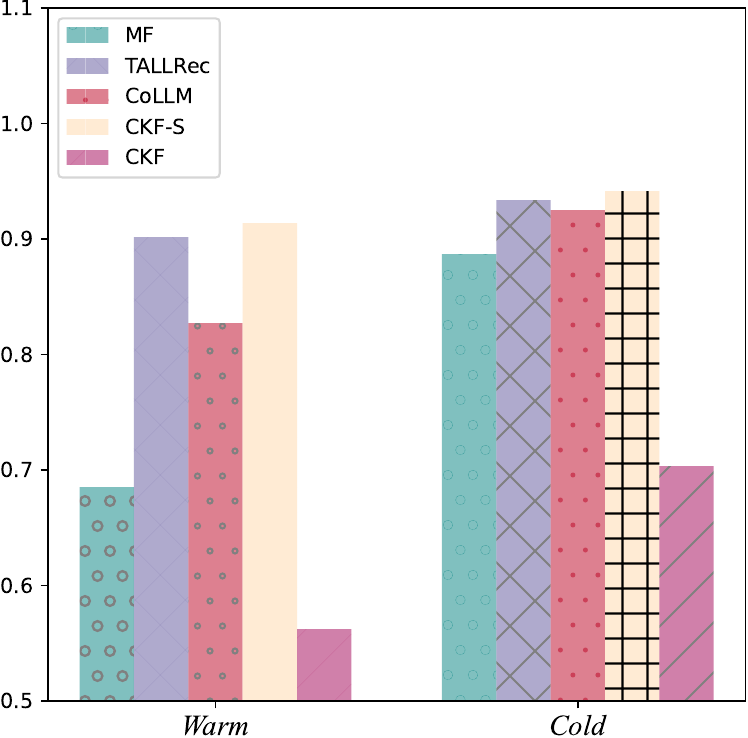}
%\caption{fig1}
\label{fig:wc-rp-a}
\end{minipage}%
}%
\subfigure[CTR (AUC $\uparrow$)]{
\begin{minipage}[t]{0.24\linewidth}
\centering
\includegraphics[width=\linewidth]{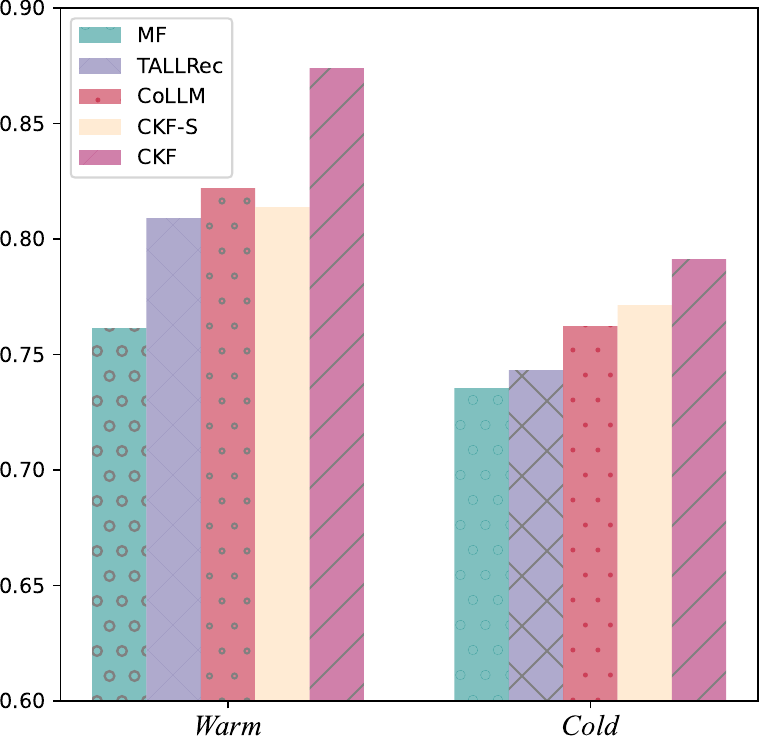}
%\caption{fig1}
\label{fig:wc-ctr-a}
\end{minipage}%
}%
\subfigure[Top-K (Hit@1-E $\uparrow$)]{
\begin{minipage}[t]{0.24\linewidth}
\centering
\includegraphics[width=\linewidth]{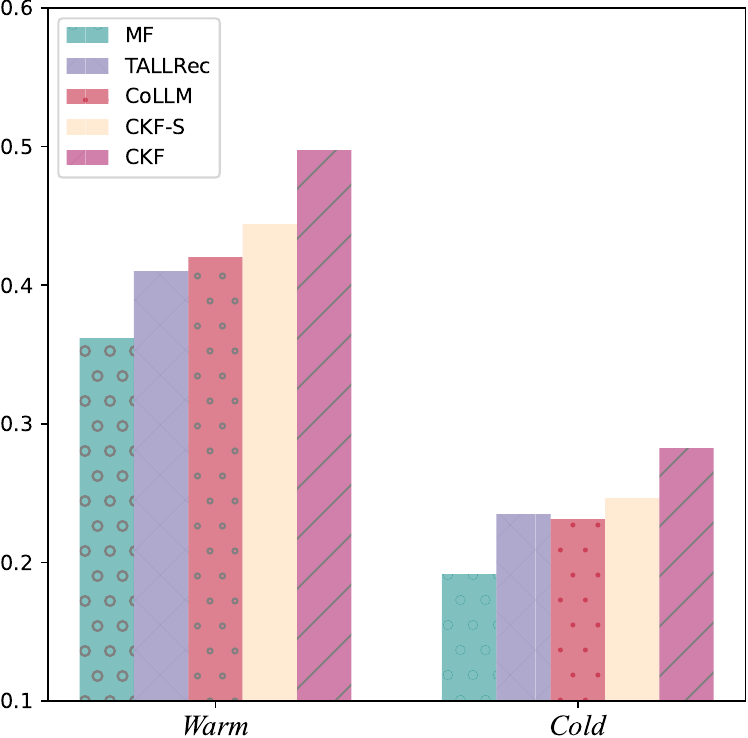}
%\caption{fig1}
\label{fig:wc-topk-a}
\end{minipage}%
}%
\subfigure[Explain (MAE $\downarrow$)]{
\begin{minipage}[t]{0.24\linewidth}
\centering
\includegraphics[width=\linewidth]{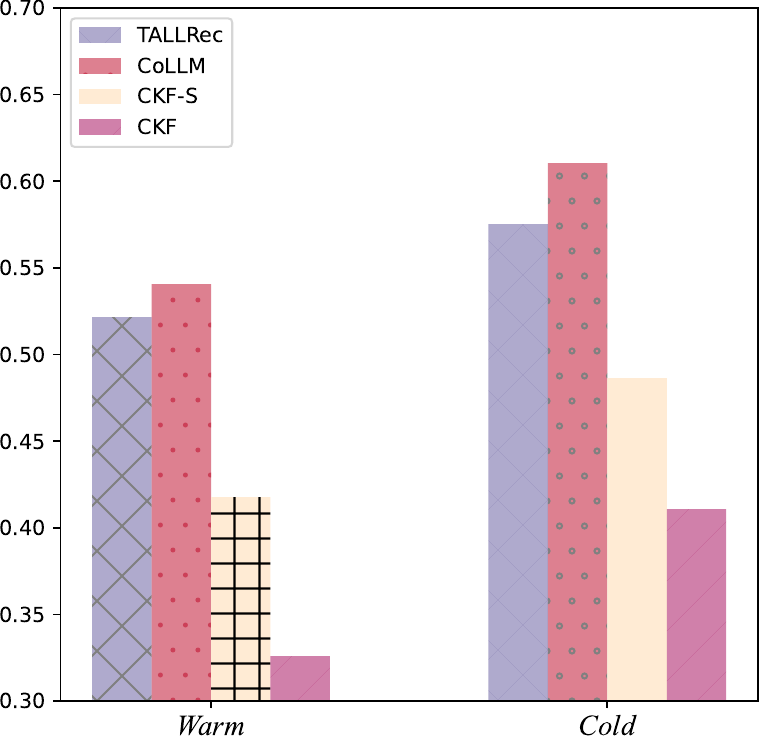}
%\caption{fig1}
\label{fig:wc-exp-a}
\end{minipage}%
}%
\centering
\caption{Warm-cold scenarios (Amazon Movies-TV data set). In (d), the comparison is limited to LLM baselines, as the CF model lacks suitability for Explainable recommendations.}
\label{fig:wc-a}
\end{figure*}

\begin{figure*}[!ht]
\centering
\subfigure[RP (MAE $\downarrow$)]{
\begin{minipage}[t]{0.24\linewidth}
\centering
\includegraphics[width=\linewidth]{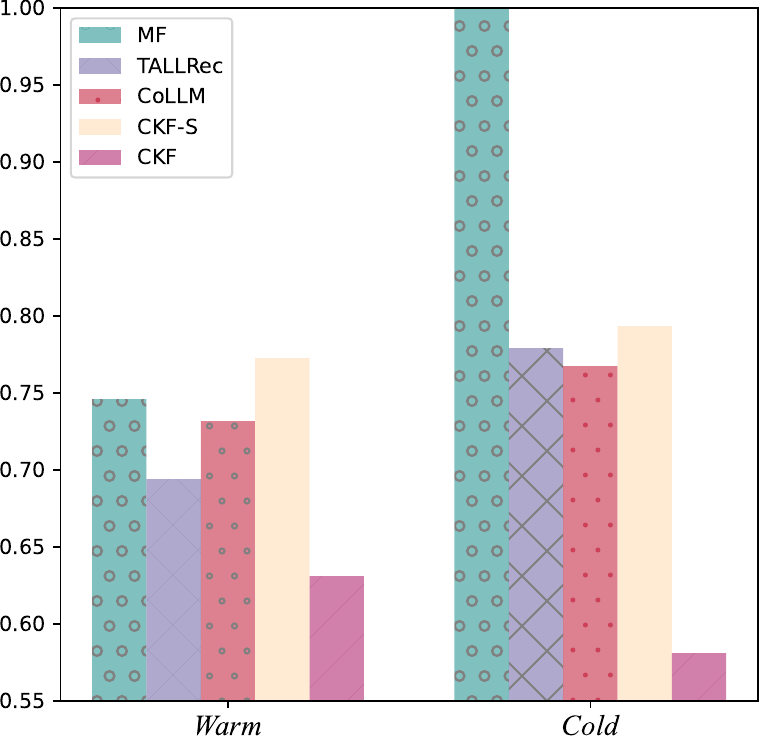}
%\caption{fig1}
\label{fig:wc-rp}
\end{minipage}%
}%
\subfigure[{CTR (AUC $\uparrow$)}]{
\begin{minipage}[t]{0.24\linewidth}
\centering
\includegraphics[width=\linewidth]{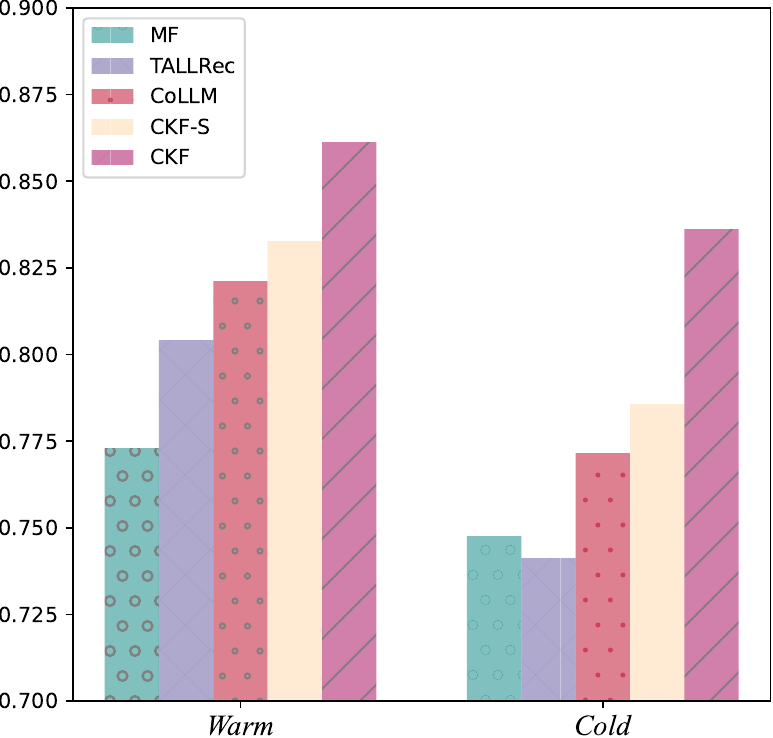}
%\caption{fig1}
\label{fig:wc-ctr}
\end{minipage}%
}%
\subfigure[Top-K (Hit@1-E $\uparrow$)]{
\begin{minipage}[t]{0.24\linewidth}
\centering
\includegraphics[width=\linewidth]{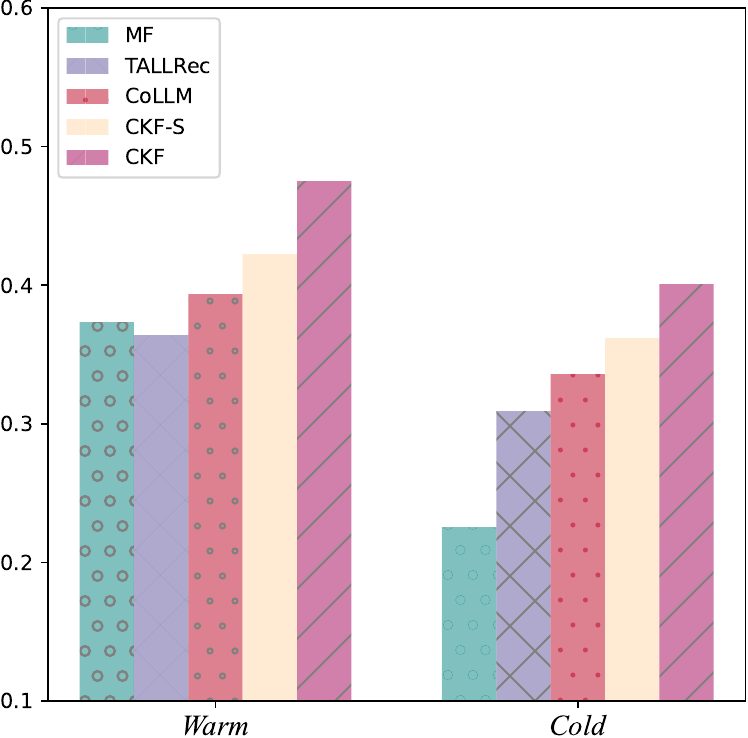}
%\caption{fig1}
\label{fig:wc-topk}
\end{minipage}%
}%
\subfigure[Explain (MAE $\downarrow$)]{
\begin{minipage}[t]{0.24\linewidth}
\centering
\includegraphics[width=\linewidth]{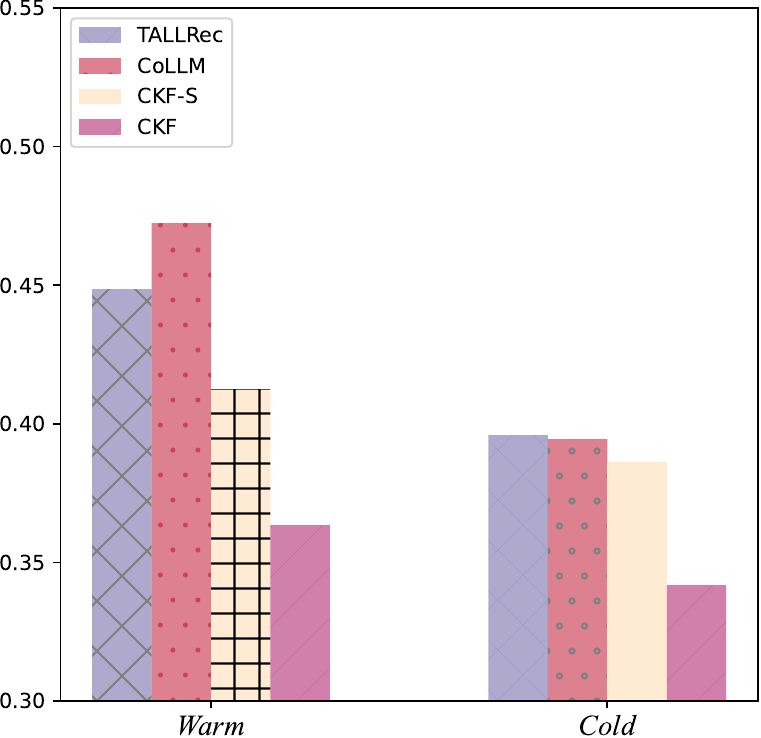}
%\caption{fig1}
\label{fig:wc-exp}
\end{minipage}%
}%
\centering
\caption{Warm-cold scenarios (Amazon Books data set).}
\label{fig:wc}
\end{figure*}

\begin{figure*}[!ht]
\centering
\subfigure[{RP (MAE $\downarrow$)}]{
\begin{minipage}[t]{0.24\linewidth}
\centering
\includegraphics[width=\linewidth]{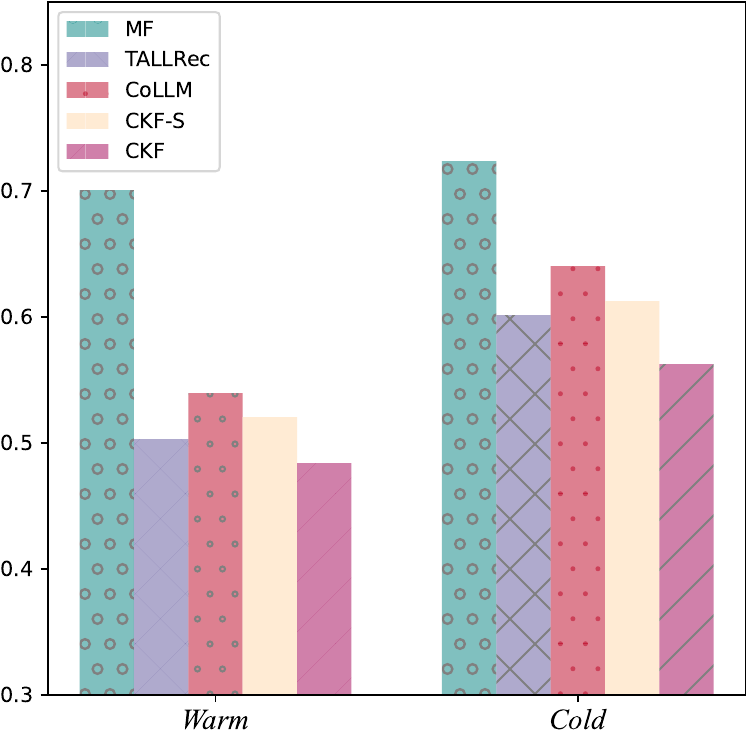}
%\caption{fig1}
\label{fig:sup:wc-rp}
\end{minipage}%
}%
\subfigure[{CTR (AUC $\uparrow$)}]{
\begin{minipage}[t]{0.24\linewidth}
\centering
\includegraphics[width=\linewidth]{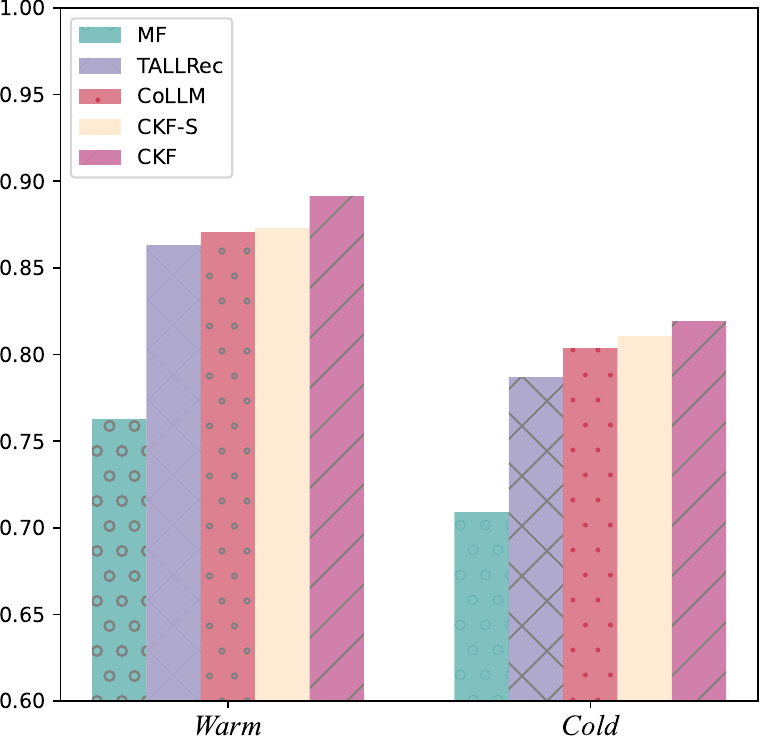}
%\caption{fig1}
\label{fig:sup:wc-ctr}
\end{minipage}%
}%
\subfigure[{Top-K (Hit@1-E $\uparrow$)}]{
\begin{minipage}[t]{0.24\linewidth}
\centering
\includegraphics[width=\linewidth]{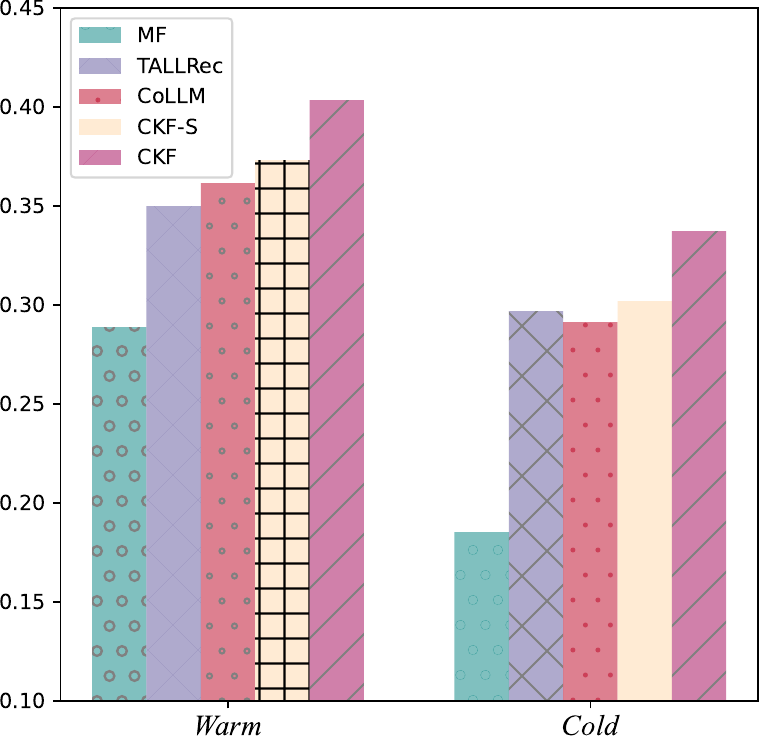}
%\caption{fig1}
\label{fig:sup:wc-topk}
\end{minipage}%
}%
\subfigure[{Explain (MAE $\downarrow$)}]{
\begin{minipage}[t]{0.24\linewidth}
\centering
\includegraphics[width=\linewidth]{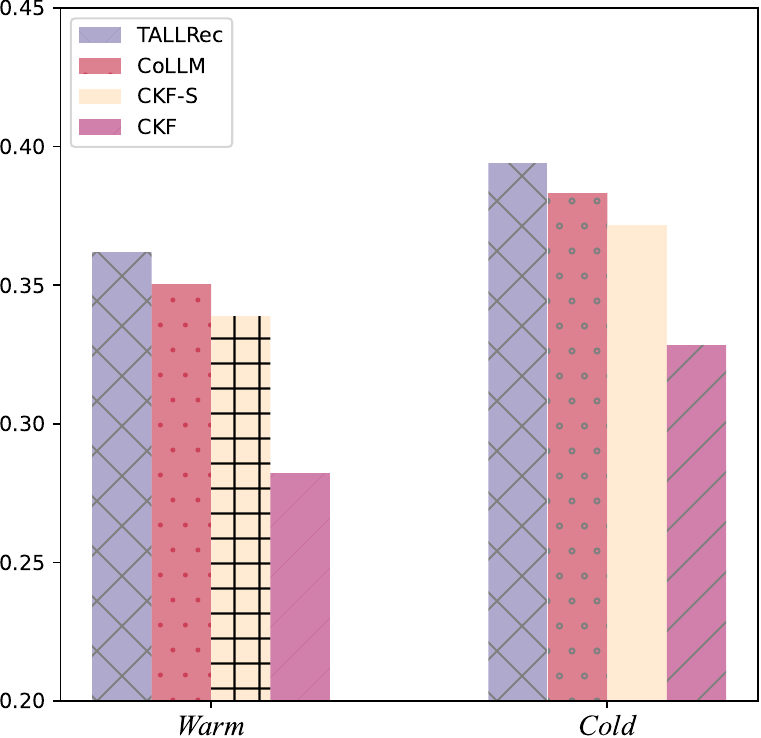}
%\caption{fig1}
\label{fig:sup:wc-exp}
\end{minipage}%
}%
\centering
\caption{{Warm-cold scenarios (Amazon Kindles data set).}}
\label{fig:sup:wc}
\end{figure*}

As mentioned in~\cite{zhang2023collm,talrec}, LLMs-based recommender systems may perform worse compared to CF-based models in information-rich scenarios.  To validate our approach's effectiveness, we rigorously evaluate it in two distinct contexts: the `warm start' and `cold start' scenarios, as shown in Figure~\ref{fig:wc-m}-\ref{fig:sup:wc}.
In the `warm start' scenario, the test users have prior representation in the training data set, while in the `cold start' scenario, the test users are entirely new and not included in the training set.
In line with intuition, algorithm performance in warm scenes generally surpasses that in cold scenes, suggesting that necessary interactions assist the model in recalling user preferences.
% Consistent with intuition, in the warm-start scenario, the performance disparity between MF and our algorithm is significantly reduced, highlighting the substantial impact and necessity of incorporating collaborative knowledge. 
{G}iven the LLM's capacity to discern intricate semantic nuances from text, the benefits of LLM-based algorithms become increasingly pronounced in cold-start scenarios.
{Notably, CoLLM's performance is not significantly better than TALLRec's for cold-start users. This limitation likely stems from its simple linear mapping equation, which hinders effective collaborative signal conversion for users who fall outside the training distribution. In contrast, CKF-S performs better in the cold start scenario, which illustrates the advantage of personalized mapping. 
Even in Amazon Movies-TV (RP / CTR), the overall performance of CKF-S is weak, with a significantly smaller performance gap in cold scenes compared to warm scenes.
We observe that in certain cases, such as the Amazon Books (Explain task), the performance of the LLM-based baseline in cold conditions is lower than in warm conditions. This is because LLM predictions primarily depend on the knowledge acquired during pretraining, rather than merely relying on co-occurrence signals.}
It is worth noting that CKF achieves a minimum of 4 \% and  10 \% improvement over other LLM baselines in warm and cold scenarios, respectively. 
In a nutshell, CKF embodies the merits of both approaches {and leverages the flexibility of the personalized mapping function}, rendering it adaptable to an extensive variety of recommendation scenarios.

\subsection{Few-shot Application}\label{sec:5.2}
\begin{figure}[!h] % exclude hyper
\centering
\subfigure[RP (MAE $\downarrow$)]{
\begin{minipage}[t]{0.24\linewidth}
\centering
\includegraphics[width=\linewidth]{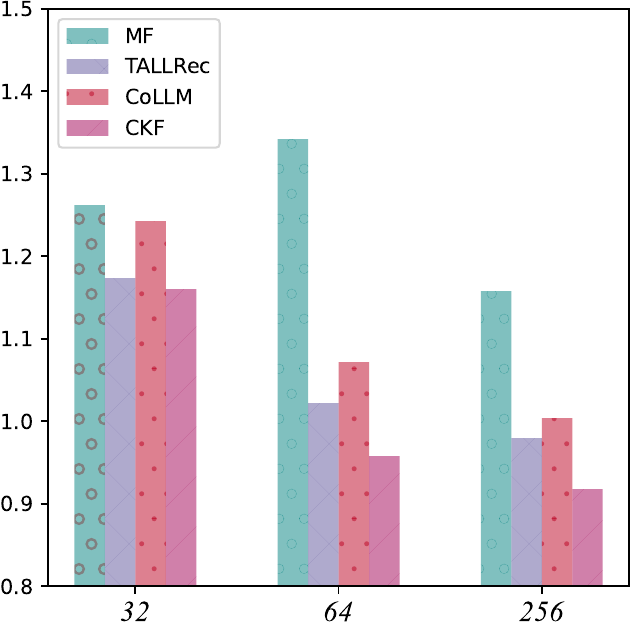}
%\caption{fig1}
\label{fig:few-rps-m}
\end{minipage}%
}%
\subfigure[CTR (AUC $\uparrow$)]{
\begin{minipage}[t]{0.24\linewidth}
\centering
\includegraphics[width=\linewidth]{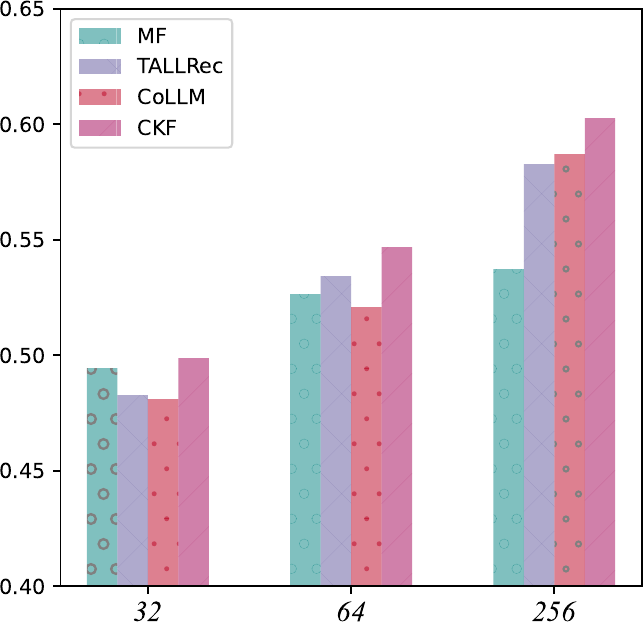}
%\caption{fig1}
\label{fig:few-ctrs-m}
\end{minipage}%
}%
\subfigure[Top-K (Hit@1-E $\uparrow$)]{
\begin{minipage}[t]{0.24\linewidth}
\centering
\includegraphics[width=\linewidth]{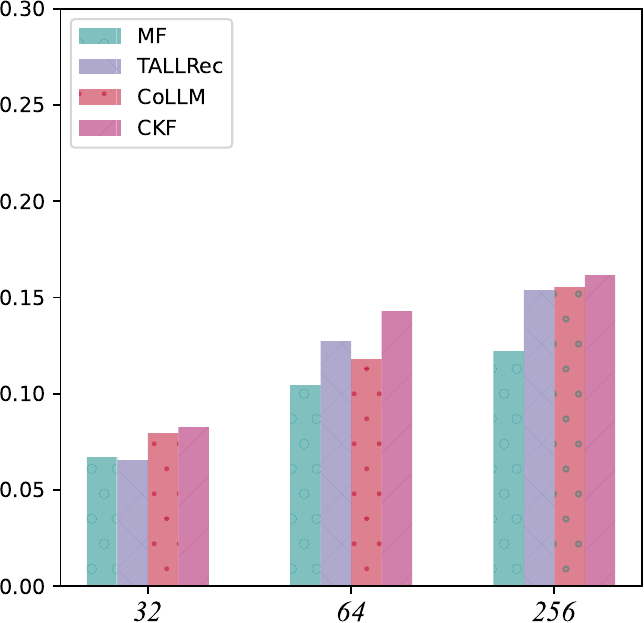}
%\caption{fig1}
\label{fig:few-topks-m}
\end{minipage}%
}%
% \subfigure[Explain (MAE $\downarrow$)]{
% \begin{minipage}[t]{0.24\linewidth}
% \centering
% \includegraphics[width=\linewidth]{fig/few-exps-m.pdf}
% %\caption{fig1}
% \label{fig:few-exps-m}
% \end{minipage}%
% }%
\centering
\caption{Few-shot {a}pplication (Movie-Lens data set). 32, 64, {and} 256 correspond to N-shot in the few-shot scenario, which represents the number of training data. A smaller N indicates a greater need for the model's rapid adaptation. We select the strong baselines MF, TALLRec, CoLLM{,} and CKF for comparison.}
\label{fig:few-m}
\end{figure}

\begin{figure}[!h] % exclude hyper
\centering
\subfigure[RP (MAE $\downarrow$)]{
\begin{minipage}[t]{0.24\linewidth}
\centering
\includegraphics[width=\linewidth]{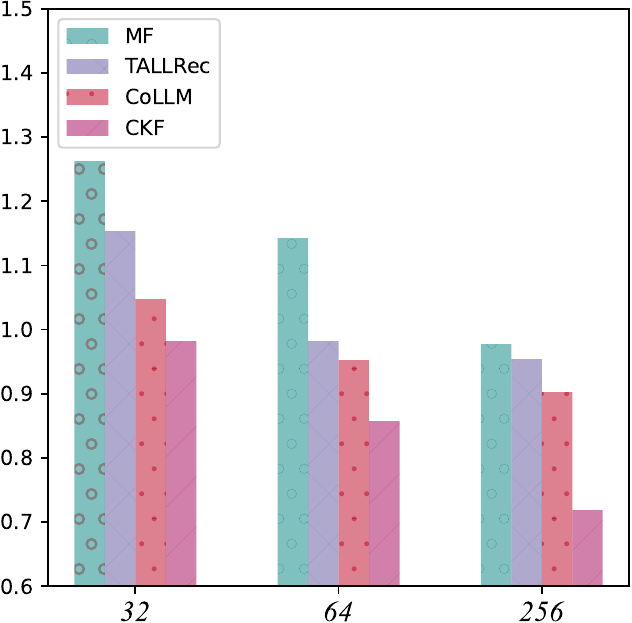}
%\caption{fig1}
\label{fig:few-rps-a}
\end{minipage}%
}%
\subfigure[CTR (AUC $\uparrow$)]{
\begin{minipage}[t]{0.24\linewidth}
\centering
\includegraphics[width=\linewidth]{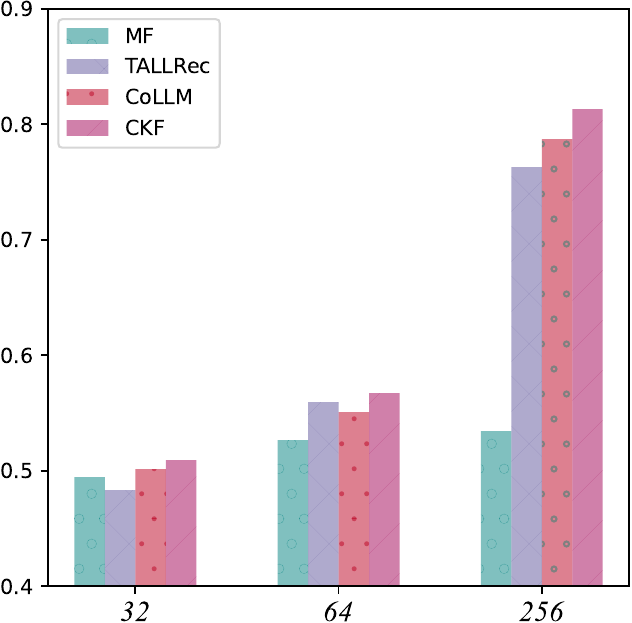}
%\caption{fig1}
\label{fig:few-ctrs-a}
\end{minipage}%
}%
\subfigure[Top-K (Hit@1-E $\uparrow$)]{
\begin{minipage}[t]{0.24\linewidth}
\centering
\includegraphics[width=\linewidth]{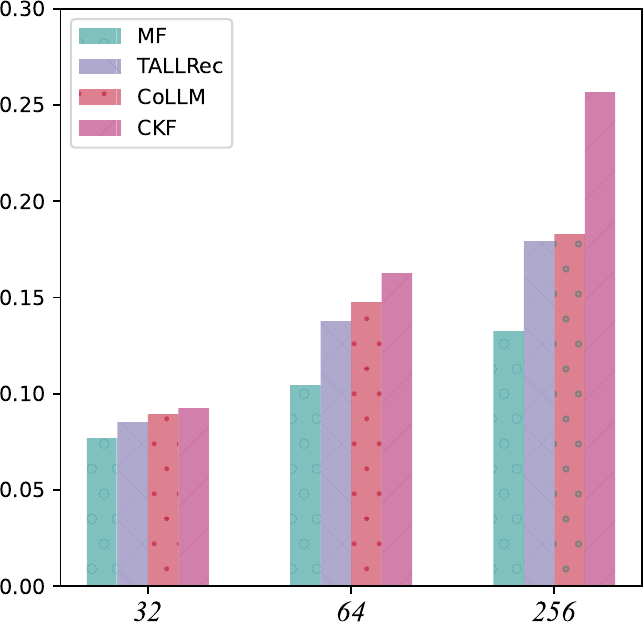}
%\caption{fig1}
\label{fig:few-topks-a}
\end{minipage}%
}%
\subfigure[Explain (MAE $\downarrow$)]{
\begin{minipage}[t]{0.24\linewidth}
\centering
\includegraphics[width=\linewidth]{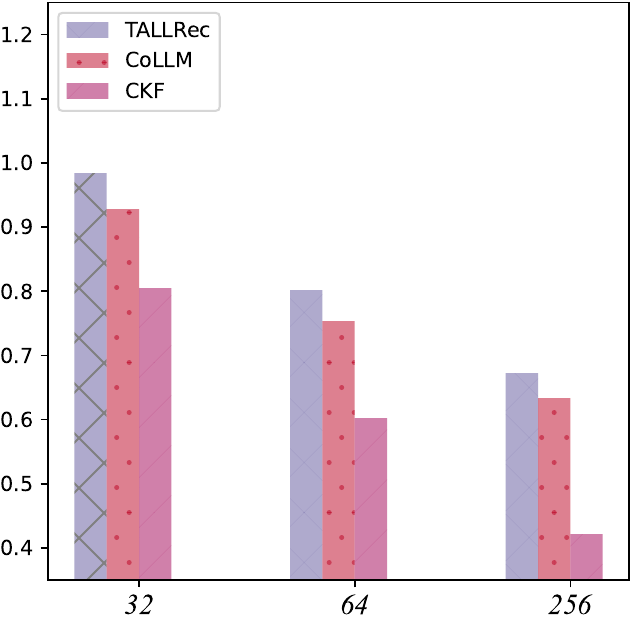}
%\caption{fig1}
\label{fig:few-exps-a}
\end{minipage}%
}%
\centering
\caption{Few-shot {a}pplication (Amazon Movies-TV data set). In (d), the comparison is limited to LLM baselines, as the CF model lacks suitability for Explainable recommendations.}
\label{fig:few-a}
\end{figure}

\begin{figure}[!h] % exclude hyper
\centering
\subfigure[RP (MAE $\downarrow$)]{
\begin{minipage}[t]{0.24\linewidth}
\centering
\includegraphics[width=\linewidth]{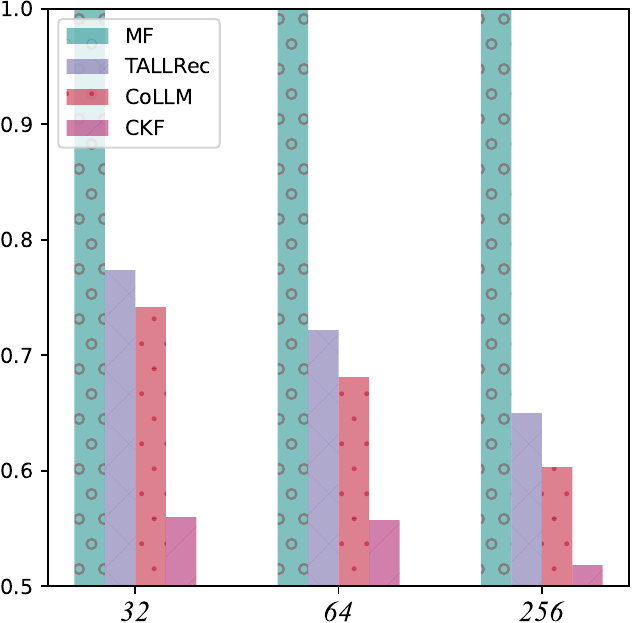}
%\caption{fig1}
\label{fig:few-rps}
\end{minipage}%
}%
\subfigure[CTR (AUC $\uparrow$)]{
\begin{minipage}[t]{0.24\linewidth}
\centering
\includegraphics[width=\linewidth]{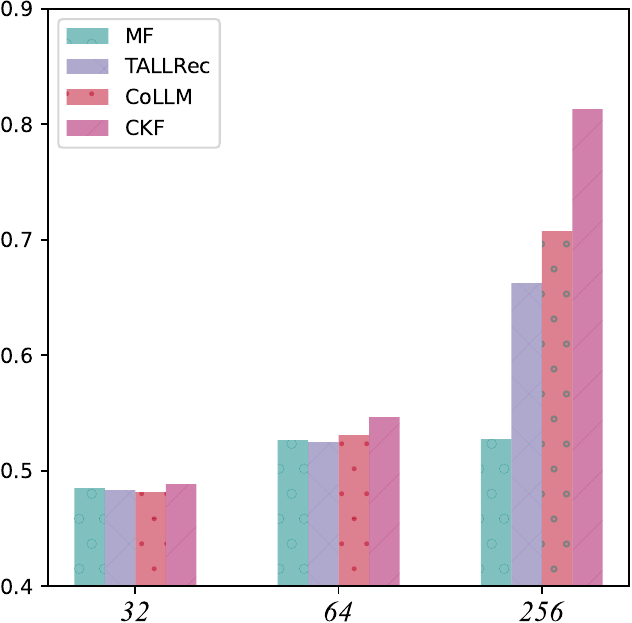}
%\caption{fig1}
\label{fig:few-ctrs}
\end{minipage}%
}%
\subfigure[Top-K (Hit@1-E $\uparrow$)]{
\begin{minipage}[t]{0.24\linewidth}
\centering
\includegraphics[width=\linewidth]{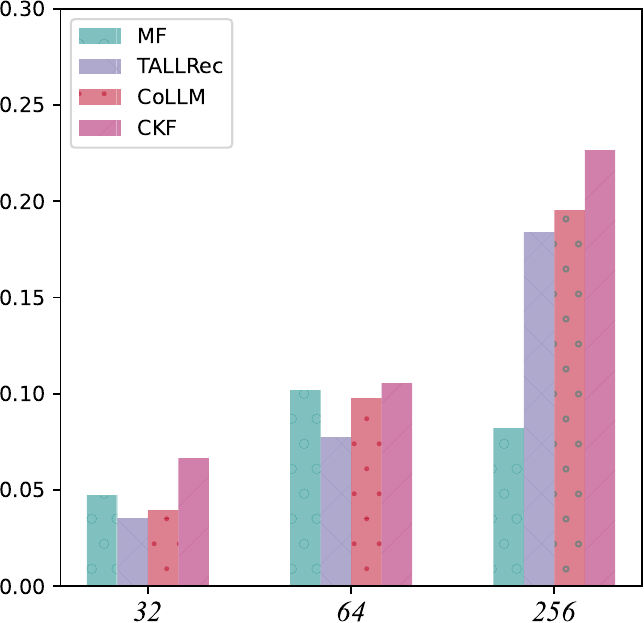}
%\caption{fig1}
\label{fig:few-topks}
\end{minipage}%
}%
\subfigure[Explain (MAE $\downarrow$)]{
\begin{minipage}[t]{0.24\linewidth}
\centering
\includegraphics[width=\linewidth]{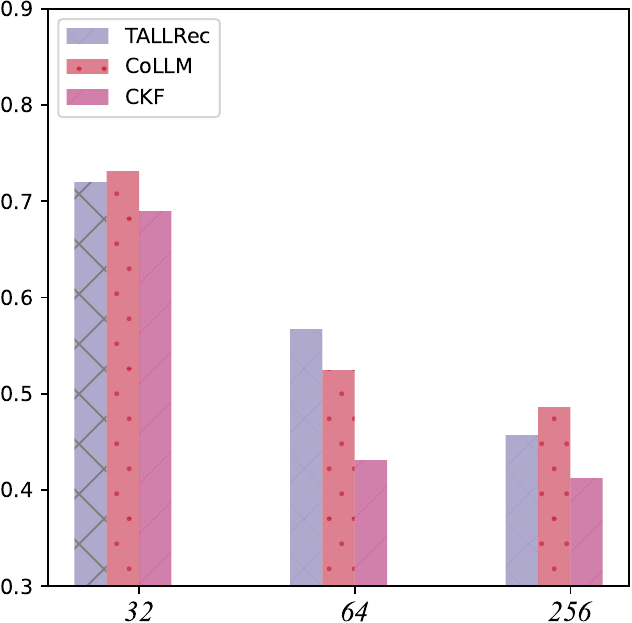}
%\caption{fig1}
\label{fig:few-exps}
\end{minipage}%
}%
\centering
\caption{Few-shot {a}pplication (Amazon Books data set). }
\label{fig:few}
\end{figure}

\begin{figure}[!h] % exclude hyper
\centering
\subfigure[{RP (MAE $\downarrow$)}]{
\begin{minipage}[t]{0.24\linewidth}
\centering
\includegraphics[width=\linewidth]{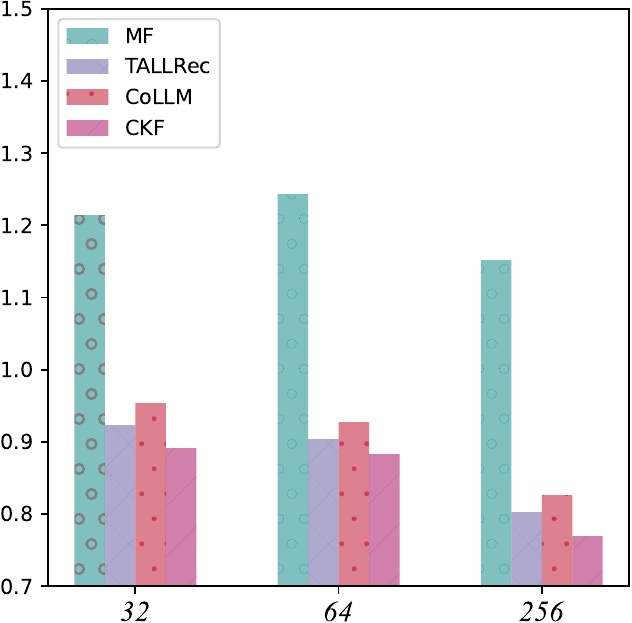}
%\caption{fig1}
\label{fig:sup:few-rps}
\end{minipage}%
}%
\subfigure[{CTR (AUC $\uparrow$)}]{
\begin{minipage}[t]{0.24\linewidth}
\centering
\includegraphics[width=\linewidth]{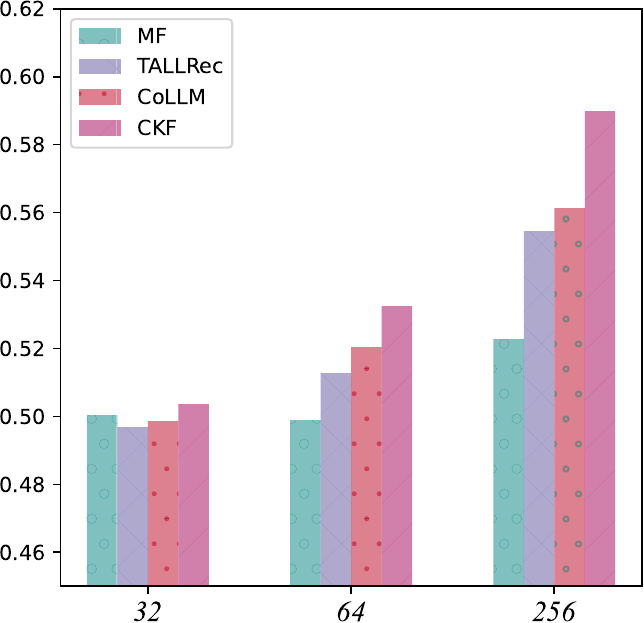}
%\caption{fig1}
\label{fig:sup:few-ctrs}
\end{minipage}%
}%
\subfigure[{Top-K (Hit@1-E $\uparrow$)}]{
\begin{minipage}[t]{0.24\linewidth}
\centering
\includegraphics[width=\linewidth]{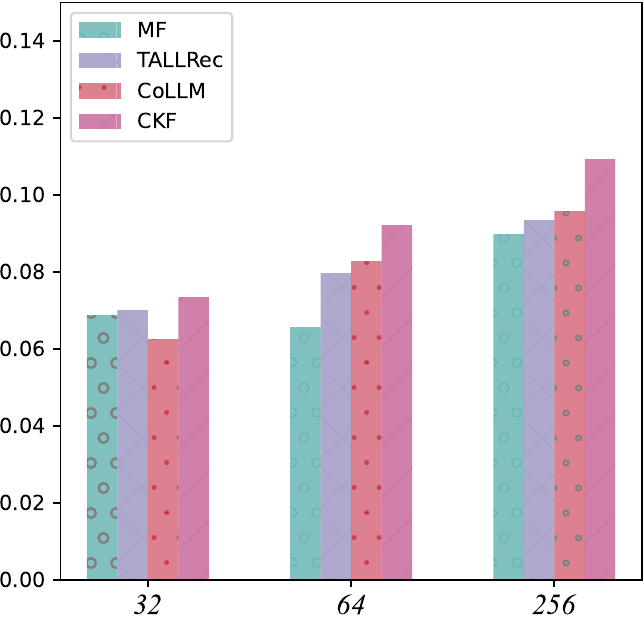}
%\caption{fig1}
\label{fig:sup:few-topks}
\end{minipage}%
}%
\subfigure[{Explain (MAE $\downarrow$)}]{
\begin{minipage}[t]{0.24\linewidth}
\centering
\includegraphics[width=\linewidth]{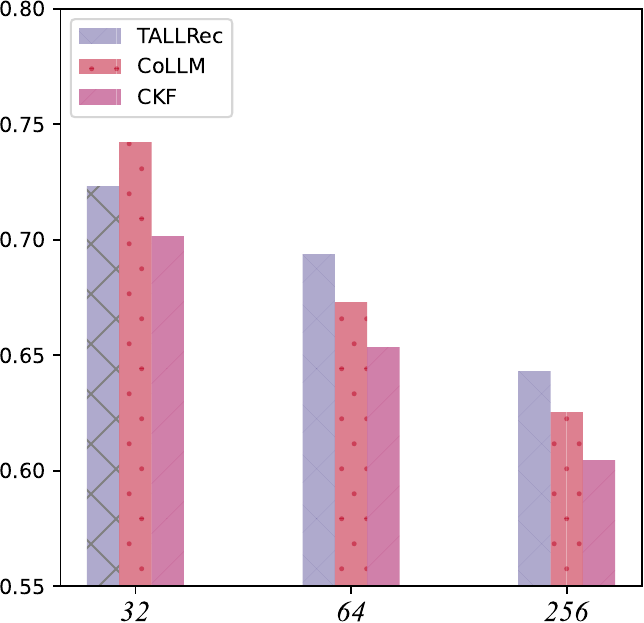}
%\caption{fig1}
\label{fig:sup:few-exps}
\end{minipage}%
}%
\centering
\caption{{Few-shot application (Amazon Kindles data set). }}
\label{fig:sup:few}
\end{figure}

Considering the strong {i}n-context learning capabilities of LLM-based recommendation systems, we extend our testing to evaluate algorithms' performance in various few-shot scenarios, as shown in Figure~\ref{fig:few-m}-\ref{fig:sup:few}.
With the increase in data volume, there is a corresponding enhancement in algorithmic performance, aligning well with intuitive expectations.
Obviously, the LLM-based baseline demonstrates significant progress; for instance, with just 32 pieces of data, TALLRec achieves a MAE of 0.78 (RP) in the Amazon Books data set. This notable performance is largely attributed to the extensive knowledge embedded in LLMs during the pre-training phase, facilitating efficient and rapid adaptation to tasks.
Conversely, under all few-shot training settings, except for Top-K (32, 64), MF consistently demonstrate{s} subpar performance. This suggests that the CF-based approach may lack the ability to rapidly acquire recommendation proficiency when faced with a limited number of training samples. This also suggests that collaborative knowledge might be less effective due to the insufficiency of training data.
Remarkably, even in a scenario with a mere 64 data points and exceedingly sparse signals, our model attains up to 19\%, 4\%, 1\%, and 13\% improvements in RP, CTR, Top-K, and Explain tasks (Amazon Books) respectively. We also observe considerable improvements {in} the other {three} data sets.
We posit that this improvement primarily stems from the enhanced Multi-Lora strategy, which substantially augments the source of supervision signals for the proposed CKF.

Overall, LLM-based baselines exhibit robust recommendation capabilities in few-shot contexts, with CKF emerging as the most sophisticated among them. This underscores CKF's practical worth in scenarios characterized by sparse data.

\subsection{Contexual Examination}
\begin{figure}[!h] % exclude hyper
\centering
\subfigure[RP (MAE $\downarrow$)]{
\begin{minipage}[t]{0.24\linewidth}
\centering
\includegraphics[width=\linewidth]{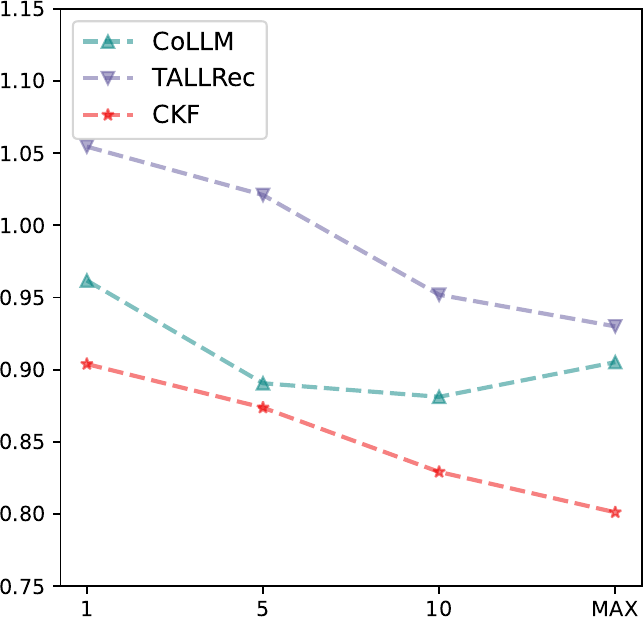}
%\caption{fig1}
\label{fig:con-rps-m}
\end{minipage}%
}%
\subfigure[CTR (AUC $\uparrow$)]{
\begin{minipage}[t]{0.24\linewidth}
\centering
\includegraphics[width=\linewidth]{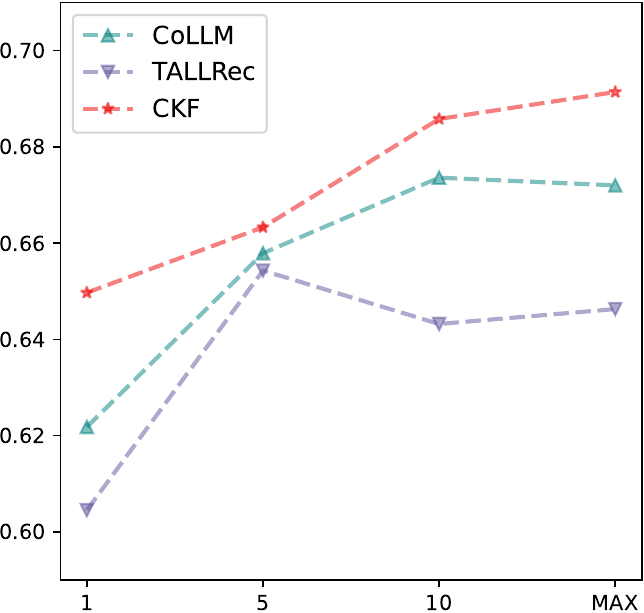}
%\caption{fig1}
\label{fig:con-ctrs-m}
\end{minipage}%
}%
\subfigure[Top-K (Hit@1-E $\uparrow$)]{
\begin{minipage}[t]{0.24\linewidth}
\centering
\includegraphics[width=\linewidth]{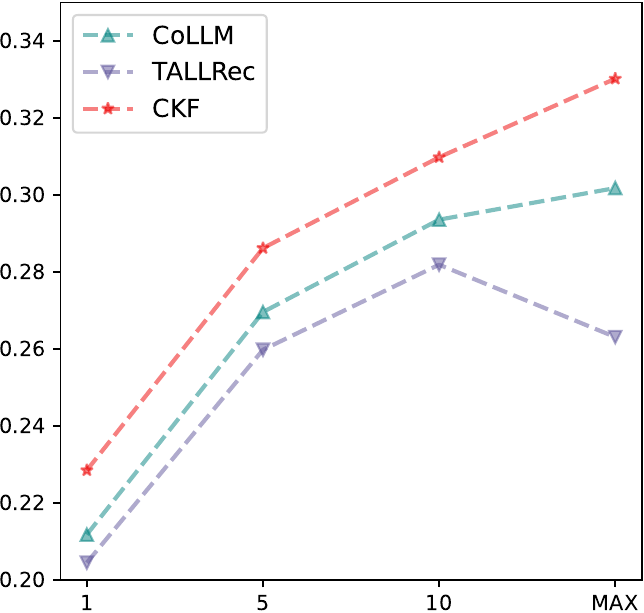}
%\caption{fig1}
\label{fig:con-topks-m}
\end{minipage}%
}%
% \subfigure[Explain (MAE $\downarrow$)]{
% \begin{minipage}[t]{0.24\linewidth}
% \centering
% \includegraphics[width=\linewidth]{fig/con-eps-m.pdf}
% %\caption{fig1}
% \label{fig:con-exps-m}
% \end{minipage}%
% }%
\centering
\caption{Contextual {e}xamination (Movie-lens data set). During the inference phase, we intentionally limit the length of the historical interaction sequence to specific endpoints [1, 5, 10, Max], with "Max" denoting the longest historical interaction sequence available. We select the strong LLM-based baselines TALLRec, CoLLM {,} and CKF for comparison.}
\label{fig:con-m}
\end{figure}

\begin{figure}[!h] % exclude hyper
\centering
\subfigure[RP (MAE $\downarrow$)]{
\begin{minipage}[t]{0.24\linewidth}
\centering
\includegraphics[width=\linewidth]{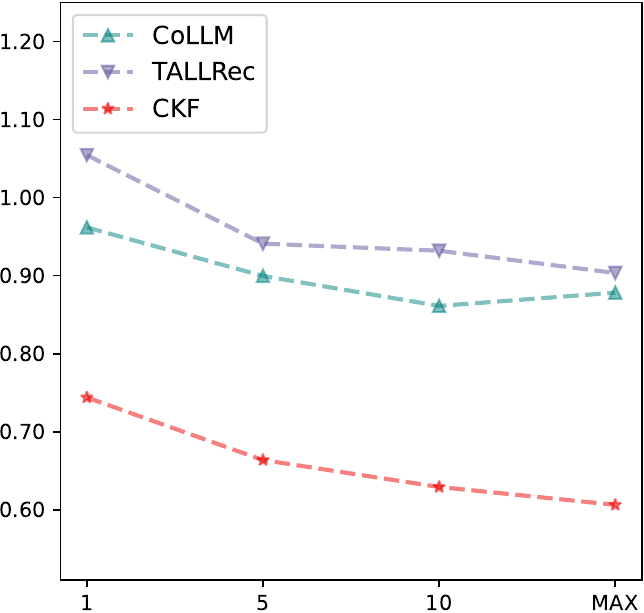}
%\caption{fig1}
\label{fig:con-rps-a}
\end{minipage}%
}%
\subfigure[CTR (AUC $\uparrow$)]{
\begin{minipage}[t]{0.24\linewidth}
\centering
\includegraphics[width=\linewidth]{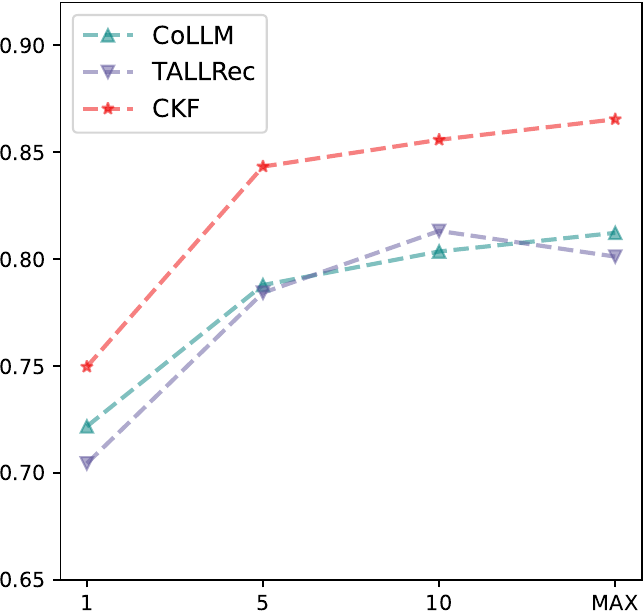}
%\caption{fig1}
\label{fig:con-ctrs-a}
\end{minipage}%
}%
\subfigure[Top-K (Hit@1-E $\uparrow$)]{
\begin{minipage}[t]{0.24\linewidth}
\centering
\includegraphics[width=\linewidth]{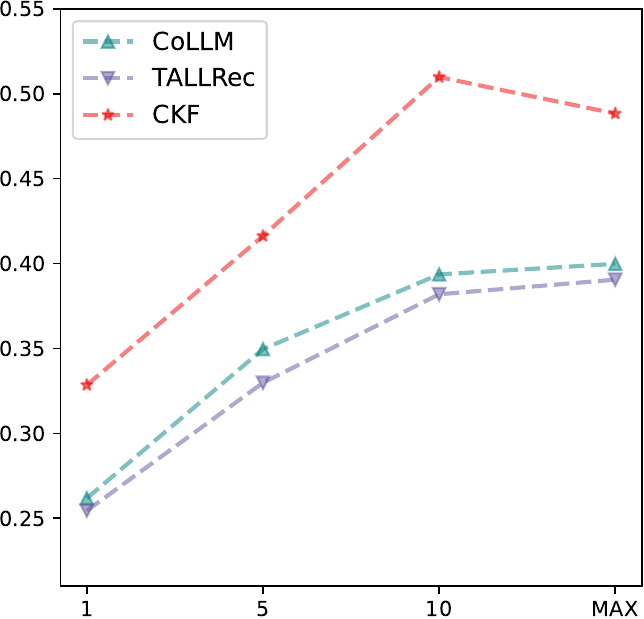}
%\caption{fig1}
\label{fig:con-topks-a}
\end{minipage}%
}%
\subfigure[Explain (MAE $\downarrow$)]{
\begin{minipage}[t]{0.24\linewidth}
\centering
\includegraphics[width=\linewidth]{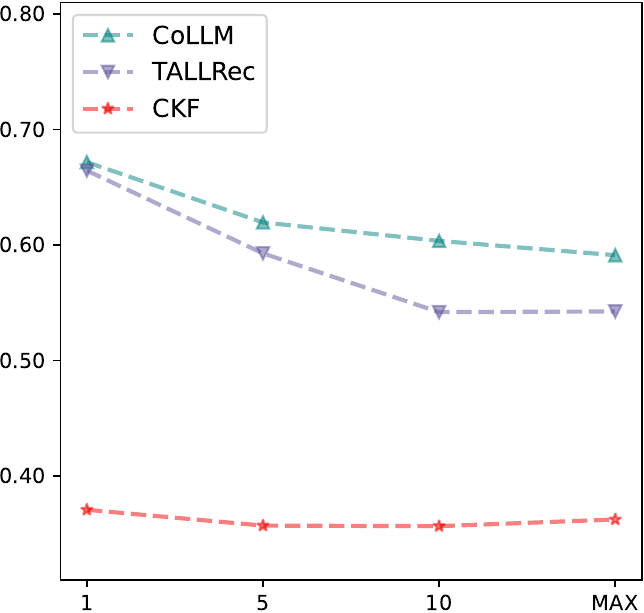}
%\caption{fig1}
\label{fig:con-exps-a}
\end{minipage}%
}%
\centering
\caption{Contextual {e}xamination (Amazon Movies-TV data set).}
\label{fig:con-a}
\end{figure}

\begin{figure}[!h] % exclude hyper
\centering
\subfigure[RP (MAE $\downarrow$)]{
\begin{minipage}[t]{0.24\linewidth}
\centering
\includegraphics[width=\linewidth]{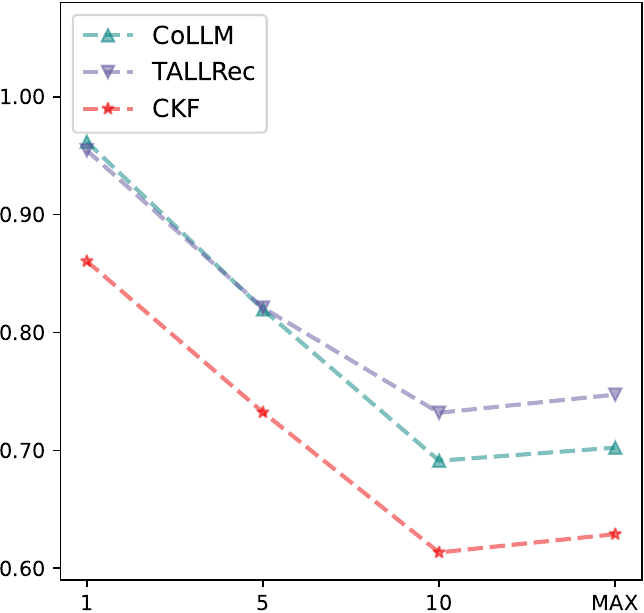}
%\caption{fig1}
\label{fig:con-rps}
\end{minipage}%
}%
\subfigure[CTR (AUC $\uparrow$)]{
\begin{minipage}[t]{0.24\linewidth}
\centering
\includegraphics[width=\linewidth]{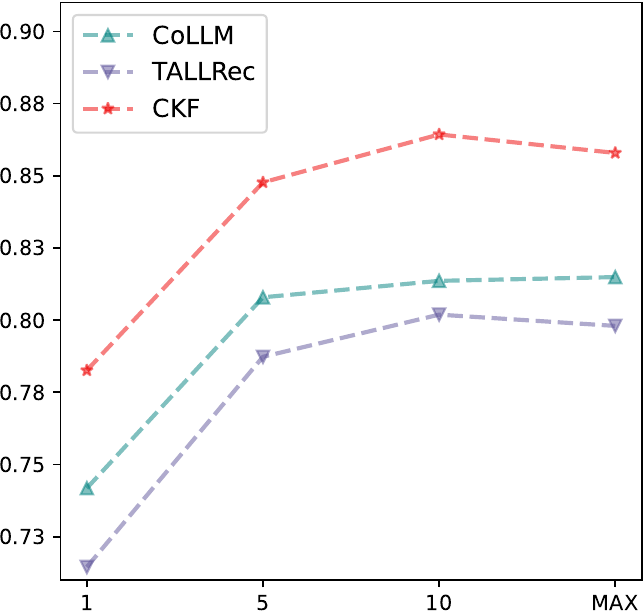}
%\caption{fig1}
\label{fig:con-ctrs}
\end{minipage}%
}%
\subfigure[Top-K (Hit@1-E $\uparrow$)]{
\begin{minipage}[t]{0.24\linewidth}
\centering
\includegraphics[width=\linewidth]{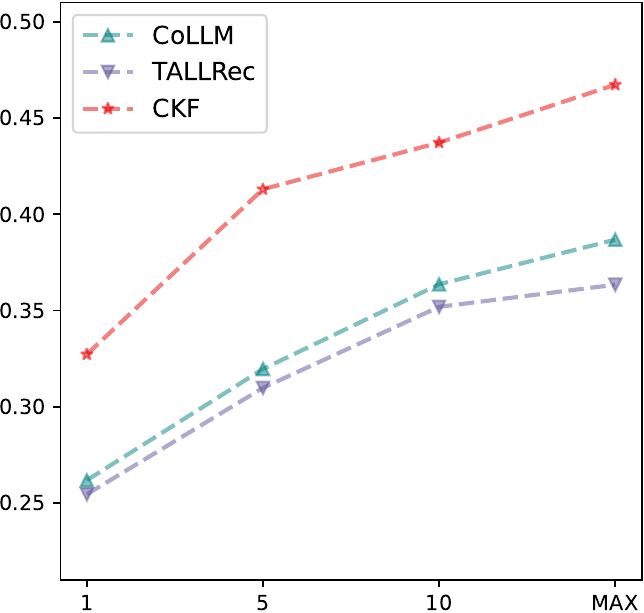}
%\caption{fig1}
\label{fig:con-topks}
\end{minipage}%
}%
\subfigure[Explain (MAE $\downarrow$)]{
\begin{minipage}[t]{0.24\linewidth}
\centering
\includegraphics[width=\linewidth]{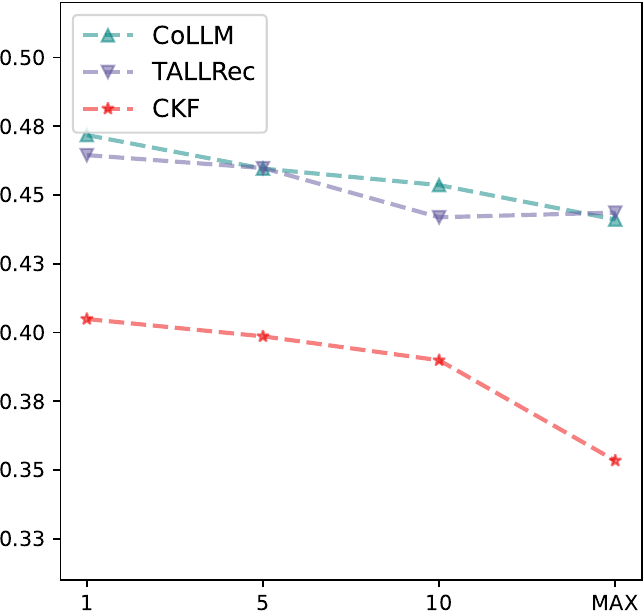}
%\caption{fig1}
\label{fig:con-exps}
\end{minipage}%
}%
\centering
\caption{Contextual {e}xamination (Amazon Books data set).}
\label{fig:con}
\end{figure}

\begin{figure}[!h] % exclude hyper
\centering
\subfigure[{RP (MAE $\downarrow$)}]{
\begin{minipage}[t]{0.24\linewidth}
\centering
\includegraphics[width=\linewidth]{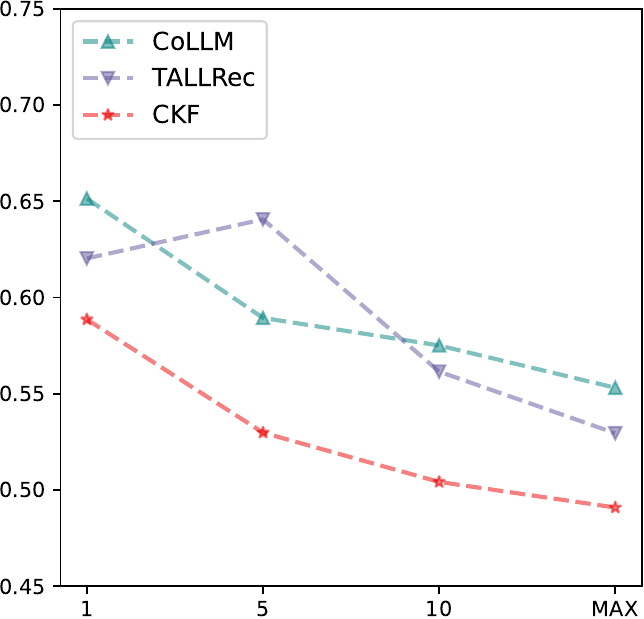}
%\caption{fig1}
\label{fig:sup:con-rps}
\end{minipage}%
}%
\subfigure[{CTR (AUC $\uparrow$)}]{
\begin{minipage}[t]{0.24\linewidth}
\centering
\includegraphics[width=\linewidth]{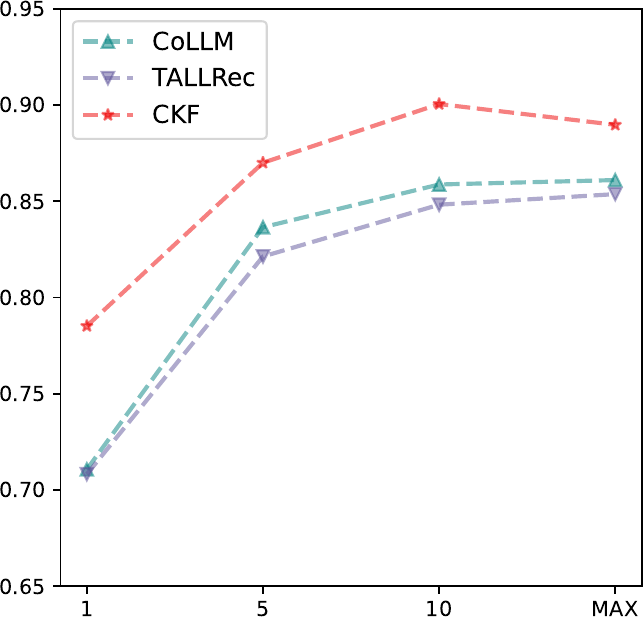}
%\caption{fig1}
\label{fig:sup:con-ctrs}
\end{minipage}%
}%
\subfigure[{Top-K (Hit@1-E $\uparrow$)}]{
\begin{minipage}[t]{0.24\linewidth}
\centering
\includegraphics[width=\linewidth]{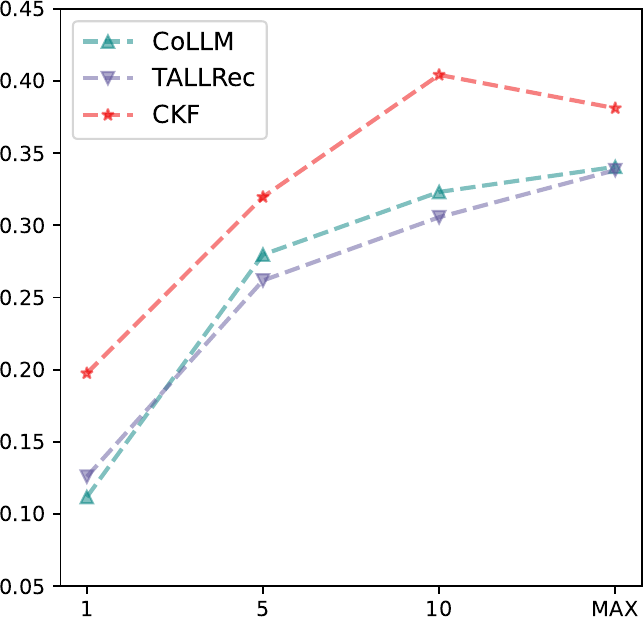}
%\caption{fig1}
\label{fig:sup:con-topks}
\end{minipage}%
}%
\subfigure[{Explain (MAE $\downarrow$)}]{
\begin{minipage}[t]{0.24\linewidth}
\centering
\includegraphics[width=\linewidth]{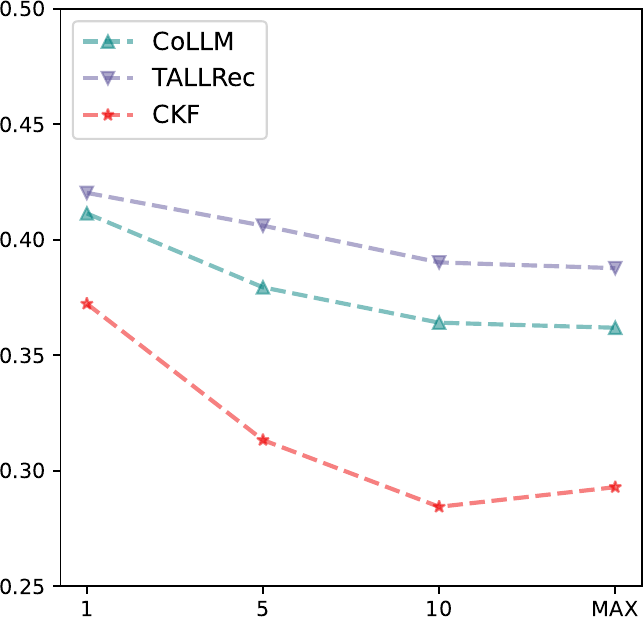}
%\caption{fig1}
\label{fig:sup:con-exps}
\end{minipage}%
}%
\centering
\caption{{Contextual examination (Amazon Kindles data set).}}
\label{fig:sup:con}
\end{figure}

% 把使用的Length也进行检测。
We further explore the effect of historical sequence length during the inference phase, {as it determines the scope of context accessible to the LLM}. A truncated history sequence curtails the context available for the model's use, ostensibly testing its capacity to leverage pre-existing LLM knowledge. As illustrated in Figure~\ref{fig:con-m}-\ref{fig:sup:con}, across various tasks, diminishing the length of historical sequences detrimentally affects all examined algorithms.  
% task
We additionally observe that the other two baseline{s} experience  significant detriment in performance on the Top-K task, attributable to their insufficiency in negative samples and a heightened requirement for more extensive data to effectively conduct ranking.
{Conversely, CKF improves ranking by incorporating users' positive and negative preferences from other tasks.}
% comparison
Against the LLM-based benchmark, our model registers at least a 3\% and 4\% advancement over TALLRec and CoLLM when only one item is observed. This underscores the substantial knowledge enhancement facilitated by our model's fine-tuning of the LLM. Even as historical sequence lengths extend, the disparity between our algorithm and other benchmarks remains relatively unchanged, suggesting that the enriched semantics derived from multiple tasks adeptly aid the LLM in grasping contextual nuances, thereby fostering personalized recommendations.
{To sum up, this experiment demonstrates that CKF has strong performance across users with varying levels of information richness, highlighting its practical value.}

\subsection{{Varied Pre-processing Protocols}}\label{sec:5.4}
\begin{figure}[!h] % exclude hyper
\centering
\subfigure[{RP (MAE $\downarrow$)}]{
\begin{minipage}[t]{0.24\linewidth}
\centering
\includegraphics[width=\linewidth]{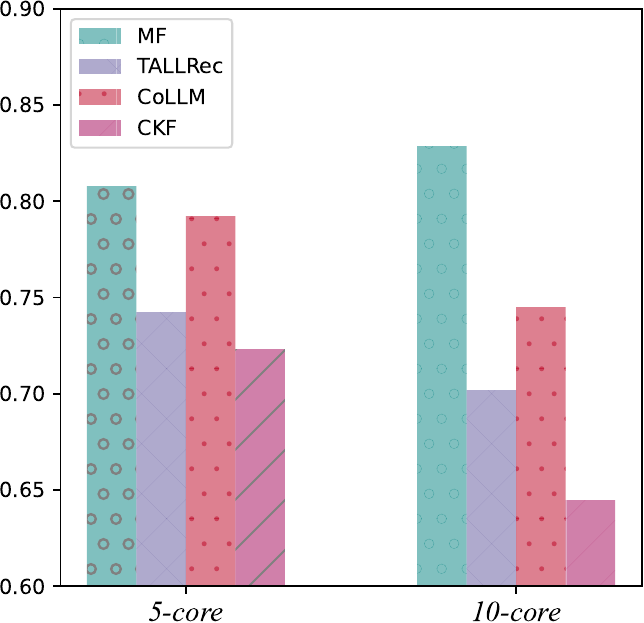}
%\caption{fig1}
\label{fig:sup:5:pre-rps}
\end{minipage}%
}%
\subfigure[{CTR (AUC $\uparrow$)}]{
\begin{minipage}[t]{0.24\linewidth}
\centering
\includegraphics[width=\linewidth]{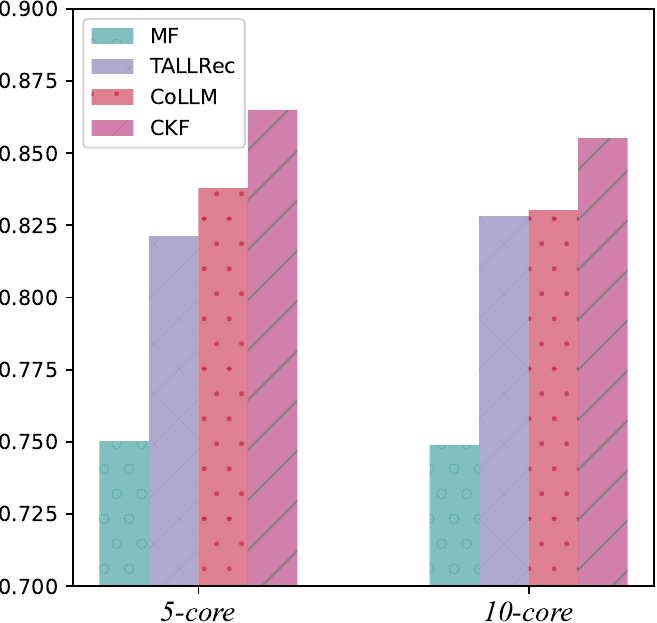}
%\caption{fig1}
\label{fig:sup:5:pre-ctrs}
\end{minipage}%
}%
\subfigure[{Top-K (Hit@1-E $\uparrow$)}]{
\begin{minipage}[t]{0.24\linewidth}
\centering
\includegraphics[width=\linewidth]{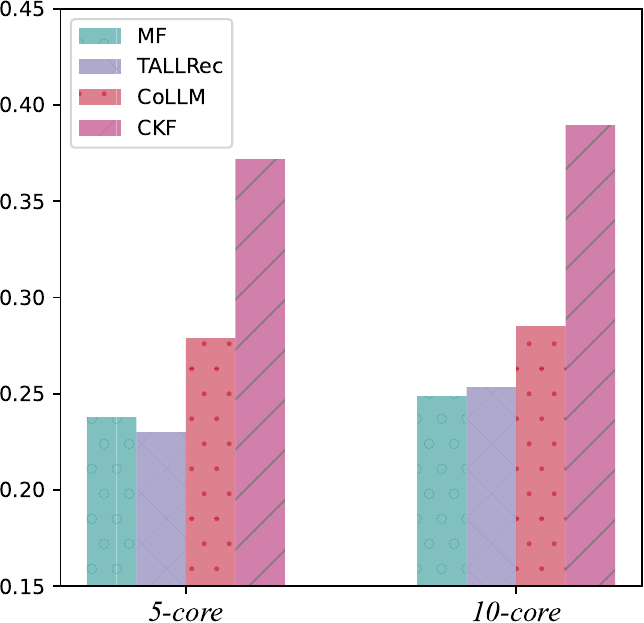}
%\caption{fig1}
\label{fig:sup:5:pre-topks}
\end{minipage}%
}%
\subfigure[{Explain (MAE $\downarrow$)}]{
\begin{minipage}[t]{0.24\linewidth}
\centering
\includegraphics[width=\linewidth]{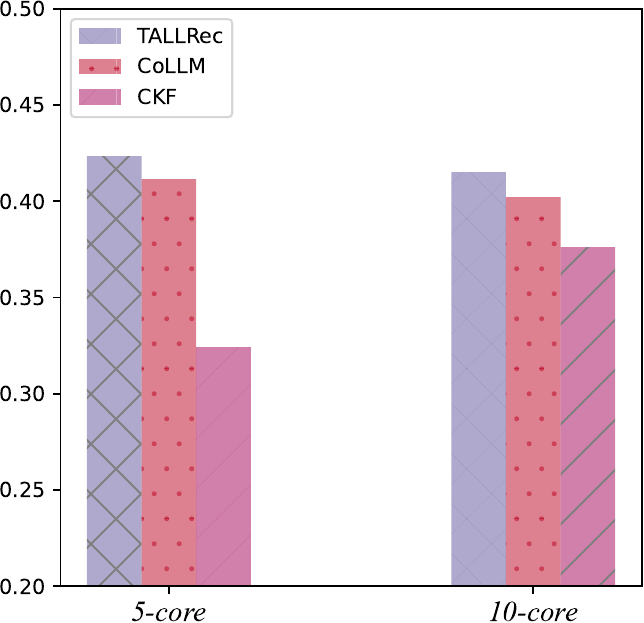}
%\caption{fig1}
\label{fig:sup:5:pre-exps}
\end{minipage}%
}%
\centering
\caption{{Varied pre-processing protocols (5-core / 10-core) (Amazon Books data set).}}
\label{fig:pre:5-10}
\end{figure}

{Following the previous LLM-based algorithms~\cite{zhang2023collm,liao2023llara}, we initially adopt a 20-core filtering method. To better align with real-world scenarios, we further evaluate the performance under varying pre-processing protocols using the Amazon Book data set.
Specifically, we adopt commonly used 5-core / 10-core criteria, retaining users and items with more than 5 / 10 interactions, respectively. 
The results are presented in Figure~\ref{fig:pre:5-10}}.

% \begin{figure}[!h] % exclude hyper
% \centering
% \subfigure[{RP (MAE $\downarrow$)}]{
% \begin{minipage}[t]{0.24\linewidth}
% \centering
% \includegraphics[width=\linewidth]{fig/pre-rps.pdf}
% %\caption{fig1}
% \label{fig:sup:10:pre-rps}
% \end{minipage}%
% }%
% \subfigure[{CTR (AUC $\uparrow$)}]{
% \begin{minipage}[t]{0.24\linewidth}
% \centering
% \includegraphics[width=\linewidth]{fig/pre-ctrs.pdf}
% %\caption{fig1}
% \label{fig:sup:10:pre-ctrs}
% \end{minipage}%
% }%
% \subfigure[{Top-K (Hit@1-E $\uparrow$)}]{
% \begin{minipage}[t]{0.24\linewidth}
% \centering
% \includegraphics[width=\linewidth]{fig/pre-topks.pdf}
% %\caption{fig1}
% \label{fig:sup:10:pre-topks}
% \end{minipage}%
% }%
% \subfigure[{Explain (MAE $\downarrow$)}]{
% \begin{minipage}[t]{0.24\linewidth}
% \centering
% \includegraphics[width=\linewidth]{fig/pre-exps.pdf}
% %\caption{fig1}
% \label{fig:sup:10:pre-exps}
% \end{minipage}%
% }%
% \centering
% % \setlength{\abovecaptionskip}{-0.01cm}   %调整图片标题与图距离
% % \setlength{\belowcaptionskip}{-0.1cm}   %调整图片标题与下文距离
% \caption{{Varied pre-processing protocols (10-core).}}
% \label{fig:pre:10}
% \end{figure}
{
We have two interesting findings.
First, we observe that although the 5-core pre-processing introduces more data points, it leads to a degradation of sequence quality. This degradation significantly impacts the performance of LLM-based algorithms. More precisely, LLM performance in Top-K and Explain tasks is most adversely affected, as users with fewer interactions typically exhibit unclear intentions and provide low-quality comments. This lack of data makes it challenging for LLMs to accurately grasp comprehensive user preferences, consequently leading to a decline in performance.
In contrast, the CF-based algorithm is less impacted by this issue. For example, in the RP task, MF's MAE decrease with an increased number of co-occurrence relationships.
Second, CKF and CoLLM show better results across all scenarios. This stability stems from their ability to leverage the dual advantages of CF and LLM methodologies. While shorter user sequences present challenges for the LLM components, the enhanced collaborative signals from their CF components effectively compensate for these limitations. Additionally, we observe CKF extending its advantage over CoLLM. On the one hand, the decline in data quality makes the model more dependent on collaborative signals. A more sophisticated mapping design helps preserve users' preferences and personalized characteristics within the collaborative space, as detailed in Section~\ref{sec:5.1}. On the other hand, the integration of multi-task signals effectively facilitates the understanding of user preferences by enabling mutual knowledge transfer across tasks.}
{To sum up, these tests help validate our framework's adaptability and superiority, confirming that it can effectively handle real-world data sets that may exhibit varying degrees of sparsity.}

\begin{figure}[!h] % exclude hyper
\centering
\subfigure[{RP (MAE $\downarrow$)}]{
\begin{minipage}[t]{0.24\linewidth}
\centering
\includegraphics[width=\linewidth]{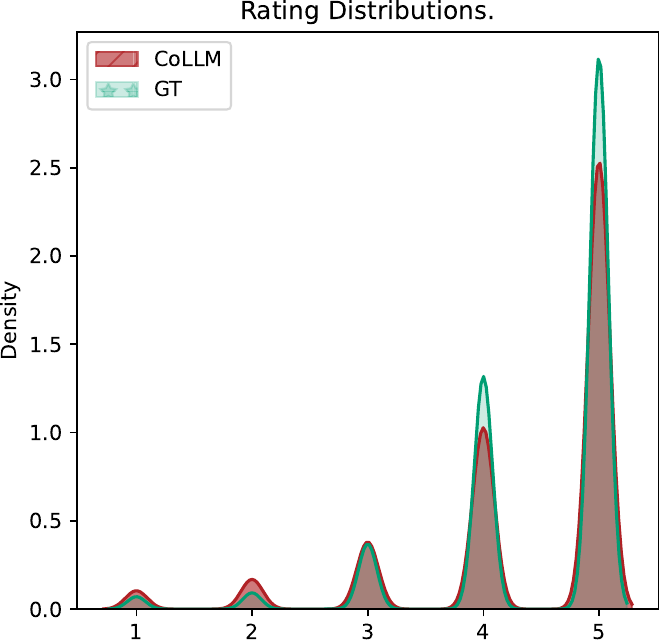}
%\caption{fig1}
\label{fig:sup:rp:collm}
\end{minipage}%
}%
\subfigure[{Explain (MAE $\downarrow$)}]{
\begin{minipage}[t]{0.24\linewidth}
\centering
\includegraphics[width=\linewidth]{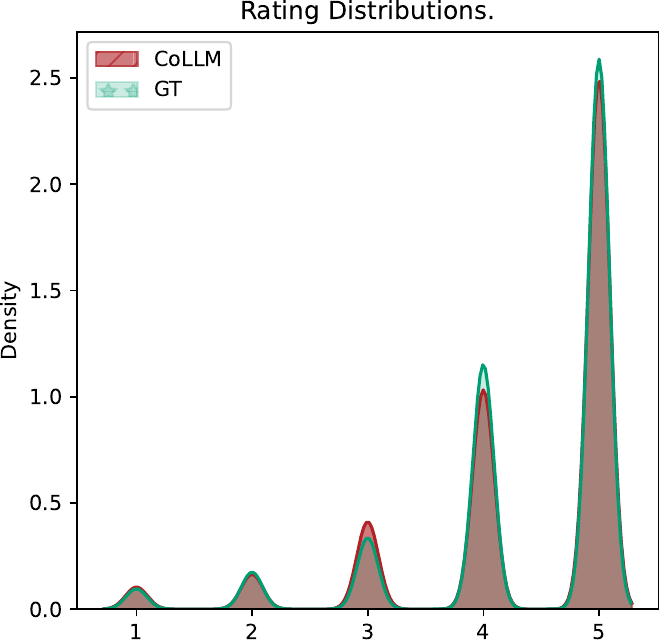}
%\caption{fig1}
\label{fig:sup:rp:ckf}
\end{minipage}%
}%
\subfigure[{RP (MAE $\downarrow$)}]{
\begin{minipage}[t]{0.24\linewidth}
\centering
\includegraphics[width=\linewidth]{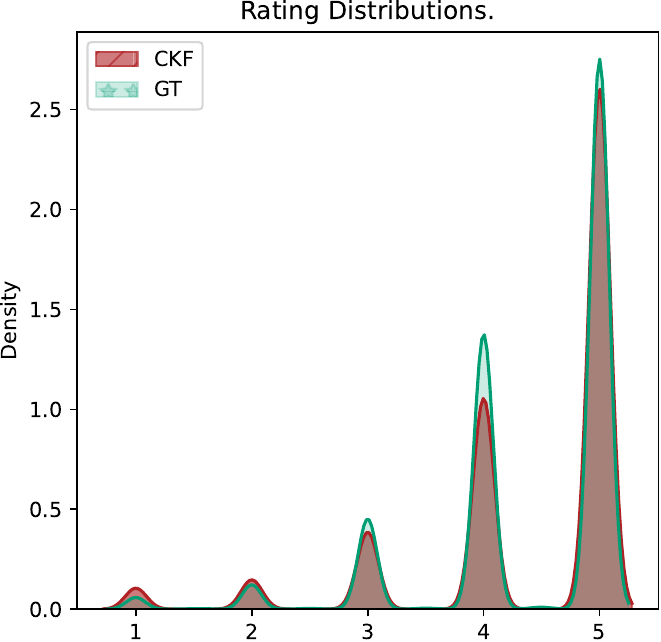}
\label{fig:sup:exp:collm}
\end{minipage}%
}%
\subfigure[{Explain (MAE $\downarrow$)}]{
\begin{minipage}[t]{0.24\linewidth}
\centering
\includegraphics[width=\linewidth]{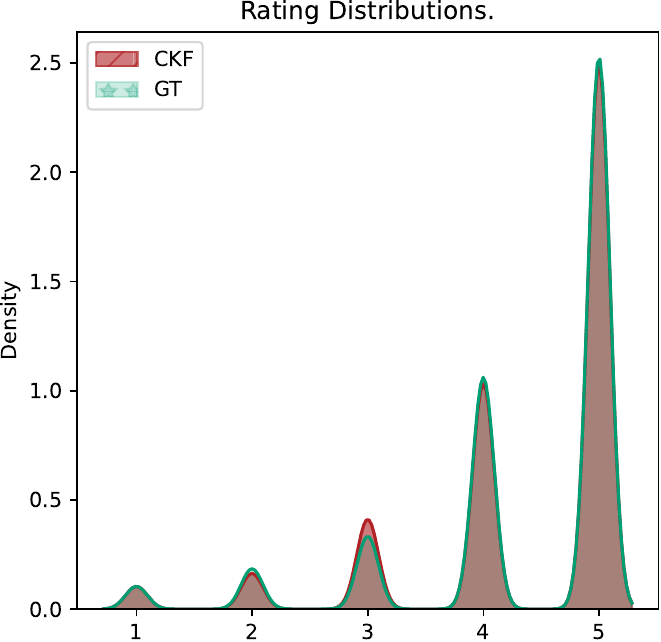}
%\caption{fig1}
\label{fig:sup:exp:ckf}
\end{minipage}%
}%
\centering
\caption{{Rating distribution. (a) and (b) show the rating distribution of CoLLM on RP and Explain tasks, while (c) and (d) represent the rating distribution of CKF on RP and Explain tasks.}}
\label{fig:sup:rp}
\end{figure}

\subsection{{Task Specificity Examination}}
{In this subsection, we enhance the discussion on different tasks to provide deeper insights. Without loss of generality, we test them on the Amazon Books data set.}
\subsubsection{{Rating Distribution}}\label{sec:5.5.1}
{
Since rating data is not randomly missing as interaction data is, users typically rate items they particularly like or dislike~\cite{ying2006leveraging}. This bias can lead the model to depend on an intermediate score to minimize loss rather than the actual predicted score. To investigate the underlying mechanisms of various algorithms, we visualize the prediction distribution for some competitive algorithms on RP and Explain tasks, leading to two key findings. 
First, as expected, the distribution of ratings is indeed skewed towards the 3-5 range, which could potentially bias the model's predictions towards this popular range.
Second, our analysis indicates that CoLLM is more adept at predicting scores within the majority range, while its performance worsens in the 1-2 range. This confirms that the biased distribution impacts the models.
In contrast, CKF more effectively predicts scores that closely align with the true distribution of ratings. We attribute this to CKF's multi-task design, which requires the simultaneous optimization of multiple tasks. This approach compels the model to consider the impact of representations on other tasks, thereby reducing bias towards popular ratings.}

\subsubsection{{Top-K Scalability}}\label{sec:5.5.2}
{
Unlike traditional collaborative recommender systems~\cite{zhou2018deep} and the approach described in~\cite{geng2022recommendation}, which rely solely on ID encoding, our method follows~\cite{zhang2023collm,talrec,bao2023bi} by directly inputting the item titles. This allows the LLM to extract semantic associations from the text, aiming for a deeper understanding of content relevance~\cite{talrec,zhang2023collm}. 
However, the ability to include a larger candidate set is limited by the LLM's capabilities, as it would necessitate processing longer text sequences. This requirement is constrained by the LLM's window size and increased computational burden. Nonetheless, in the Top-K recommendation, the number of negative samples in the candidate set significantly impacts the results. An increased number of negative samples $\mathcal{N}_{neg}$ can introduce potentially similar entities, which may obscure the recommendation outcomes. Therefore, we incorporate tests focusing on Top-K scalability. 
}

\begin{figure}[!h] % exclude hyper
\centering
\subfigure[{RP (MAE $\downarrow$)}]{
\begin{minipage}[t]{0.24\linewidth}
\centering
\includegraphics[width=\linewidth]{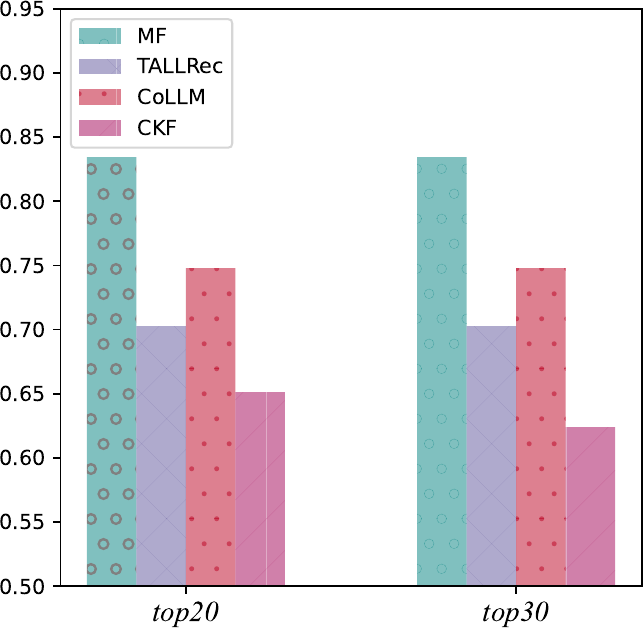}
%\caption{fig1}
\label{fig:sup:top-k:rps}
\end{minipage}%
}%
\subfigure[{CTR (AUC $\uparrow$)}]{
\begin{minipage}[t]{0.24\linewidth}
\centering
\includegraphics[width=\linewidth]{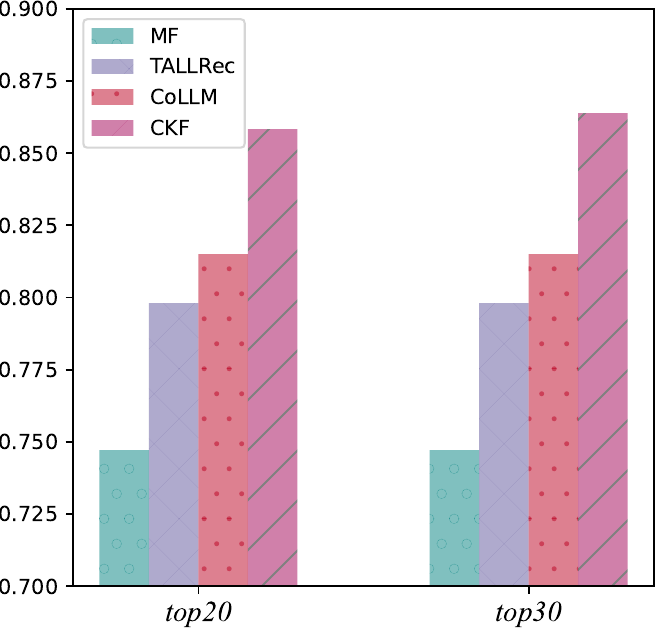}
%\caption{fig1}
\label{fig:sup:top-k:ctrs}
\end{minipage}%
}%
\subfigure[{Top-K (Hit@1-E $\uparrow$)}]{
\begin{minipage}[t]{0.24\linewidth}
\centering
\includegraphics[width=\linewidth]{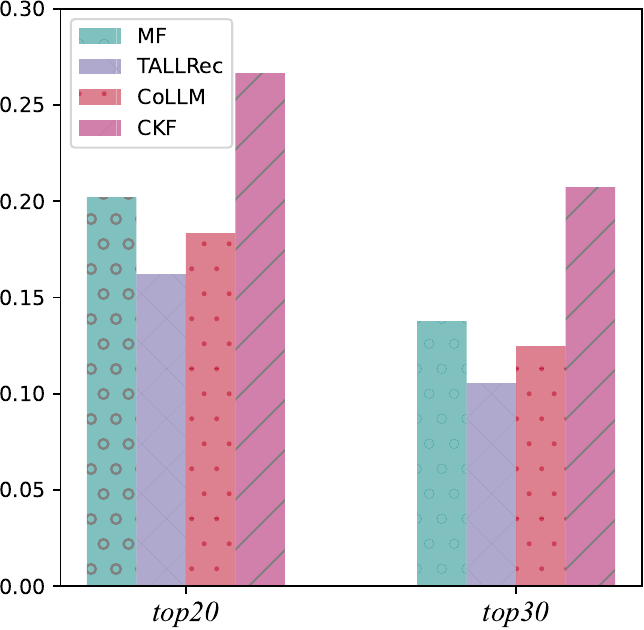}
%\caption{fig1}
\label{fig:sup:top-k:topks}
\end{minipage}%
}%
\subfigure[{Explain (MAE $\downarrow$)}]{
\begin{minipage}[t]{0.24\linewidth}
\centering
\includegraphics[width=\linewidth]{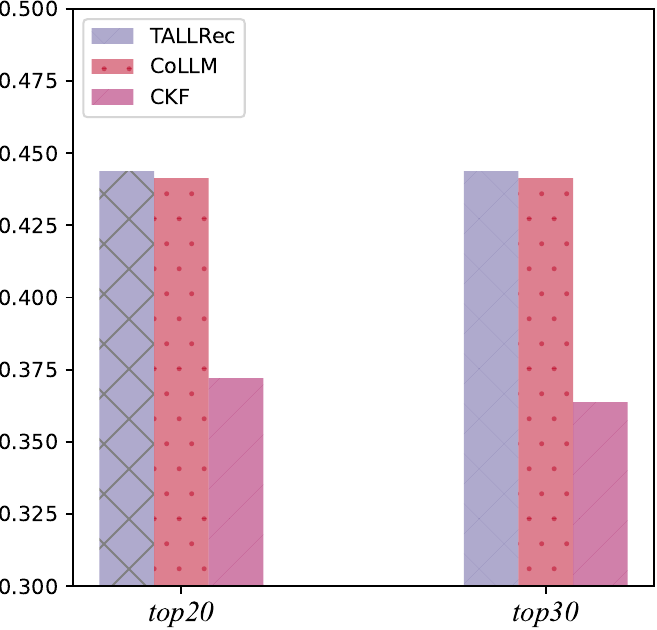}
%\caption{fig1}
\label{fig:sup:top-k:exps}
\end{minipage}%
}%
\centering
\caption{{Top-K scalability ($\mathcal{N}_{neg}$=20 / 30, in line with~\cite{bao2023bi,liao2023llara}). }}
\label{fig:sup:top-k}
\end{figure}

{
As illustrated in Figure~\ref{fig:sup:top-k}, we have three intriguing observations. First, we note that as the size of the candidate set expands (i.e., as $\mathcal{N}_{neg}$ increases), the performance of all algorithms in the Top-K evaluations declines. This decline could be attributed to the larger number of negative samples, which likely introduces more samples with similar semantics. This scenario demands that the model possess enhanced capabilities for semantic discrimination among samples.
Second, we observe a significant improvement of CKF for the CTR and RP tasks as the size of $\mathcal{N}_{neg}$ increases. This indicates a strong correlation between the Top-K task and the other tasks. In the context of multi-task training, the exposure to more negative samples in the Top-K task helps it gain an advantage in other tasks, such as Yes or No decisions.}

% \begin{figure}[!h] % exclude hyper
% \centering
% \subfigure[{RP (MAE $\downarrow$)}]{
% \begin{minipage}[t]{0.24\linewidth}
% \centering
% \includegraphics[width=\linewidth]{fig/top30-rp.pdf}
% %\caption{fig1}
% \label{fig:sup:30:top-k:rps}
% \end{minipage}%
% }%
% \subfigure[{CTR (AUC $\uparrow$)}]{
% \begin{minipage}[t]{0.24\linewidth}
% \centering
% \includegraphics[width=\linewidth]{fig/top30-ctr.pdf}
% %\caption{fig1}
% \label{fig:sup:30:top-k:ctrs}
% \end{minipage}%
% }%
% \subfigure[{Top-K (Hit@1-E $\uparrow$)}]{
% \begin{minipage}[t]{0.24\linewidth}
% \centering
% \includegraphics[width=\linewidth]{fig/top30-topk.pdf}
% %\caption{fig1}
% \label{fig:sup:30:top-k:topks}
% \end{minipage}%
% }%
% \subfigure[{Explain (MAE $\downarrow$)}]{
% \begin{minipage}[t]{0.24\linewidth}
% \centering
% \includegraphics[width=\linewidth]{fig/top30-exp.pdf}
% %\caption{fig1}
% \label{fig:sup:30:top-k:exps}
% \end{minipage}%
% }%
% \centering
% % \setlength{\abovecaptionskip}{-0.01cm}   %调整图片标题与图距离
% % \setlength{\belowcaptionskip}{-0.1cm}   %调整图片标题与下文距离
% \caption{{Top-K scalability ($\mathcal{N}_{neg}$=30).}}
% \label{fig:sup:30:top-k}
% \end{figure}
\subsubsection{{Diverse Explainable Settings}}\label{sec:5.5.3}
\begin{figure*}[!ht]
  \centering
   \includegraphics[width=0.8\linewidth]{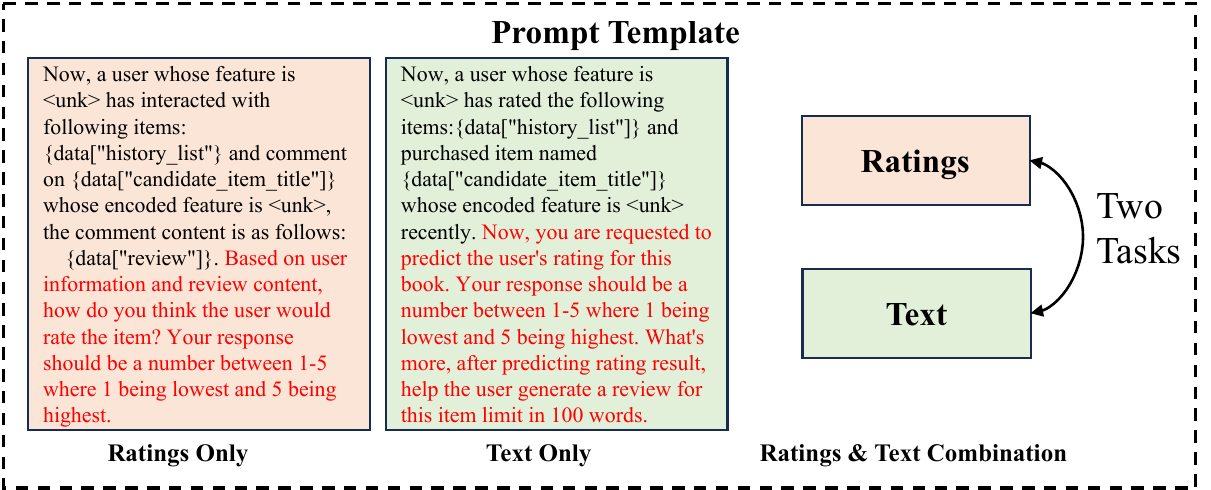}
   \caption{{Task Prompt. Note that the definition of "ratings only" is the same as that of the Explain task in Figure~\ref{fig:taskp}. "Combination" refers to the use of two Explain tasks simultaneously.}}
   \label{fig:taskp2}
\end{figure*}

{Our initial approach on the Explain task primarily focuses on using comments to predict ratings. However, we note that other methods may employ different definitions of the Explain task. For example,
in~\cite{geng2022recommendation}, two tasks are employed as criteria for evaluating explanation recommendation: using comments to predict ratings and generating explainable texts to the ratings. 
Therefore, we have broadened our analysis, exploring three distinct approaches: predicting ratings alone, generating texts alone, and simultaneously generating both ratings and texts. Their prompts are shown in Figure~\ref{fig:taskp2}. Please note that in all settings, following~\cite{geng2022recommendation}, we use ROUGE-L as the uniform evaluation metric, where a higher value indicates better generation performance.}

\begin{figure}[!h] % exclude hyper
\centering
\subfigure[{RP (MAE $\downarrow$)}]{
\begin{minipage}[t]{0.24\linewidth}
\centering
\includegraphics[width=\linewidth]{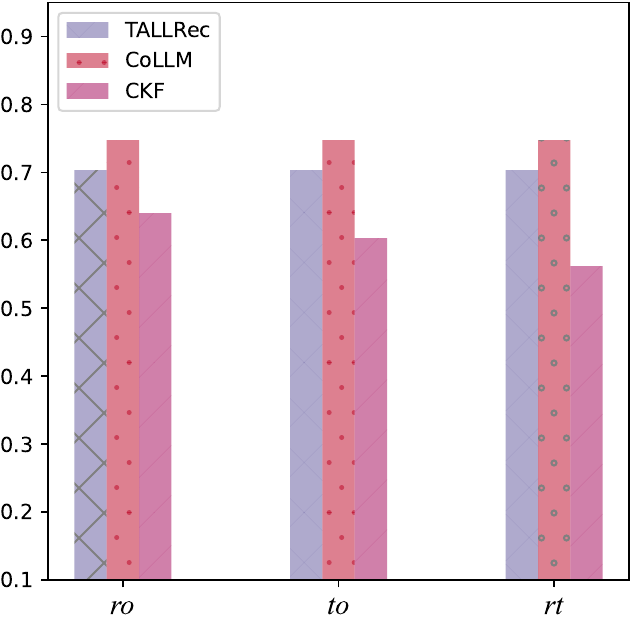}
%\caption{fig1}
\label{fig:sup:exp:ro:rps}
\end{minipage}%
}%
\subfigure[{CTR (AUC $\uparrow$)}]{
\begin{minipage}[t]{0.24\linewidth}
\centering
\includegraphics[width=\linewidth]{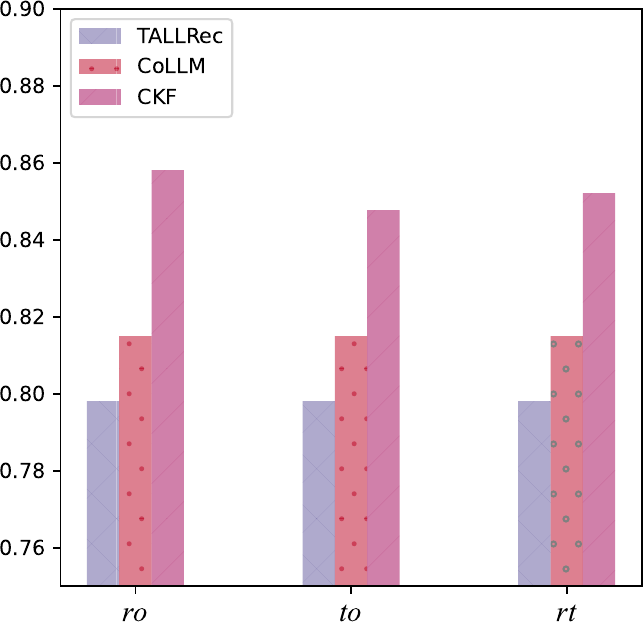}
%\caption{fig1}
\label{fig:sup:exp:ro:ctrs}
\end{minipage}%
}%
\subfigure[{Top-K (Hit@1-E $\uparrow$)}]{
\begin{minipage}[t]{0.24\linewidth}
\centering
\includegraphics[width=\linewidth]{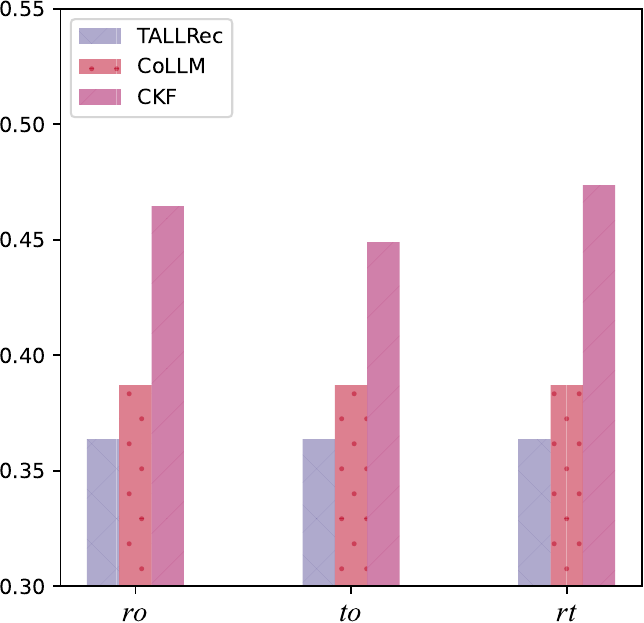}
%\caption{fig1}
\label{fig:sup:exp:ro:topks}
\end{minipage}%
}%
\subfigure[{Explain (ROUGE-L $\uparrow$)}]{
\begin{minipage}[t]{0.24\linewidth}
\centering
\includegraphics[width=\linewidth]{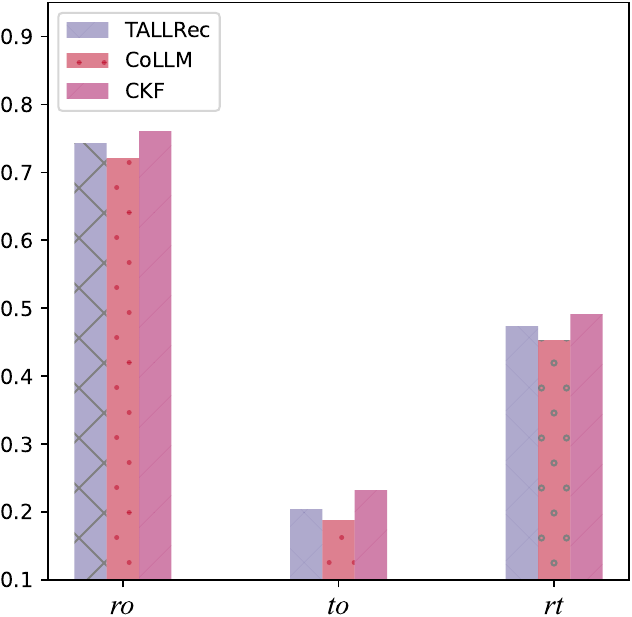}
%\caption{fig1}
\label{fig:sup:exp:ro:exps}
\end{minipage}%
}%
\centering
\caption{{Diverse explainable settings (rating only (ro) / text only (to) / rating \& text combination (rt)). }} 
\label{fig:sup:exp:ro}
\end{figure}

{According to the results in Figure~\ref{fig:sup:exp:ro}, we identify three significant findings. First, the combined use of ratings and explanations enhances performance compared to the text-only approach. On one hand, this combination compels the model to simultaneously analyze the semantic relationships within items and the sentiments expressed in the comments. On the other hand, this extended multi-task approach can also serve as a regularization mechanism [2], helping to prevent the model from overfitting.
Second, using text generation as the sole Explain task may lead to a deterioration in overall task performance. This issue may stem from a semantic gap between the text space and the item space. 
% Without the directional guidance provided by the rating task, the model might prioritize minimizing loss based on token associations within the text, rather than delving into the semantics of the items.
In this scenario, a significant disparity in the definition of Explain loss compared to other tasks can result in negative transfer, further impacting performance. The losses of other recommendation tasks are directly related to the relationship between items, while text generation only focuses on the co-occurrence between tokens.
CKF achieves leading performance in all three task settings by leveraging collaborative signals from multiple tasks and review data, creating a more complete user interest profile. This approach enhances the model's ability to resonate with its recommendation results. Additionally, we provide a case study in Section~\ref{sec:5.10} to intuitively demonstrate our superiority.
}

\subsection{{Recommendation Coverage}} %https://blog.csdn.net/weixin_42327752/article/details/123919143
{Beyond effectiveness, we also explore the potential of our algorithm for enhancing recommendation coverage~\cite{wu2016relevance,kunaver2017diversity}. This is crucial for the practicality of recommendation frameworks, as they can be susceptible to popularity bias.
We employ three common metrics~\cite{zhao2023fairness}: Coverage Ratio (CR), Shannon Entropy (SE), and Gini Index (Gini).
}

\begin{figure}[!h] % exclude hyper
\centering
\subfigure[{Coverage Ratio ($\uparrow$)}]{
\begin{minipage}[t]{0.32\linewidth}
\centering
\includegraphics[width=\linewidth]{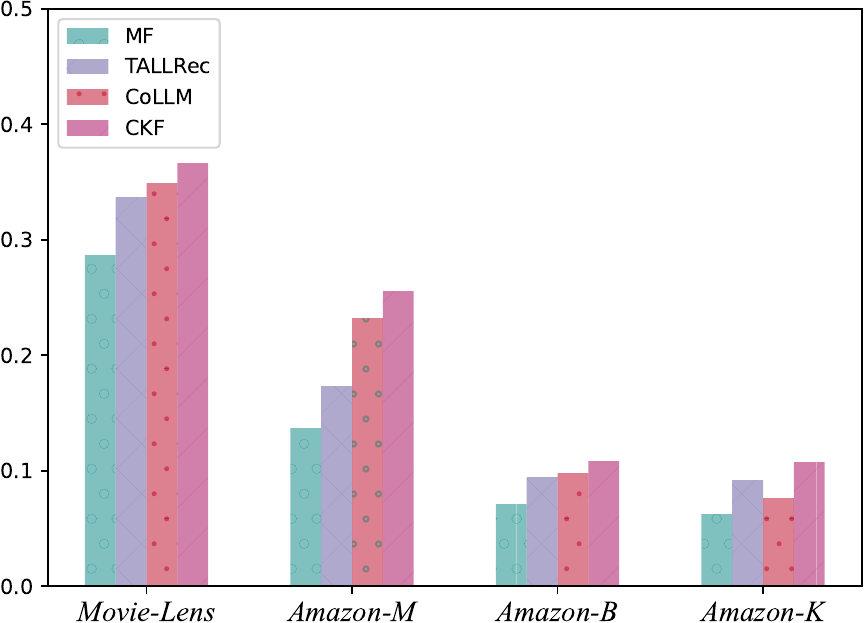}
%\caption{fig1}
\label{fig:sup:div:co}
\end{minipage}%
}%
\subfigure[{Shannon Entropy ($\uparrow$)}]{
\begin{minipage}[t]{0.32\linewidth}
\centering
\includegraphics[width=\linewidth]{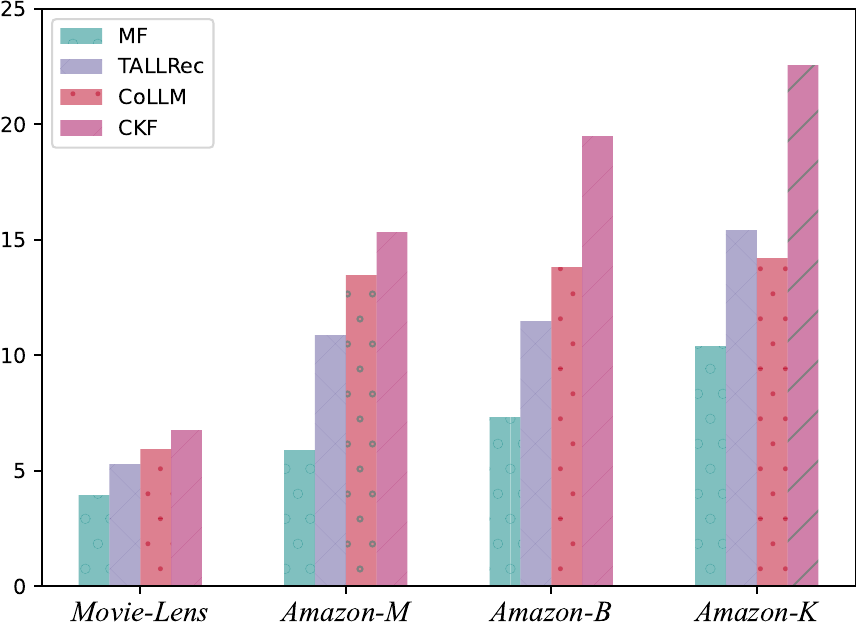}
%\caption{fig1}
\label{fig:sup:div:se}
\end{minipage}%
}%
\subfigure[{Gini Index ($\downarrow$)}]{
\begin{minipage}[t]{0.32\linewidth}
\centering
\includegraphics[width=\linewidth]{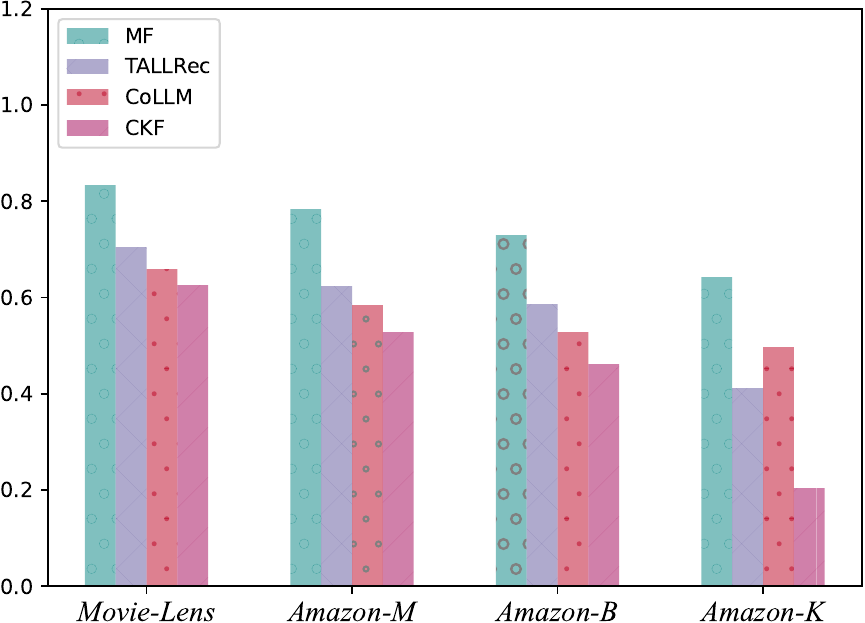}
%\caption{fig1}
\label{fig:sup:div:gn}
\end{minipage}%
}%
\centering
\caption{{Recommendation coverage. Note that the Coverage ratio here represents the ratio of all recommended items to the total number of items. In Shannon entropy and Gini index, probability is defined as the likelihood that a recommended item appears across all user recommendations.}}
\label{fig:sup:div}
\end{figure}
{According to Figure~\ref{fig:sup:div}, we identify two notable findings.  First, compared to traditional  CF-based methods that rely solely on collaborative signals, the LLM-based algorithms recommend more diverse items but lose accuracy. This is attributed to the extensive pre-trained knowledge within LLMs, which encompasses rich semantic associations. While this content-based association offers more diverse recommendations, it lacks personalized perception, which can lead to reduced accuracy.
Second, both CoLLM and our method successfully maintain diversity and accuracy, achieving SE improvements of up to 20\% on the Amazon Books data set. However, CoLLM is constrained by its focus on user preferences within a specific task and the mixed mapping function, which limits the representation semantics. This results in its limited diversity within the Amazon Kindles data set.
In contrast, our approach distinctly separates the mapping functions for each user and item and captures diverse user interests by incorporating user-item interactions across a broader spectrum of tasks.  This ensures our leading performance.
}

\subsection{Plug-in Application}\label{sec:5.7}
{ Our framework is flexible and can be improved as CF and LLM models advance. To demonstrate this, we create multiple} plug-in applications. Without loss of generality, we choose Amazon Books for robustness experiments as it meets the data requirements for the four tasks.

\begin{figure*}[!h]
\centering
\subfigure[RP (MAE $\downarrow$)]{
\begin{minipage}[t]{0.24\linewidth}
\centering
\includegraphics[width=\linewidth]{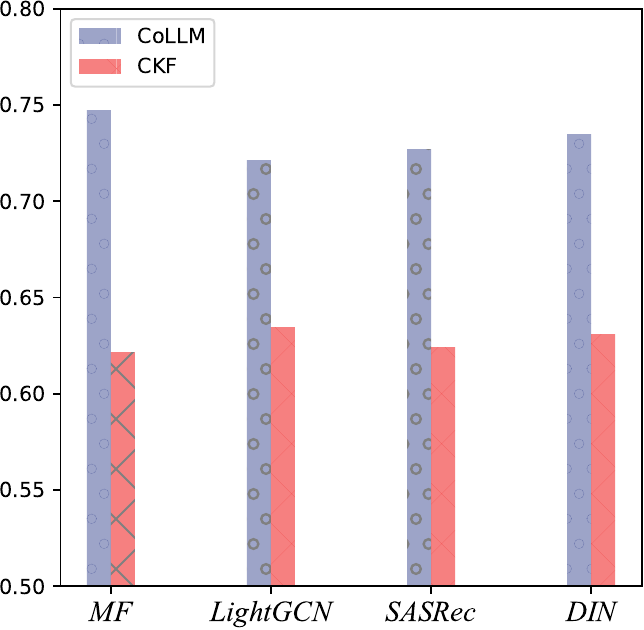}
%\caption{fig1}
\end{minipage}%
}%
\subfigure[CTR (AUC $\uparrow$)]{
\begin{minipage}[t]{0.24\linewidth}
\centering
\includegraphics[width=\linewidth]{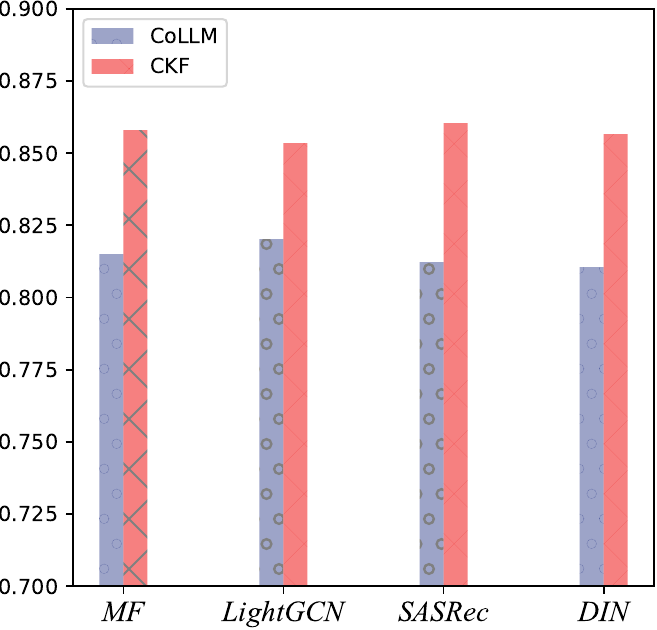}
%\caption{fig1}
\end{minipage}%
}%
\subfigure[Top-K (Hit@1-E $\uparrow$)]{
\begin{minipage}[t]{0.24\linewidth}
\centering
\includegraphics[width=\linewidth]{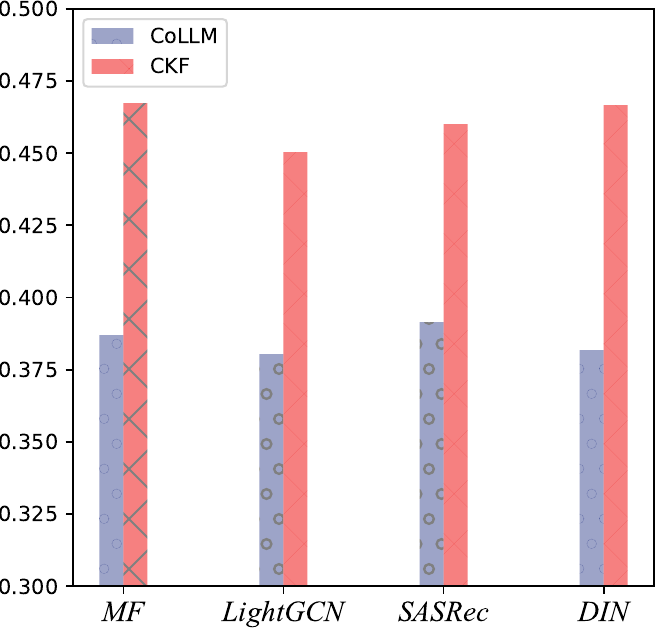}
%\caption{fig1}
\end{minipage}%
}%
\subfigure[Explain (MAE $\downarrow$)]{
\begin{minipage}[t]{0.24\linewidth}
\centering
\includegraphics[width=\linewidth]{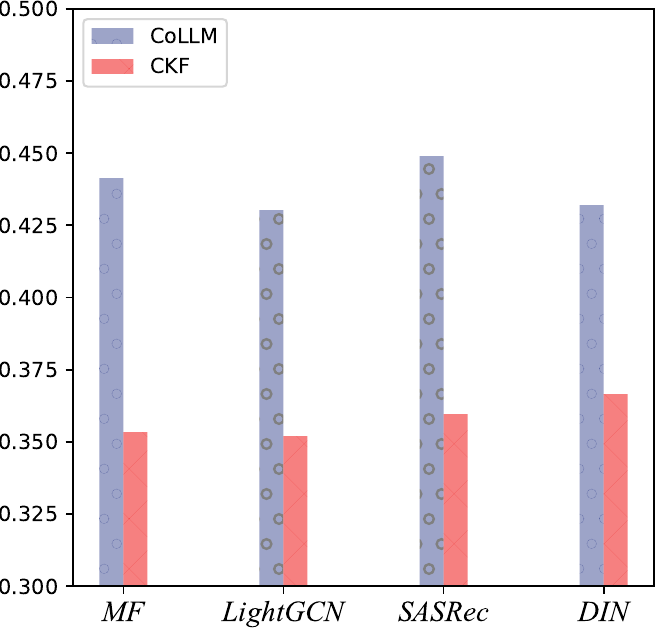}
%\caption{fig1}
\end{minipage}%
}%
\centering
\caption{Plug-in application (Different CF models). We select MF, LightGCN, SASRec, and DIN as the CF model for CoLLM and CKF, as they both use collaborative knowledge to enhance LLM.}
\label{fig:plugcf}
\end{figure*}

% \vskip\baselineskip
\begin{figure*}[!h]
\centering
\subfigure[RP (MAE $\downarrow$)]{
\begin{minipage}[t]{0.24\linewidth}
\centering
\includegraphics[width=\linewidth]{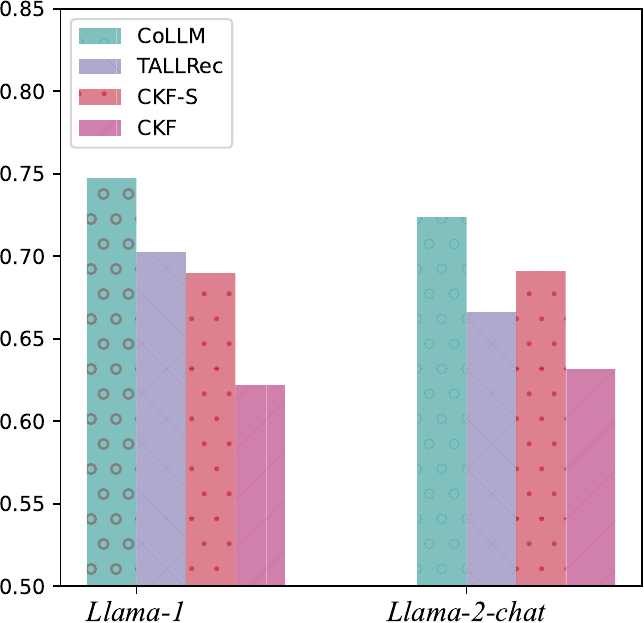}
%\caption{fig1}
\end{minipage}%
}%
\subfigure[CTR (AUC $\uparrow$)]{
\begin{minipage}[t]{0.24\linewidth}
\centering
\includegraphics[width=\linewidth]{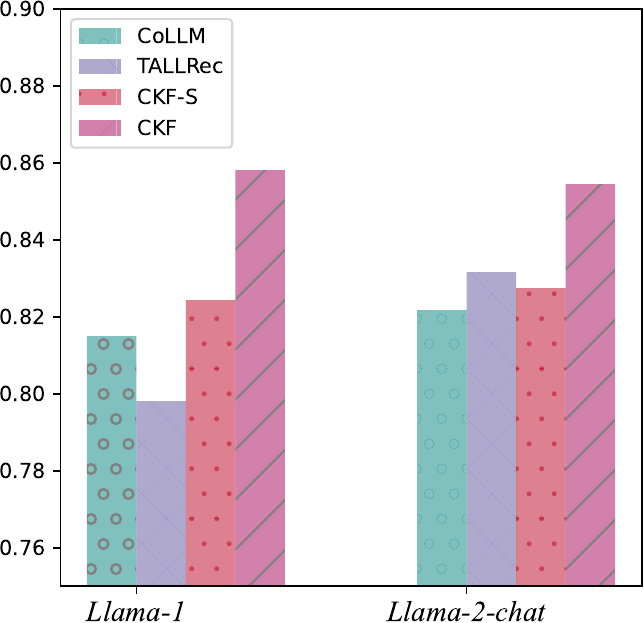}
%\caption{fig1}
\end{minipage}%
}%
\subfigure[Top-K (Hit@1-E $\uparrow$)]{
\begin{minipage}[t]{0.24\linewidth}
\centering
\includegraphics[width=\linewidth]{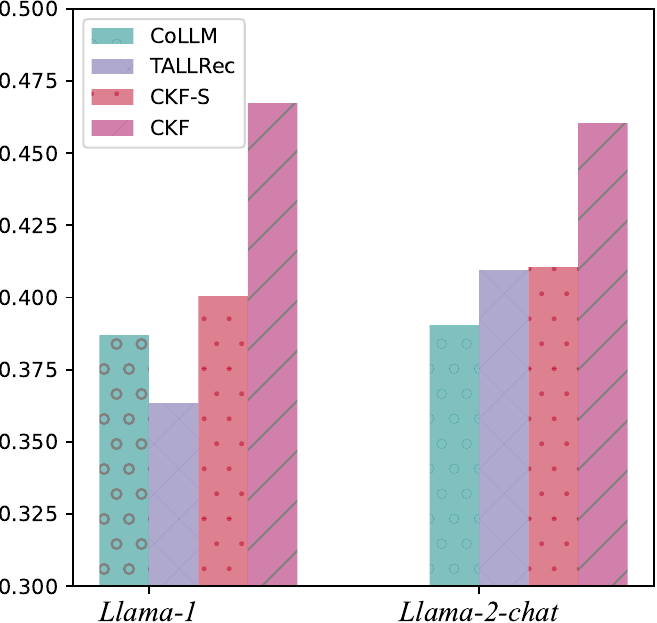}
%\caption{fig1}
\end{minipage}%
}%
\subfigure[Explain (MAE $\downarrow$)]{
\begin{minipage}[t]{0.24\linewidth}
\centering
\includegraphics[width=\linewidth]{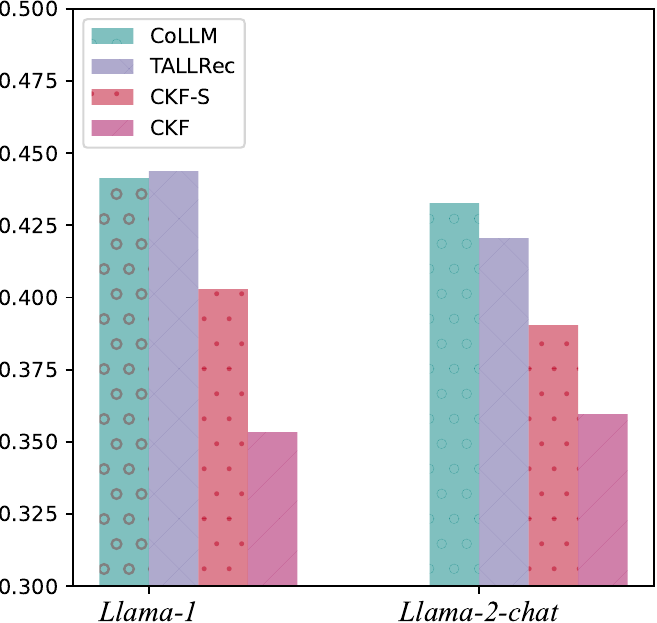}
%\caption{fig1}
\end{minipage}%
}%
\centering
\caption{Plug-in application (Different LLMs). We select Llama1 and Llama2-chat as the LLM backbone. We opt for TALLRec, CoLLM, CKF-S, and CKF for our comparative analysis due to their robust performance among LLM-based baselines.}
\label{fig:plugllm}
\end{figure*}

\subsubsection{Diverse CF Models}
% As technological advancements progress, the capability to extract collaborative knowledge will be further enhanced, owing to the improved representational performance of CF models.
To test the robustness to different CF models, we choose to work with both classic static CF models, including MF and LightGCN, as well as dynamic CF models like SASRec and DIN.
As shown in Figure~\ref{fig:plugcf}, the dynamic variant shows marginal improvement over the static one, with DIN even underperforming compared to MF in CTR and Top-K. This could be attributed to the insufficient timing length and a dearth of timing information.
SASRec-based variant exhibits superior performance in most tasks, which can be attributed to the powerful self-attention mechanism.
It is worth noting that CKF has a greater improvement than CoLLM in various CF model scenarios, showing our superiority. 
% This plug-in experiment demonstrates the universality and potential application value of CKF.

\subsubsection{Diverse LLMs}
Our framework imposes no constraints on the LLM; however, the representational capabilities of the LLM do set the upper bound for the model's performance. We choose Llama1~\cite{llama} and Llama2-chat~\cite{llama2} as the respective backbones for our study, with the results presented in Figure~\ref{fig:plugllm}.
Owing to Llama2-chat's use of a greater number of parameters and a broader spectrum of pre-training data, it possesses a more robust understanding of text semantics, which translates into superior performance.
The rapid advancements in LLMs have led us to consider employing more powerful models for future testing. This adaptability underscores the broad applicability of our framework across various scenarios.

\subsection{Ablation Study}\label{sec:5.3}
To investigate the efficacy of each sub-module, we conduct a series of ablation studies in the Amazon Books data set, isolating each module while maintaining all other components {consistently}.
Table~\ref{tab:aba} clearly illustrates that the CKF model surpasses all ablation variants across multiple tasks, consistently achieving a minimum improvement of approximately 1\%.
CKF-NML shows the most significant performance drop, likely due to task-specific semantic differences and mutual interference from a uniform fine-tuning scheme. CKF-NPM and CKF-NCK exhibit minimum performance degradations of 3\% and 2\%, respectively. CKF-NPM's decline is due to projecting users and items embeddings using the same mapping function, overlooking semantic differences, whereas CKF-NCK's reduction stems from failing to perceive collaborative knowledge. {CKF-TLM explicitly distinguishes user and item semantics but shows degraded performance. This shortcoming is attributed to the limited mapping power of the linear layer and its reduced ability to generalize to new users or items in the test set.}
% CKF-NEN stands out as the strongest baseline; however, it adopts two stages like CoLLM, which may cause the optimization goals to be inconsistent.
{CKF-NEN is a strong baseline, but its two-stage training leads to a significant drop in performance on the RP \& Explain tasks.
As noted in Section~\ref{sec:intro}, the absence of a weighting strategy may enable the model to circumvent rigorous training of LLM and instead rely directly on CF signals for scoring. However, CF signals are inherently coarse-grained, primarily reflecting co-occurrence relationships, which are more suited for straightforward tasks like CTR and Top-K, where the goal is to distinguish between positive and negative preferences. In contrast, collaborative signals are less direct for RP and Explain tasks, necessitating  LLM to accurately grasp text semantics and analyze user preferences in a more nuanced manner.
}
CKF-S in Table~\ref{tab:exp1}-\ref{tab:exp-sup} could be regarded as a variant using different Lora for different tasks, but it falls short in effectively leveraging the abundant supervisory signals offered by multi-task learning and {employing} more parameters.

To sum up, each sub-module plays an indispensable role in CKF, contributing to its final superiority {over other baselines}.

\begin{table}[!h]
\centering
\caption{Ablation study. CKF-NCK does not incorporate collaborative knowledge and only performs multi-task tuning. CKF-NPM employs a unified mapping function for both the user and item rather than {treating} them separately. 
{CKF-TLM distinguishes between users and items and employs two distinct linear layers for collaborative signal mapping.}
CKF-NML leverages the same Lora for each task. CKF-NEN jointly trains Multi-Lora and two meta networks without applying the smooth weighting strategy.}
\label{tab:aba}
\resizebox{0.75\textwidth}{!}{
\begin{tabular}{c|c|ccccc||c} 
\hline
Task                     & Metric & CKF-NCK & CKF-NPM & {CKF-TLM} & CKF-NML & CKF-NEN & CKF  \\ 
\hline
\multirow{2}{*}{RP}      & MAE $\downarrow$   &   0.6894   & 0.7313  & {0.6530}   & 0.7095     &  0.7206     &   \textbf{0.6289}   \\
                         & MSE $\downarrow$   &    1.1856     &     1.2165  &{1.0387}  &     1.0601    &    1.2006   & \textbf{0.9841}      \\ 
\hline\hline
\multirow{2}{*}{CTR}     & AUC $\uparrow$    &    0.8433     &      0.8332 &{0.8506}  &  0.8203       &     0.8467    &  \textbf{0.8579}    \\
                         & U-AUC $\uparrow$   &    0.8319     &    0.8219  &{0.8397}   &  0.8123       &    0.8423     &  \textbf{0.8452}    \\ 
\hline\hline
\multirow{2}{*}{Top-K}   & Hit@1-E $\uparrow$   &    0.4146     &    0.4047   &{0.4593}  &      0.4494   &    0.4634     &  \textbf{0.4673}    \\
                         & Hit@1-H $\uparrow$  &   0.2610      &    0.2602   &{0.2877}  &   0.2676      &   0.2710      & \textbf{0.2954}     \\ 
\hline\hline
\multirow{2}{*}{Explain} & MAE $\downarrow$   &     0.4063    &      0.4122  &{0.3871} &       0.4026  &   0.4002      & \textbf{0.3534}     \\
                         & MSE $\downarrow$   &      0.4901   &    0.4986 &{0.4732}     &    0.4823     &    0.4801     &  \textbf{0.4269}    \\
\hline
\end{tabular}}
\end{table}

\subsection{T-sne Visualization}
\begin{figure*}[!h]
\centering
\subfigure[CoLLM-mapping]{
\begin{minipage}[t]{0.24\linewidth}
\centering
\includegraphics[width=\linewidth]{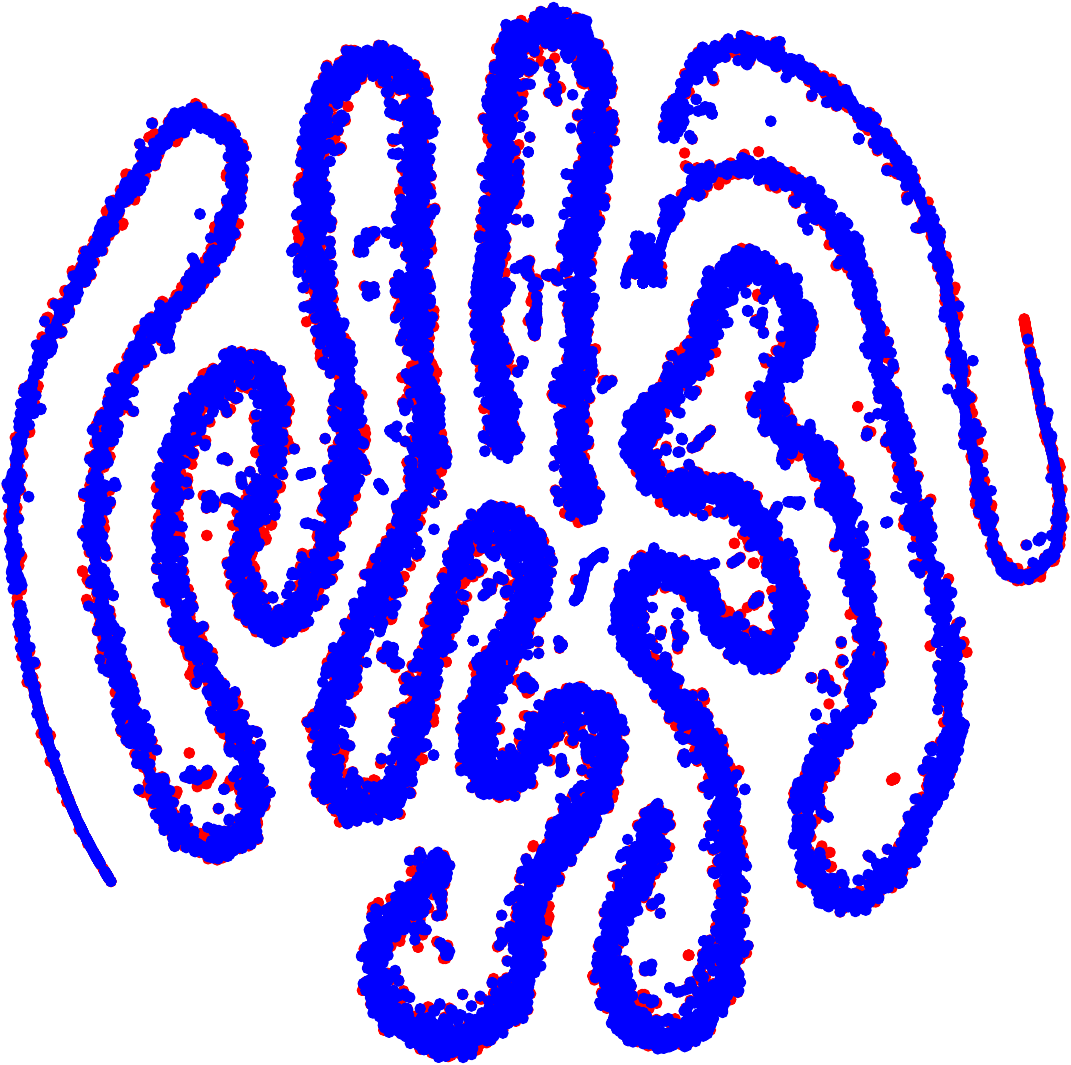}
\label{fig:tsne:co}
\end{minipage}%
}%
\subfigure[CKF-mapping]{
\begin{minipage}[t]{0.24\linewidth}
\centering
\includegraphics[width=\linewidth]{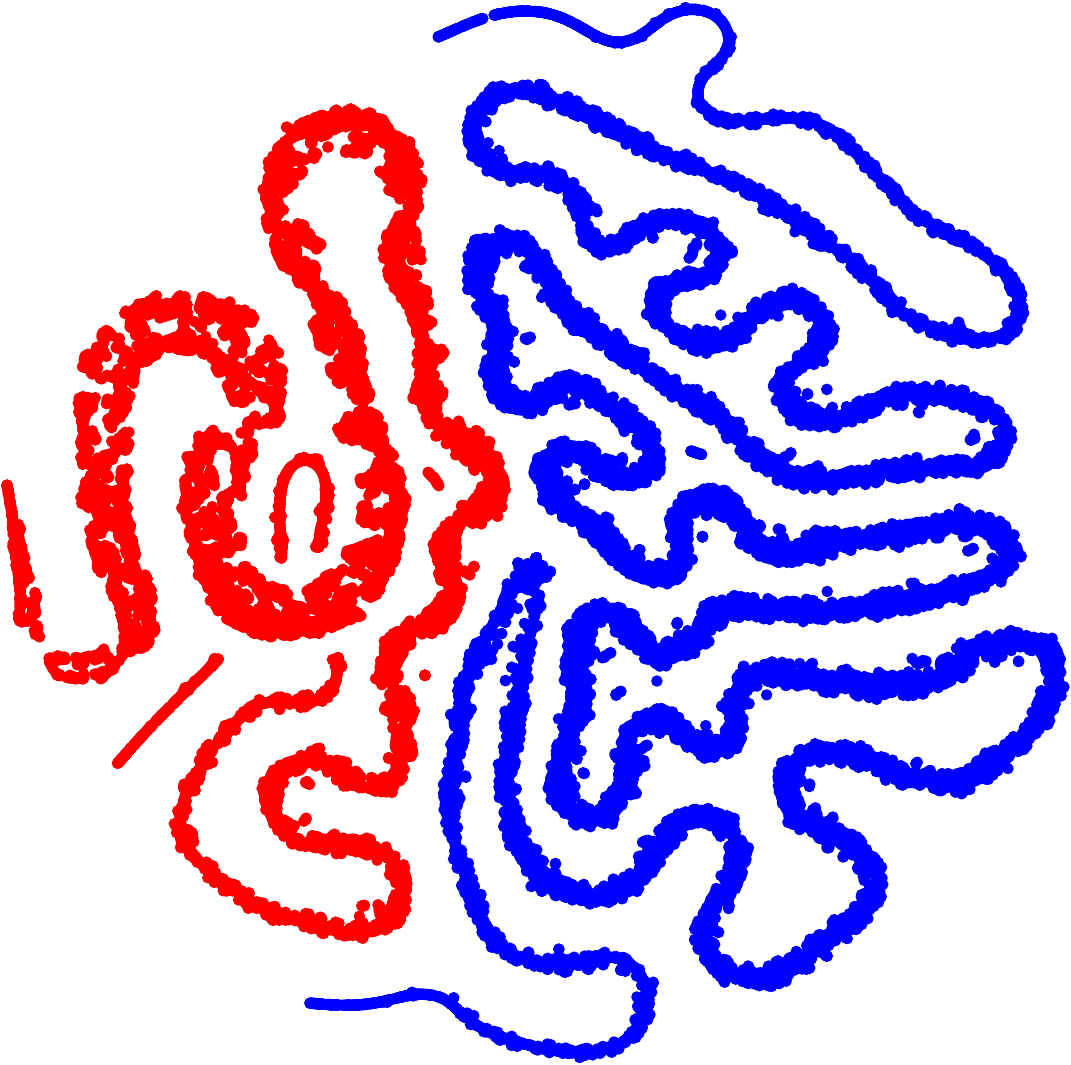}
\label{fig:tsne:ck}
\end{minipage}%
}%
\subfigure[Lora-CKF]{
\begin{minipage}[t]{0.24\linewidth}
\centering
\includegraphics[width=\linewidth]{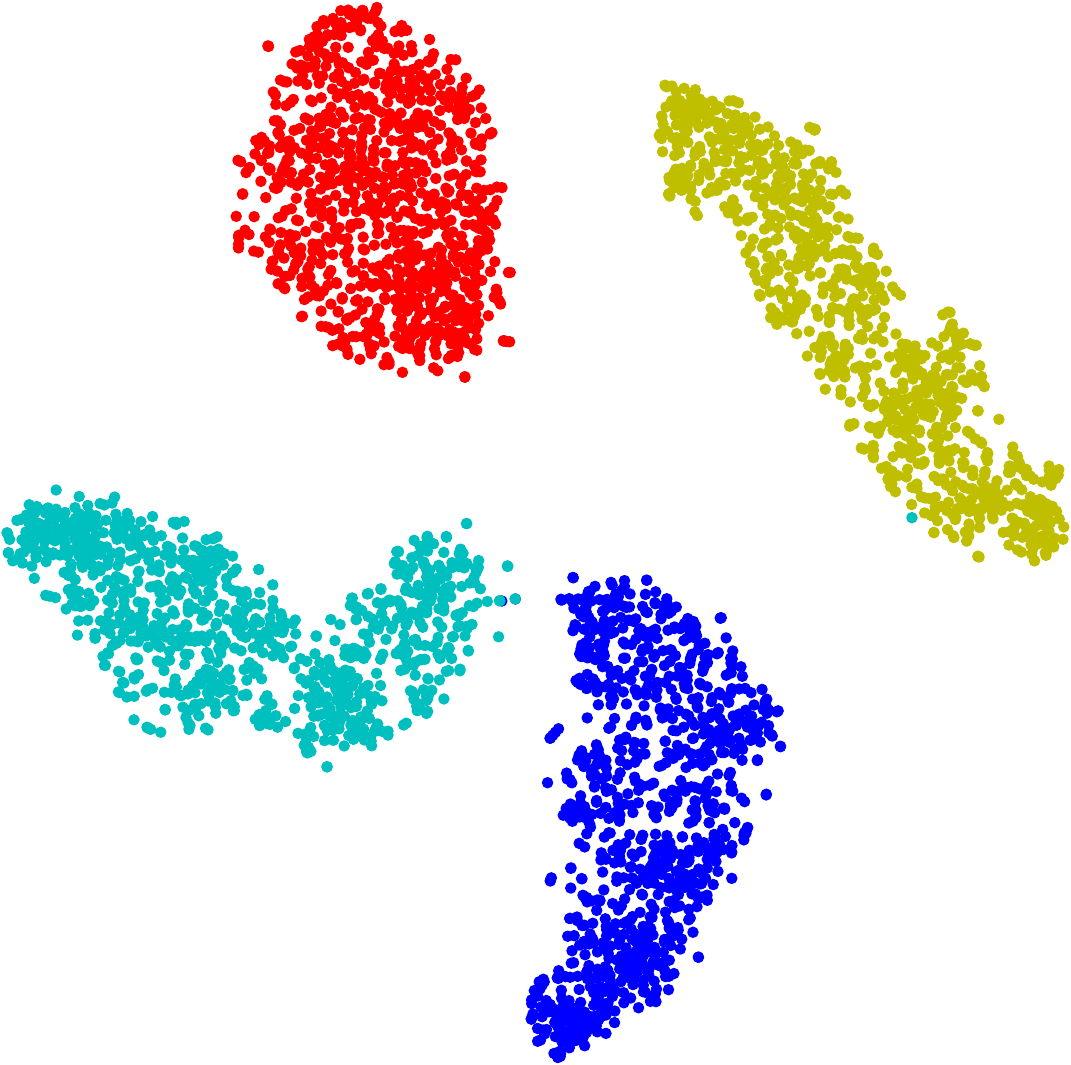}
\label{fig:tsne:losh}
\end{minipage}%
}%
\subfigure[Lora-same Lora]{
\begin{minipage}[t]{0.24\linewidth}
\centering
\includegraphics[width=\linewidth]{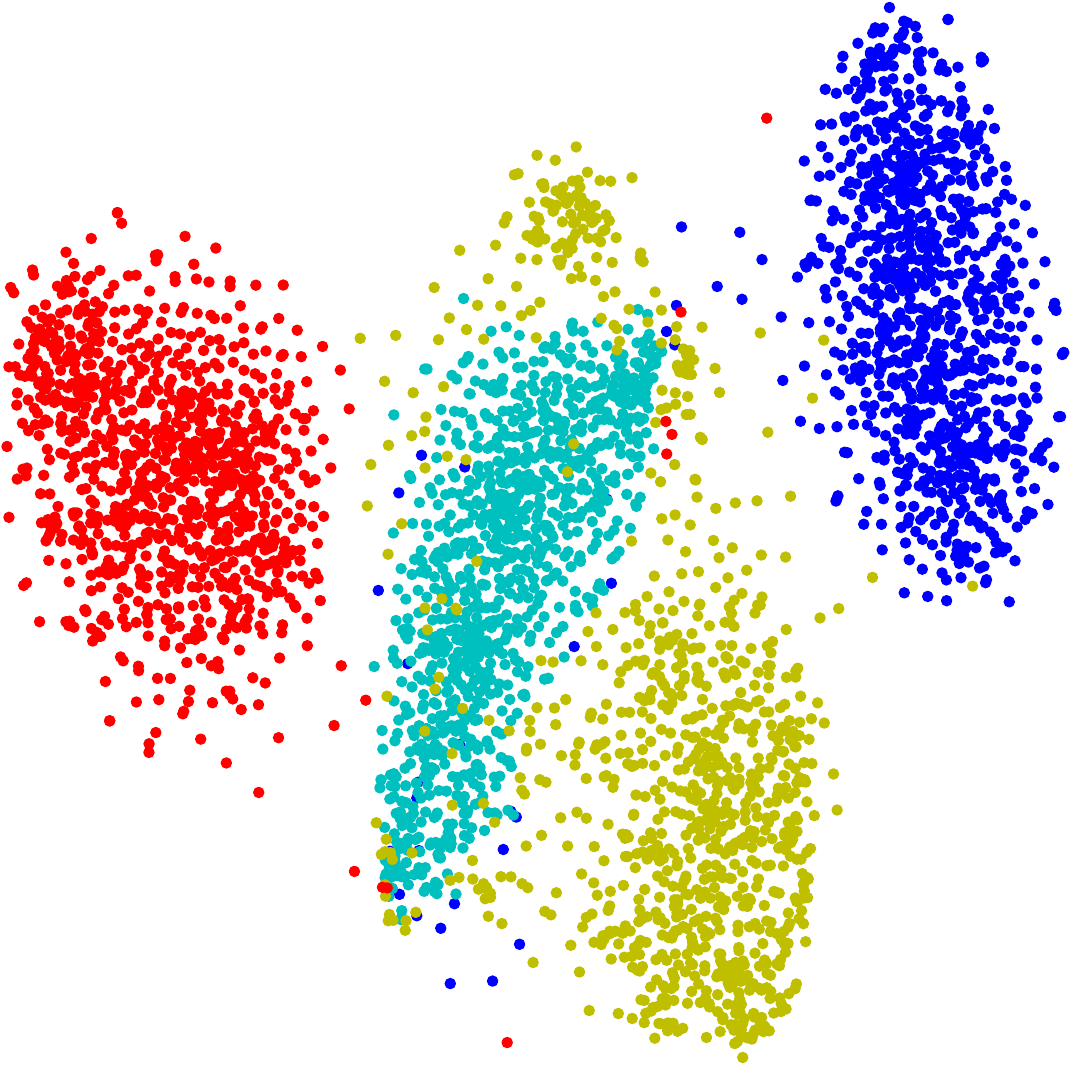}
\label{fig:tsne:losp}
\end{minipage}%
}%
\centering
\caption{T-sne visualization. For (a) and (b), the colors red and blue represent users and items, respectively. For (c) and (d), colors correspond to the prompt representations of distinct tasks (Green for RP, Red for CTR, Blue for Top-K, and Yellow for Explain).}
\label{fig:tsne}
\end{figure*}

In order to augment the model's interpretability and offer deeper insights into the function of its modules, we conduct T-sne visualizations~\cite{zhao2023sequential} in the Amazon Books data set to illustrate the user-item mapping both before and after the implementation of CKF and CoLLM.
Figure~\ref{fig:tsne:ck} demonstrates that within CKF, users and items exhibit semantic segregation and form smaller clusters. This phenomenon stems from CKF's deliberate utilization of distinct meta-networks for the personalized mapping of collaborative knowledge. As shown in Figure~\ref{fig:tsne:co}, CoLLM, utilizing a singular mapping function, blurs the distinction between item and user spaces, introducing noise into the system. We contend that this ambiguity can lead LLMs to potentially misinterpret the semantics of textual prompts, resulting in inaccurate recommendation outcomes.

We further visualize the user prompts post-Lora layer processing, as illustrated in Figure~\ref{fig:tsne:losh}, to assess the efficacy of the multi-Lora approach. Evidently, this strategy enables a more distinct differentiation of task-level representations compared to the same Lora application depicted in Figure~\ref{fig:tsne:losp}. This intuitively demonstrates that employing a uniform Lora may lead to interference among tasks' signals.

% We additionally conduct a visualization of the embeddings following task-shared $\Delta\varphi_{share}$ and task-specific $\Delta\varphi_{k}$ Lora projections. It is evident that the representations from the former are more uniform, suggesting its effectiveness in extracting commonalities between tasks. Post-$\Delta\varphi_{k}$, the expanded representation distances among tasks suggest their efficacy in capturing each task's distinct characteristics.
\subsection{Case Study}\label{sec:5.10}
\begin{figure*}[!ht]
  \centering
   \includegraphics[width=\linewidth]{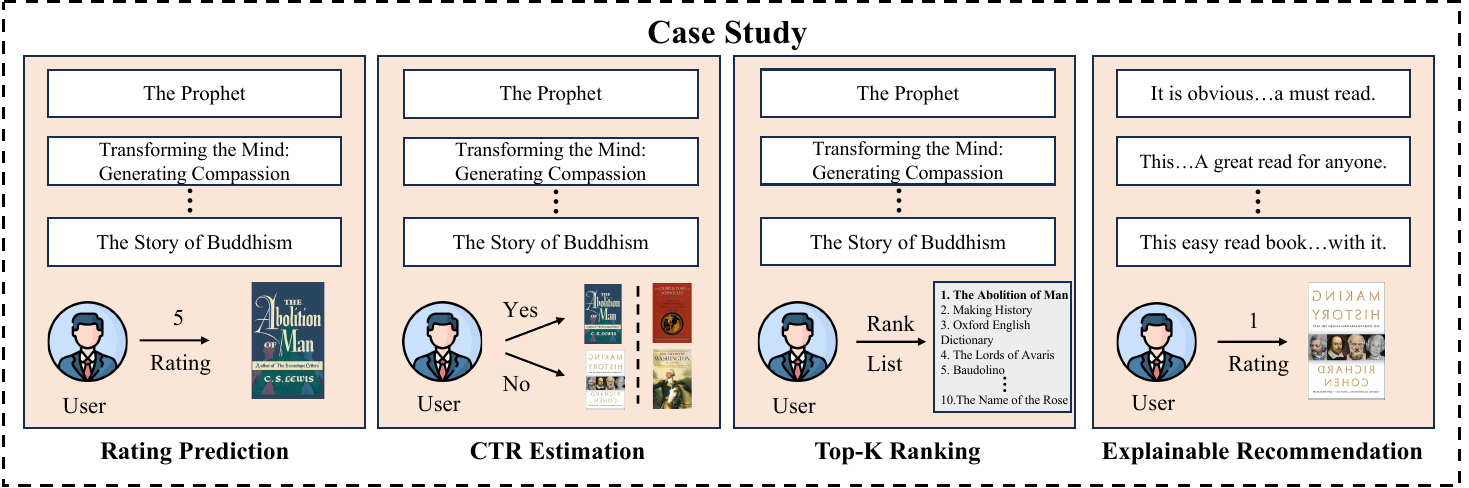}
   \caption{Case Study. We randomly selected a user in the Amazon Books data set for multi-task prediction.
   The first three tasks employ identical title sequences, yet they exhibit substantial differences in their task instructions. For the explainable recommendation task, the input utilized is the sequence of comments.}
   \label{fig:case}
\end{figure*}

% 这里需要
To effectively illustrate the efficacy of our recommendation system, we conduct a demonstration involving a randomly chosen user, presenting the outcomes from four distinct recommendation tasks, as shown in Figure~\ref{fig:case}. In the illustrated case, the user displays a marked inclination towards books within the religious and philosophical genres, as evidenced by his interaction history with titles like "The Prophet" and "The Story of Buddhism." Our approach adeptly suggest{s} books of a similar nature, such as "The Abolition of Man," and accurately distinguish{es} between preferred genres and less relevant ones, such as historical books, in its Top-K Ranking and CTR estimations. This showcases our method's ability to discern correlations between items by analyzing user-item interactions and the semantic content of item titles. Significantly, our algorithm's capacity to analyze user interests through review sequences for the generation of explainable recommendations allows for the precise prediction of user preferences at a granular level, highlighting its adeptness at recognizing user preferences across varied contexts.

\begin{figure*}[!ht]
  \centering
   \includegraphics[width=\linewidth]{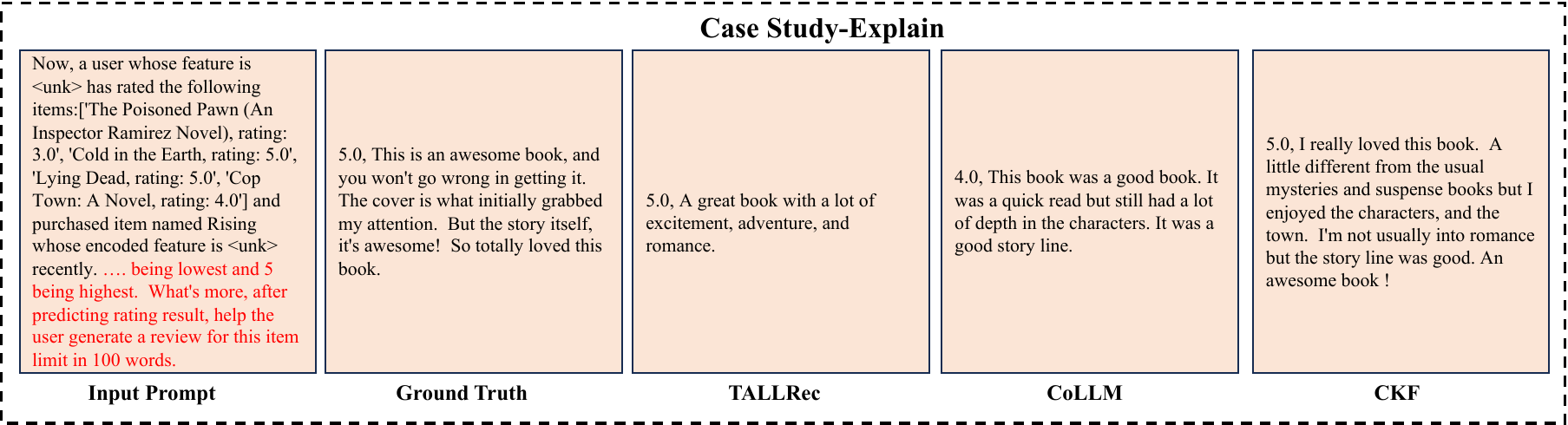}
   \caption{{Explanation generation. "Ground Truth" refers to the real comments written by users on the predicted item.}}
   \label{fig:case:sup}
\end{figure*}

{To further elucidate CKF’s insights into user interests, we analyze the explanations generated when both ratings and text are considered, as proposed in Section~\ref{sec:5.5.3}. As shown in Figure~\ref{fig:case:sup}, our model not only generates accurate ratings but also provides more detailed and precise explanations. From the text generated by CKF, we can see that the main reason for choosing the book "Rising" is its story, even though the user does not like the romance genre. This demonstrates that CKF is capable of understanding the user's true intentions rather than merely relying on co-occurrence relationships. In contrast, TALLRec's explanation is unreasonable, and CoLLM's rating prediction is incorrect. This confirms that CKF can trace the reasons behind its recommendations, demonstrating greater transparency and credibility. }

Overall, this case validates that our algorithm is capable of handling various recommendation tasks based on a single inference, demonstrating the effectiveness {and reasoning power} of our fine-tuning approach and highlighting the benefits of multi-task recommendation strategies.

\section{Hyper Parameter Testing (RQ3)}\label{sec:6}
\begin{figure}[!h] % exclude hyper
\centering
\subfigure[Meta Network size $d$]{
\begin{minipage}[t]{0.3\linewidth}
\centering
\includegraphics[width=\linewidth]{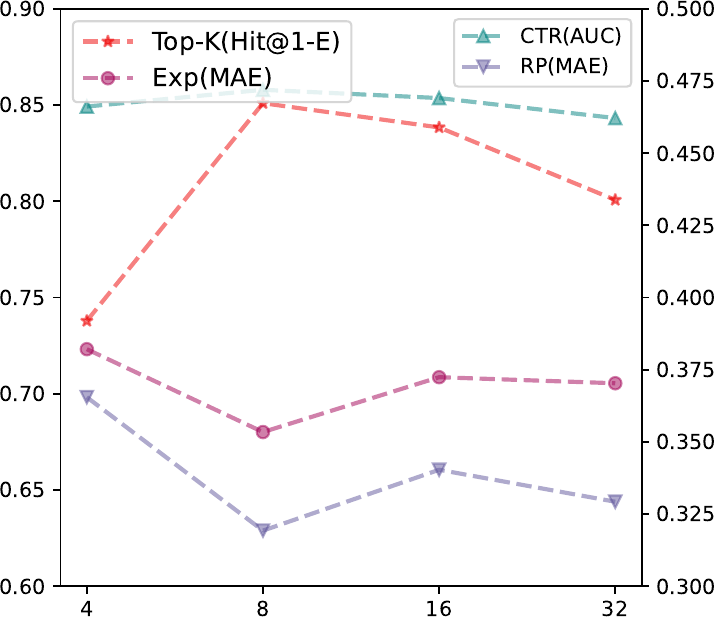}
%\caption{fig1}
\label{fig:hyper-d}
\end{minipage}%
}%
\subfigure[Lora rank $\tilde{r}$]{
\begin{minipage}[t]{0.3\linewidth}
\centering
\includegraphics[width=\linewidth]{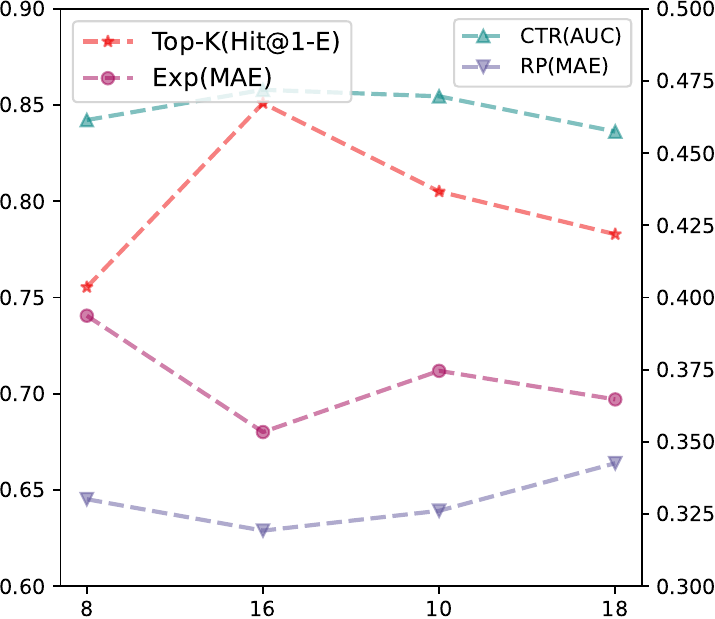}
%\caption{fig1}
\label{fig:hyper-r}
\end{minipage}%
}%
\subfigure[Temperature $\tau$]{
\begin{minipage}[t]{0.3\linewidth}
\centering
\includegraphics[width=\linewidth]{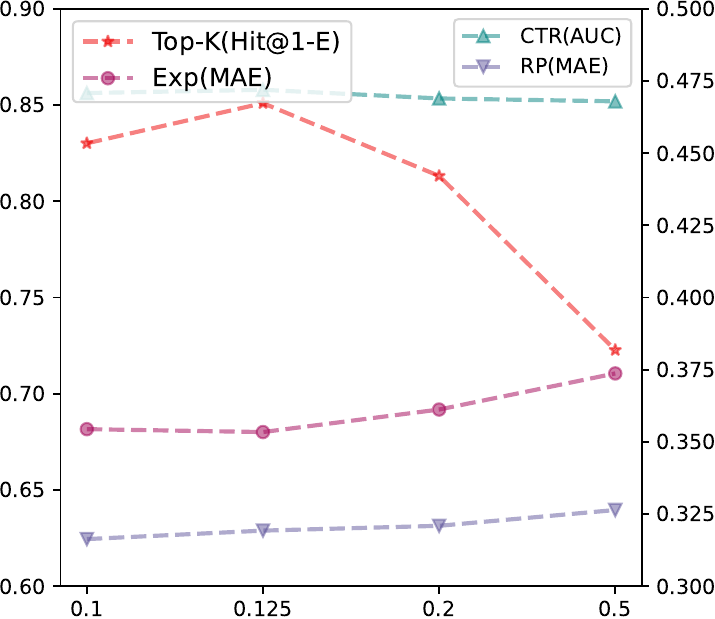}
%\caption{fig1}
\label{fig:hyper-tau}
\end{minipage}%
}%
\centering
\caption{Hyper testings in four recommendation tasks. The tuning range of $d$ is \{4,8,16,32\}, $\tilde{r}$ is \{8,10,16,18\}, and $\tau$ is \{0.1, 0.125,0.2,0.5\}}
\label{fig:hyper}
\end{figure}

To obtain the best performance and observe the impact of necessary parameters, we perform necessary tuning on key parameters, as shown in Figure~\ref{fig:hyper}.
{Given the constraints of training time and 
computational costs, following~\cite{bao2023bi,zhang2024m3oe,he2020lightgcn}, we adjust one parameter at a time.}

\subsection{Meta Network Size $d$} 
The size of the meta-network's embeddings signifies the scope of the semantic space where the model generates mapping functions for collaborative knowledge. A larger embedding size indicates a more robust generative capability, as shown in Figure~\ref{fig:hyper-d}. Optimal results are achieved when $d=8$. Excessively large sizes, however, risk leading to overfitting.

\subsection{Lora Rank $\tilde{r}$} 
The essential idea of Lora is to change the update{d} matrix of the model into the multiplication of two low-rank matrices to optimize the model effectively and cost-{efficiently}. The larger the rank, the greater the memory burden it will bring. 
% Figure~\ref{fig:hyper-r} demonstrates that beyond the point of $\tilde{r}=16$, elevating the rank of Lora does not yield substantial improvements in performance.
{Considering computational constraints, we test four Lora parameters and find that CKF performs best when $\tilde{r}=16$, as shown in Figure~\ref{fig:hyper-r}.
Please note that this does not contradict the intuition that increasing the number of Lora parameters generally improves performance. The model's performance is influenced by a combination of all hyperparameters, not solely by the setting of the Lora parameter. We acknowledge that a larger Lora rank may lead to better performance, but this exceeds our computing power and goes against our goal of efficiency.
% It is crucial to understand that the performance of the model is influenced by a combination of all hyperparameters, not solely by the setting of the LoRA parameter. Isolated adjustments to the LoRA rank, without considering the interplay with other parameters, could result in overfitting or a degradation in performance.
}

\subsection{Temperature $\tau$}
The temperature coefficient exerts a considerable influence on the curve change of smooth weighting function. As it increases, the corresponding curve steepens, reflecting an increased step degree and allocating greater weight to the training signal integrated with collaborative knowledge later. Experimentally, as shown in Figure~\ref{fig:hyper-tau}, we set $\tau=0.125$.
\section{Conclusion}\label{section7}
% overall
In this paper, we propose CKF, a cutting-edge LLMs-based recommendation framework designed to assimilate collaborative knowledge effectively.
We engineer two innovative meta-networks to align collaborative knowledge with the semantic space of LLM, and subsequently develop a novel fine-tuning approach termed Multi-Lora, designed to discern the inherent interrelations among various recommendation tasks.
Additionally, we introduce a smooth weighting strategy to balance the training process of LLM and meta-networks.
Extensive experiments and comprehensive robustness analysis on four tasks derived from {four} data sets confirm the effectiveness and state-of-the-art of our approach.
Nevertheless, CKF still has some limitations, which we defer to future research. {First, extending our multi-task framework to multiple scenarios is an interesting topic. One potential solution is to use transfer learning to mitigate domain gaps and facilitate knowledge transfer between different scenarios. Another direction is to integrate multimodal or knowledge graph information into this framework. This can be achieved by employing large-scale pre-trained visual models or advanced graph neural networks to extract representations and incorporate them into the new prompt template, similar to the current approach.}
% \begin{acks}
% To Robert, for the bagels and for explaining CMYK and color spaces.
% \end{acks}

\bibliographystyle{ACM-Reference-Format}
\bibliography{main}

% \appendix

\end{document}